\documentclass[fleqn,usenatbib]{mnras}

\usepackage{newtxtext,newtxmath}

\usepackage[T1]{fontenc}
\usepackage{ae,aecompl}

\usepackage{graphicx}	
\usepackage{amsmath}	
\usepackage{amssymb}	

\newcommand{\qso}{SDSS~J074749.17+115352.4}	

\newcommand{\sbline}{$\rm erg~s^{-1}~cm^{-2}~arcsec^{-2}$}

\title[Into the jungle]{Into the Ly$\alpha$ jungle: exploring the circumgalactic medium of galaxies at $z\sim4-5$ with MUSE}

\author[R. M. Bielby et al.]{Richard M. Bielby,$^{1,2}$\thanks{E-mail: rmbielby@gmail.com (RMB)}
Michele Fumagalli,$^{1,3,4}$
Matteo Fossati,$^{1,3}$
Marc Rafelski,$^{5,6}$ \and
Benjamin Oppenheimer,$^{7}$
Sebastiano Cantalupo,$^{8}$
Lise Christensen,$^{9}$
J. P. U. Fynbo,$^{10,11}$ \and
Sebastian Lopez,$^{12}$
Simon L. Morris,$^{1}$
Valentina D'Odorico,$^{13,14}$
Celine Peroux,$^{15,16}$
\\
$^{1}$Centre for Extragalactic Astronomy, Durham University, South Road, DH1 3LE\\
$^{2}$Institute for Data Science, Durham University, South Road, DH1 3LE\\
$^{3}$Institute for Computational Cosmology, Durham University, South Road, DH1 3LE\\
$^{4}$Dipartimento di Fisica G. Occhialini, Universit\`a degli Studi di Milano-Bicocca, Piazza della Scienza 3, 20126 Milano, Italy \\
$^{5}$Space Telescope Science Institute, 3700 San Martin Drive, Baltimore, MD 21218, USA\\
$^{6}$Department of Physics and Astronomy, Johns Hopkins University, Baltimore, MD 21218, USA\\
$^{7}$CASA, Department of Astrophysical and Planetary Sciences, University of Colorado, 389 UCB, Boulder, CO 80309, USA\\
$^{8}$Department of Physics, ETH Zurich, Wolgang-Pauli-Strasse 27, 8093, Zurich\\
$^{9}$Dark Cosmology Centre, Niels Bohr Institute, University of Copenhagen, Juliane Maries Vej 30, DK-2100 Copenhagen, Denmark\\
$^{10}$The Cosmic Dawn Center (DAWN), Vibenshuset, Lyngbyvej 2, DK-2100 Copenhagen, Denmark\\
$^{11}$Niels Bohr Institute, Lyngbyvej 2, DK-2100 Copenhagen, Denmark\\
$^{12}$Departamento de Astronom\'ia, Universidad de Chile, Casilla 36-D, Santiago, Chile\\
$^{13}$INAF, Osservatorio Astronomico di Trieste, Via G. B. Tiepolo 11, 34143 Trieste, Italy\\
$^{14}$Scuola Normale Superiore, Piazza dei Cavalieri 7, 56126 Pisa, Italy\\
$^{15}$European Southern Observatory (ESO), Karl-Schwarzschild-Str.2, D-85748 Garching b. M{\"u}nchen, Germany\\
$^{16}$Aix Marseille Universit\'{e}, CNRS, LAM (Laboratoire d'Astrophysique de Marseille) UMR 7326, 13388, Marseille, France
}

\begin{document}
\label{firstpage}
\pagerange{\pageref{firstpage}--\pageref{lastpage}}
\maketitle

\begin{abstract}
We present a study of the galaxy environment of 9 strong H~{\sc i}+C~{\sc iv} absorption line systems ($16.2<{\rm log}(N({\rm HI}))<21.2$) spanning a wide range in metallicity at $z\sim4-5$, using MUSE integral field and X-Shooter spectroscopic data collected in a $z\approx 5.26$ quasar field. We identify galaxies within  a 250~kpc and $\pm1000$~km~s$^{-1}$ window for 6 out of the 9 absorption systems, with 2 of the absorption line systems showing multiple associated galaxies within the MUSE field of view. The space density of Ly$\alpha$ emitting galaxies (LAEs) around the H~{\sc i} and \ion{C}{iv} systems is $\approx10-20$ times the average sky density of LAEs given the flux limit of our survey, showing a clear correlation between the absorption and galaxy populations. Further, we find that the strongest \ion{C}{iv} systems in our sample are those that are most closely aligned with galaxies in velocity space, i.e. within velocities of $\pm500$~km~s$^{-1}$. The two most metal poor systems lie in the most dense galaxy environments, implying we are potentially tracing gas that is infalling for the first time into star-forming groups at high redshift. Finally, we detect an extended Ly$\alpha$ nebula around the $z\approx 5.26$ quasar, which extends up to $\approx50$~kpc at the surface brightness limit of $3.8 \times 10^{-18}$~erg~s$^{-1}$~cm$^{-2}$~arcsec$^{-2}$. After scaling for surface brightness dimming, we find that this nebula is centrally brighter, having a steeper radial profile than the average for nebulae studied at $z\sim3$ and is consistent with the mild redshift evolution seen from $z\approx 2$.
\end{abstract}




\section{Introduction}

The flow of baryons onto, out of, and around galaxies is crucial to our understanding of galaxy evolution as a whole, dictating how galaxies form their stellar content. Star-forming galaxies predominantly occupy a relatively narrow ($\approx0.3$~dex) main sequence, relating star-formation to stellar mass \citep[e.g.][]{2004MNRAS.351.1151B, 2007ApJ...670..156D, 2007A&A...468...33E, 2007ApJ...660L..43N, 2007ApJS..173..267S}. The gas depletion timescales on this main-sequence are relatively short ($t_{\rm depl}\approx10^9$~years, e.g. \citealt{2008AJ....136.2782L, 2008AJ....136.2846B, 2011ApJ...730L..13B, 2013ApJ...768...74T}), and to sustain the observed star-formation levels, a steady flow of cold gas from a main-sequence galaxy's surroundings is necessary. Such inflows are required to co-exist with the large scale outflows commonly detected in star-forming galaxies across a wide range of cosmic time \citep[e.g.][]{1990ApJS...74..833H, 2001ApJ...554..981P, 2003ApJ...588...65S, 2005ApJ...621..227M, 2013ApJ...772..119L}. Effectively, the existence of the main sequence (alongside the observed galaxy stellar mass function) necessitates that the bulk of the star-forming galaxy population exists in a quasi-steady state of gas inflow, outflow and consumption \citep[e.g.][]{2010ApJ...718.1001B, 2012MNRAS.421...98D, 2014MNRAS.444.2071D, 2016MNRAS.457.2790T}.

The large scale transport of baryons is surmised to be intrinsically connected to the presence of strong H~{\sc i} absorption systems, and in particular Lyman limit systems (LLS) and damped Ly$\alpha$ systems (DLA) identified in quasar sightlines \citep[e.g.][]{Prochaska2009, Prochaska2010, 2016MNRAS.455.4100F}. Cold accretion, the dominant form of gas accretion at $z\gtrsim1$ in simulations \citep[e.g.][]{2005MNRAS.363....2K, 2009ApJ...703..785D}, is predicted to follow collimated filamentary gas structures which would be detected at high H~{\sc i} column densities given a background source \citep[e.g.][]{FG2011,Fumagalli2011}. Considering outflows, simulations predict that galactic winds from relatively low mass galaxies should contain entrained cold gas clumps that would similarly be detected in absorption as high H~{\sc i} column density systems, likely with higher than average metallicities \citep[e.g.][]{FG2016}. 
Further, the observed kinematics of DLAs are seen to be reproduced in simulations only when outflows are implemented \citep{2011MNRAS.416.1723B, 2014MNRAS.440.2313B, 2015MNRAS.447.1834B}.

Detecting the host galaxies of strong H~{\sc i} absorbers has long been a challenge in observational astronomy. Early studies provided few, if any, detections of galaxies in close proximity of DLAs \citep[e.g.][]{ 1993A&A...270...43M, 1994AJ....108.2046S, 1995ApJ...451..484L, 1996Natur.382..234D,  1999MNRAS.309..875B, 2000A&A...358...88F, 2001ApJ...550..585B, 2009A&A...497..689G}. Significant progress on connecting strong Ly$\alpha$ absorbers to the galaxy population was made with the introduction of integral field spectrograph (IFS) instruments. In particular, SINFONI on the Very Large Telescope (VLT) provided some of the first significant datasets based on blind searches for H$\alpha$ emission in the observed near infrared (i.e. $0.7\lesssim z\lesssim2$) from faint galaxies within a relatively small field of view around background quasars \citep[e.g.][]{2007ApJ...669L...5B, 2011MNRAS.410.2251P}. These surveys probe the absorber environment up to $\approx50$~kpc, focusing on small scale associations and so effectively probing the halos of the closest galaxies in absorption. Such surveys produced detection rates for galaxies in the proximity of $N({\rm HI})\gtrsim10^{19}$~cm$^{-2}$ absorption line systems of $\approx30\%$ at $z\sim1$, falling to just $\lesssim10\%$ at $z\sim2$ \citep{2016MNRAS.457..903P}.

Parallel with these studies, the introduction of the X-Shooter spectrograph on the VLT showed that the detection of associated galaxies was tied to the metallicity of the strong absorption system, with higher metallicity systems ([Si/H] > -1) showing significantly enhanced detection rates ($\approx60\%$) of associated galaxies \citep{2013MNRAS.436..361F, 2017MNRAS.469.2959K}.

With the MUSE IFS \citep{Bacon2010MUSE} on the VLT, systematic blind surveys are now possible over larger scales ($\approx 1$~arcmin), detecting simultaneously direct associations at small impact parameters as well as the environment at larger impact parameters. Whilst past targeted surveys have revealed a handful of galaxy groups in the proximity of strong absorption line systems \citep[e.g.][]{1986A&A...168....6B, 1998MNRAS.299..661M, 2010MNRAS.406..445K}, dedicated MUSE surveys are beginning to uncover greater numbers of associations with galaxy groups at low redshifts \citep[e.g.][]{2017MNRAS.464.2053P, 2017MNRAS.468.1373B, 2019MNRAS.486...21B, 2019MNRAS.490.1451F} as well as similarly complex environments traced by Ly$\alpha$ emitters (LAEs) at $3 \lesssim z \lesssim 4$ \citep[e.g.][]{Fumagalli2016, Fumagalli2017, 2019MNRAS.487.5070M, 2019MNRAS.tmp.2667L}. Such studies are now also being complemented by ALMA observations mapping the galaxy population in [CII] and CO emission \citep[e.g.][]{2017Sci...355.1285N, 2018ApJ...856L..12N, 2019ApJ...870L..19N,2019MNRAS.485.1595P}.  

With large surveys currently ongoing at $z\lesssim 4$, extending  blind searches to higher redshifts, approaching the epoch of re-ionization, is now critical for a complete view of the gas-galaxy connection across cosmic time. To this end, redshifts of $z\approx 5-5.5$ become particularly relevant as late re-ionization models predict that the end of re-ionization may extend to $z\approx<5.5$ \citep{2015A&A...580A.139H, 2015MNRAS.447.3402B, 2019MNRAS.485L..24K, 2019arXiv191003570N}. Moreover, differently from $z\gtrsim 6$ \citep[e.g.][]{Becker2012}, it is still possible to identify individual absorption line systems within the thick Ly$\alpha$ forest with high-resolution spectroscopy, which is crucial for a detailed characterization of the hydrogen content and hence metallicity of the systems \citep[e.g.][]{2014ApJ...782L..29R}. 

At present, only a handful of studies have focused on this redshift range. \citet{cai2017} presented a narrow band search for Ly$\alpha$ emitters at $z\approx5$ around C~{\sc iv} absorbers, finding one candidate pair within $\lesssim40$~kpc  and two further candidate pairs at 160~kpc and 200~kpc. \citet{dodorico2018} reported on a serendipitous CO detection of a $z\approx 5.9$ galaxy associated with a metal poor DLA at a distance of $\approx40$~kpc, surmising the DLA to be tracing either a satellite galaxy or filamentary gas structure. Likewise, \citet{2014MNRAS.442..946D, 2015MNRAS.448.1240D} conducted narrow band plus spectroscopic follow-up analysis of LAE galaxies around two $z>5$ \ion{C}{iv} absorption line systems, finding both systems resided in large scale regions hosting galaxy over-densities. 

Using more statistical techniques, \citet{meyer2019} argued for an excess of galaxies in the proximity of \ion{C}{iv} absorbers leading to an associated excess of Ly$\alpha$ transmission in quasar sightlines through such regions. Further to these, \citet{keating2019} discussed how LAEs should be hard to detect close to deep neutral \ion{H}{i} absorption at these redshifts, as those may be the regions not yet ionized, and that it may be instead easier to detect galaxies in the proximity of C~{\sc iv} absorbers, where the local medium is perhaps more likely to be ionized by galaxies.

Leveraging the discovery of several very bright quasars at $z\gtrsim 5$ \citep{Wang2016} for which we have collected high quality optical and NIR spectroscopy, we can now extend these previous studies by conducting a systematic search of galaxies in high-redshift quasar fields. In this paper, we present a detailed MUSE+X-Shooter analysis along the sightline of the quasar \qso\ at $z\approx 5.26$, within which we detect multiple strong H~{\sc i} absorption line systems. 

In Section~\ref{sec:observations}, we present the details of the observations and data reduction. Section~\ref{sec:musegals} presents the search for galaxies within the MUSE cube and Section~\ref{sec:strabs} presents the analysis of the absorption line systems, including associated metal transitions. Section~\ref{sec:eagle} introduces the simulation data used in this study. In Section~\ref{sec:assoc}, we present the results of combining the two datasets, whilst Section~\ref{sec:summary} provides our summary and conclusions. Throughout this paper, we assume a cosmology defined by the parameters in \citet[][i.e. $H_0=67.7$~km s$^{-1}$ Mpc$^{-1}$, $\Omega_m=0.307$ and $\Omega_\Lambda=0.693$]{Planck15} and express magnitudes in the AB system. Unless stated otherwise, distances are given in the proper coordinate frame.

\section{Observations and Data Reduction}
\label{sec:observations}

\subsection{MUSE data}

MUSE observations in the quasar field \qso\ have been collected 
as part of ESO programme 0102.A$-$0261 (PI: Bielby) between December 2018 and March 2019, using the Wide Field Mode combined with the GALACSI (Ground Atmospheric Layer Adaptive Corrector for Spectroscopic Imaging) adaptive optics system \citep{2012SPIE.8447E..37S}. Conditions were generally excellent, with clear sky and sub-arcsecond image quality in dark time. We acquired a total of 24 exposures, each of 900 second, for a total on-source exposure time of 6 hours. In between exposures, we applied small offsets ($\approx 3-4~$arcsec) and 15 deg rotations to reduce systematic errors. 

\begin{figure*}
    \centering
    \includegraphics[scale=0.7]{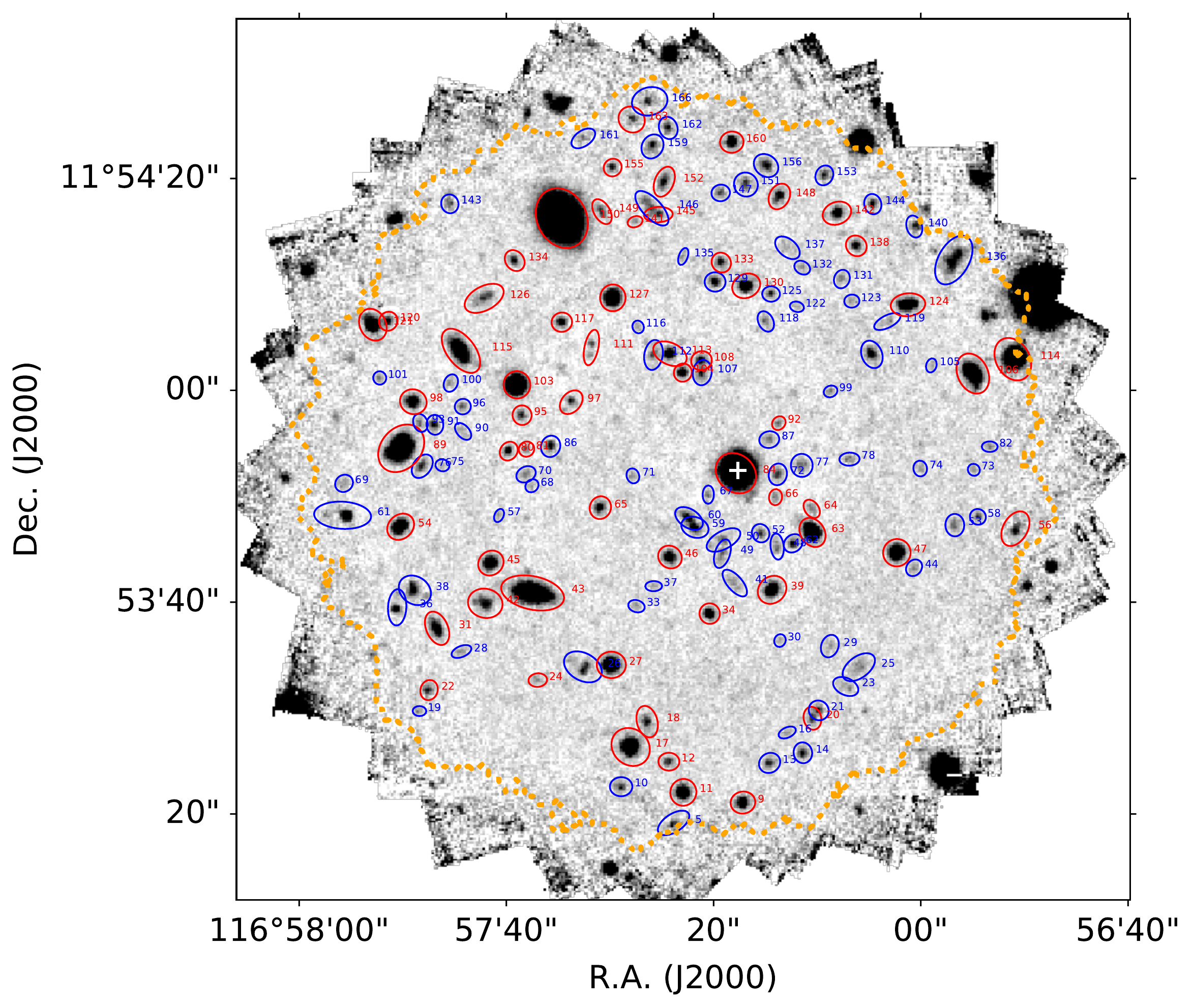}
    \caption{Reconstructed white-light image from the final data cube, showing the quasar \qso, marked with a white cross near the centre of the field. Continuum-detected sources identified within the field are marked with apertures (red  for sources with spectroscopic redshifts). The dotted orange contours enclose the region covered by at least 6 exposures.}\label{fig:fov}
\end{figure*}

Observations have been reduced following the methodology described in previous work \citep[][]{Fumagalli2016,Fumagalli2017,MF2017UVB, 2019MNRAS.tmp.2667L}. Briefly, we first use the standard ESO pipeline \citep[v2.4.1;][]{MUSEPipe2014} to perform basic reduction by applying calibrations. We then reconstruct datacubes for each individual exposure, 
which we then process with the {\sc Cubextractor} pipeline (v1.8) to improve the quality of the flat fielding and sky subtraction \citep[see][]{2019MNRAS.483.5188C}.
Finally, all exposures are combined in a single cube with pixel size of 0.2~arcsec (spatial direction) and 1.25~\AA\ (spectral direction), using both mean and median statistics. Two independent cubes, each with half the number of exposures are also produced. 
The white-light image reconstructed from the mean cube is shown in Fig.~\ref{fig:fov}. The final image quality on this white-light image is found to be $\approx 0.62\rm~arcsec$ full-width at half-maximum (FWHM) as measured by fitting Moffat profiles on point sources.

Following the procedure described in \citet{2019MNRAS.tmp.2667L}, we further re-scale the final noise cubes using bootstrap techniques of individual pixels across the 24 exposures to correctly reproduce the pixel standard deviation.  At this stage, we also derive a model for the correlated noise arising from the resampling of the pixel table onto a final grid, as described in \citet{2019MNRAS.tmp.2667L}. This model is defined as a correction that needs to be applied to the propagated error for a source in an aperture of $N$ pixels on a side, $\sigma_{\rm N}$, to recover the effective noise, $\sigma_{\rm eff}$. A second-order polynomial fit describes this correction in the form $\sigma_{\rm eff}/\sigma_{\rm N} = 1.357 + 0.128 N + 0.008 N^2$, as computed in a spectral window of 4 pixels between $6600-7900$~\AA\ (i.e. the range where we search for Ly$\alpha$ emission).

\subsection{X-Shooter data}

\begin{figure*}
    \centering
    \includegraphics[scale=0.7]{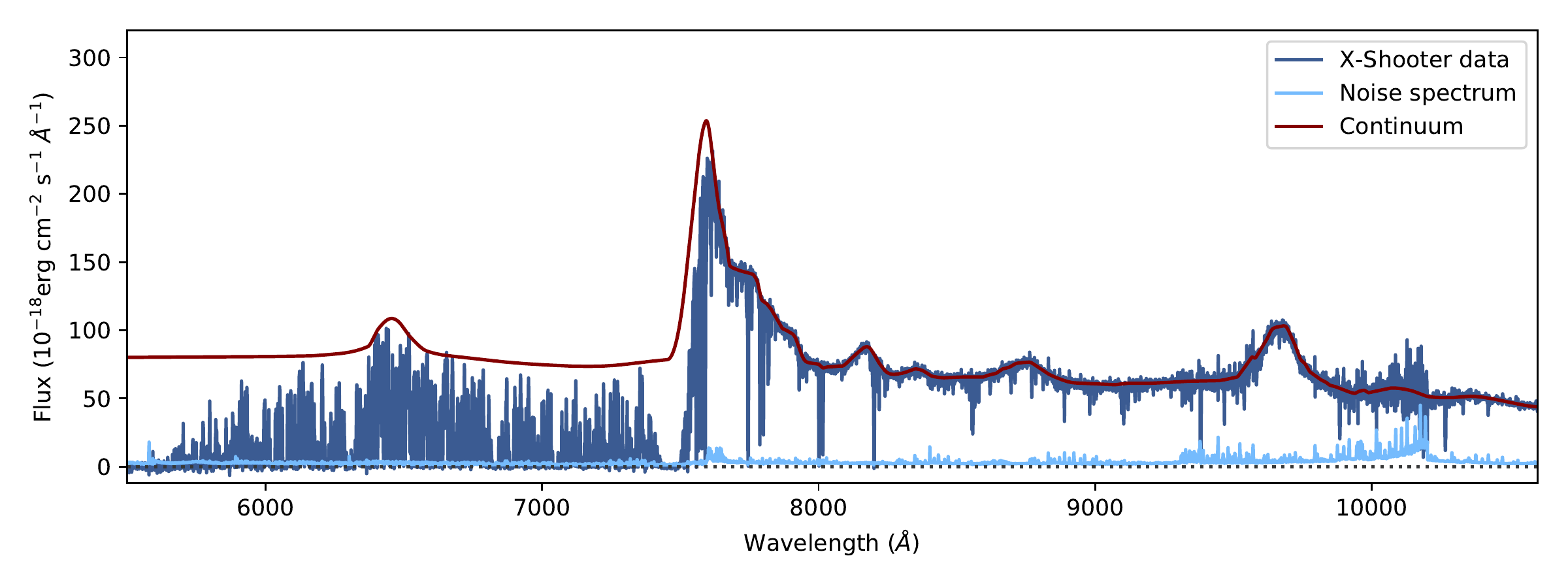}
    \caption{Reduced spectrum of \qso\ from the VLT X-Shooter spectrograph (blue curve), with the associated 1$\sigma$ noise spectrum (pale blue) and the estimated continuum profile (solid red curve).}\label{fig:xshooter}
\end{figure*}

The background quasar spectrum  (Fig.~\ref{fig:xshooter}) comes from a recent dataset obtained at the VLT using X-shooter spectrograph \citep{2011A&A...536A.105V} to observe 41 $z>5$ bright quasar spectra (PID 98.A$-$0111 and 100.A$-$0243; PI: Rafelski, \citealt{2019ApJ...883..163B}). These quasars were identified via WISE IR color selection and confirmed with low resolution spectroscopy \citep{Wang2016, Yang2016, 2017AJ....153..184Y}. X-shooter provides moderate resolution spectroscopy across three wavelength ranges: UVB (300--550~nm), VIS (550--1020~nm), and NIR (1020-2480~nm), although the UVB arm contains no to little flux for quasars at this redshift. The data typically have a signal to noise $S/N>10$ per spectral bin and we use slits of $0.9''$ in the optical and $0.6''$ in the NIR to achieve a resolution of $R\sim8000-9000$, sufficient to measure Ly$\alpha$ and accurate metallicities \citep{2012ApJ...755...89R, 2014ApJ...782L..29R}.

The X-Shooter data were reduced using standard techniques with a dedicated pipeline, as detailed in \citet{Becker2012} and \citet{2016A&A...594A..91L}. A full description of the reduction of the X-Shooter spectrum, including sky subtraction, 1D extraction and corrections for telluric absorptions is provided in \citet{2019ApJ...883..163B}. Based on the X-shooter spectrum, we measure a redshift of $z_{\rm qso}=5.265\pm0.015$ for the quasar, by fitting the \ion{Si}{IV} and \ion{C}{IV} emission lines. We note that these lines show velocity offsets from intrinsic redshifts of $1\lesssim z\lesssim6$ QSOs across a range of $\approx\pm500$~km~s$^{-1}$ \citep{2019MNRAS.487.3305M}.

\section{Search for associated galaxies}
\label{sec:musegals}

\begin{table}
	\centering
	\caption{Spectroscopic redshifts (including quality flags) and coordinates for continuum-detected sources. Only the sources at $z>4$ are listed, with the full table available as online only-material.}
	\label{tab:redshifts}
	\begin{tabular}{rccc} 
	\hline
ID & R.A./Dec. & Redshift & Q.F. \\ 
 \hline
22&J074751.16+115331.7&4.6250&3\\ 
56&J074747.39+115346.9&4.0830&3\\ 
84&J074749.19+115352.2&5.2548&4\\ 
141&J074749.84+115415.9&5.0703&2\\ 
\hline
\end{tabular}
\end{table}

\begin{figure}
    \centering
    \includegraphics[scale=0.65]{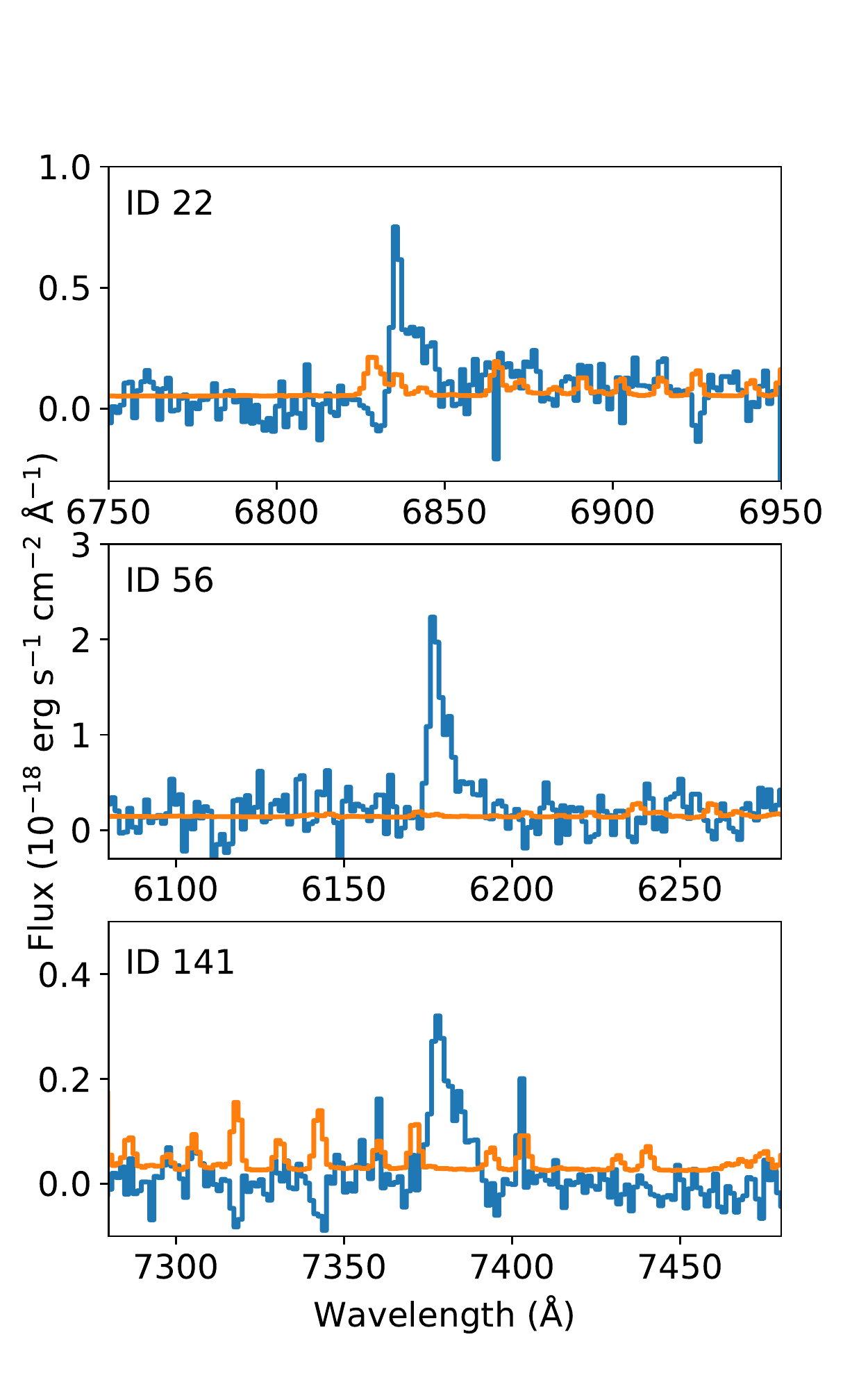}
    \caption{Spectra of all three continuum detected galaxies at $z>4$ (excluding the quasar, see also Table~\ref{tab:redshifts}). At these redshifts, spectroscopic identification is based on Ly$\alpha$ emission which often shows an asymmetric profile.}
    \label{fig:contspec}
\end{figure}

As in previous analyses of MUSE data \citep[e.g.][]{Fumagalli2017, 2019MNRAS.487.5070M, 2019MNRAS.tmp.2667L, 2019MNRAS.490.1451F}, we conduct a redshift survey of galaxies detected in the continuum together with a search for line emitters in the cube.

\begin{table*}
    \centering
    \caption{Properties of the emission line galaxies detected at $S/N \ge 6$ in the MUSE cube
        between $6100-7900~$\AA. The area is calculated for each source out to a surface brightness level of $10^{-18.25}~\rm~erg~s^{-1}~cm^{-2}~arcsec^{-2}$.}
    \label{tab:emitters}
    \begin{tabular}{rcccccccccl}
    \hline
ID & R.A. & Dec.    & $i$   & $M_{\rm UV}$ & F$_{\rm line}$ & L$_{\rm line}$ & Redshift & Area & Class & Type\\
& (J2000) & (J2000) & (mag) & (mag) & ($10^{-18}\rm erg~s^{-1}~cm^{-2}$) & ($10^{40}\rm erg~s^{-1}$) & & (kpc$^2$) & &\\
    \hline
1 &07:47:47.399&11:53:55.96 & $27.0\pm0.5$ & $-22.0\pm0.5$ & $2.87\pm0.30$&$72.36\pm7.66$&4.8384& 137 & 1& LAE \\ 
2 &07:47:47.705&11:53:36.66 & $>26.4$ & $>-22.1$ & $4.39\pm0.39$&$113.27\pm9.98$&4.8820& 225 & 1& LAE \\ 
3 &07:47:47.988&11:53:51.94 & $>26.6$ & $>-21.7$& $1.01\pm0.16$&$22.26\pm3.50$&4.5594& 155 & 2& LAE$^a$ \\ 
4 &07:47:47.991&11:53:49.79 & $>27.4$ & $>-21.4$ & $2.53\pm0.27$&$63.23\pm6.85$&4.8191& 254 & 1& LAE \\ 
5 &07:47:47.998&11:53:52.28 & $>26.6$ & --- & $1.25\pm0.17$&$0.25\pm0.03$&0.6582& - & 1& [OII]$^b$ \\ 
6 &07:47:48.188&11:54:07.18 & $>26.9$ & --- & $3.22\pm0.37$&$1.98\pm0.22$&1.0443& - & 1& [OII] \\ 
7 &07:47:48.278&11:53:59.98 & $>27.0$ & $>-21.8$ & $1.82\pm0.20$&$45.82\pm5.12$&4.8309& 149 & 1& LAE \\ 
8 &07:47:48.355&11:53:29.96 & $27.0\pm0.5$ & $-21.8\pm0.6$ & $16.44\pm0.74$&$483.64\pm21.92$&5.1655& 446 & 1& LAE \\ 
9 &07:47:48.777&11:54:23.19 & $26.7\pm0.6$ & $-21.2\pm0.6$ & $16.76\pm0.92$&$284.30\pm15.59$&4.0876& 631 & 1& LAE \\ 
10&07:47:49.283&11:53:32.31 & $>26.7$ & --- & $37.49\pm1.37$&$20.97\pm0.77$&1.0054& - & 1& [OII] \\ 
11&07:47:49.375&11:54:02.98 & $25.6\pm0.2$ & --- & $2.31\pm0.26$&$0.46\pm0.05$&0.6586& - & 1& [OII] \\ 
12&07:47:49.415&11:54:16.36 & $>27.0$ & $>-21.1$ & $11.13\pm0.69$&$212.12\pm13.23$&4.2946& 766 & 1& LAE \\ 
13&07:47:49.460&11:54:02.94 & $25.6\pm0.2$ & $-22.2\pm0.4$ & $1.09\pm0.16$&$26.20\pm3.94$&4.7473& 136 & 2& LAE$^c$ \\ 
14&07:47:49.915&11:54:15.94 & $>26.9$ & $>-22.0$ & $14.47\pm0.64$&$407.48\pm17.97$&5.0703& 513 & 1& LAE \\ 
15&07:47:49.919&11:53:49.74 & $>27.2$ & $>-21.8$ & $3.53\pm0.32$&$91.09\pm8.19$&4.8830& 249 & 1& LAE \\ 
16&07:47:50.196&11:54:18.30 & $26.3\pm0.2$ & $-21.6\pm0.2$ & $4.37\pm0.41$&$75.53\pm7.02$&4.1194& 231 & 1& LAE$^d$ \\ 
17&07:47:50.435&11:54:12.43 & $>26.9$ & $>-21.6$ & $0.80\pm0.13$&$19.42\pm3.08$&4.7516& 120 & 2& LAE \\ 
18&07:47:50.801&11:53:32.26 & $>26.9$ & --- & $1.64\pm0.25$&$0.34\pm0.05$&0.6697& - & 2& [OII] \\ 
19&07:47:50.805&11:54:18.27 & $>26.9$ & $>-21.8$ & $5.34\pm0.43$&$133.47\pm10.67$&4.8186& 211 & 1& LAE \\ 
20&07:47:51.207&11:53:49.61 & $>26.9$ & --- & $3.44\pm0.38$&$0.71\pm0.08$&0.6698& - & 1& [OII] \\ 
21&07:47:51.808&11:53:41.18 & $26.0\pm0.3$ & $-22.4\pm0.4$ & $1.35\pm0.22$&$33.72\pm5.39$&4.8161& 202 & 1& LAE \\ 
\hline
\end{tabular}
\\$^a$ Overlaps with emitter 4 in projection, but appears at different redshift.   $^b$Classification uncertain. $^c$ Overlaps with continuum source 108 in projection, but appears at different redshift. $^d$ Overlaps in part with continuum source 150 in projection, but appears at different redshift. 
\end{table*}

\subsection{Continuum-detected galaxies}

For continuum-detected galaxies, we run {\sc SExtractor} \citep{sextractor} on the deep white-light image, folding in the propagated pixel variance and masking regions where the number of exposures falls below 6 (i.e. the region outside the dashed yellow bounding line in Fig.~\ref{fig:fov}). Only sources above 5 times the propagated error and with minimum area of 10 pixels are marked as detected. For each detected source, we reconstruct a 1D spectrum using all pixels within the segmentation mask, also transforming the wavelength to vacuum. 

\begin{figure}
    \centering
    \includegraphics[scale=0.45]{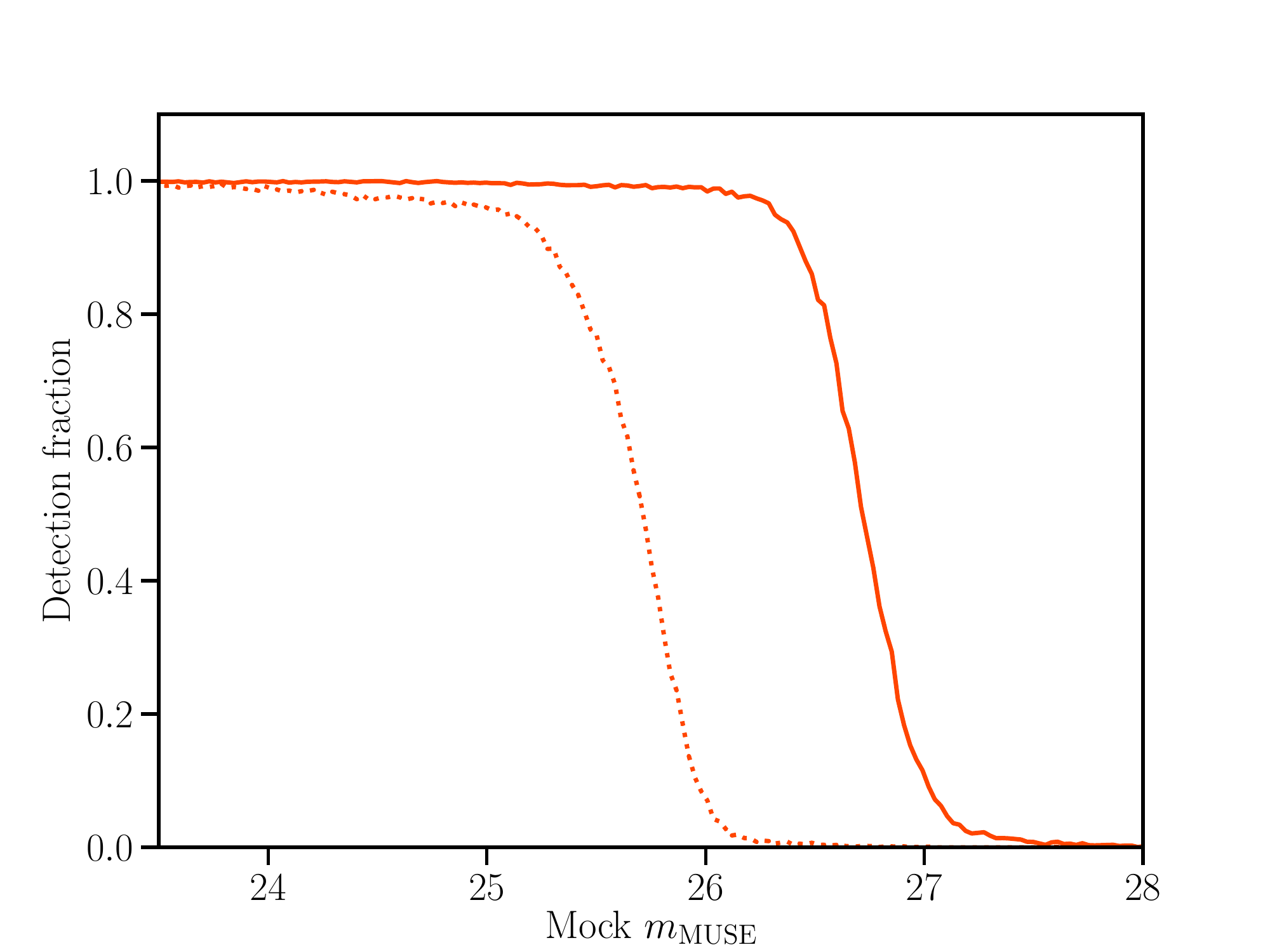}
    \caption{Completeness function for continuum sources for point sources matched to the MUSE PSF (solid line) and for extended sources with scale-length of 0.26~arcsec (dotted line). These observations are $50\%$ ($90\%$) complete at 26.7 mag (26.4 mag) for point sources, and $50\%$ ($90\%$) complete at 25.7 mag (25.3 mag) for extended sources.}
    \label{fig:contcomp}
\end{figure}

We then measure redshifts using the {\sc Marz} redshifting software \citep{Marz}, which we customize\footnote{This version is available at \url{https://matteofox.github.io/Marz}, described in \citet{2019MNRAS.490.1451F}} with high-resolution synthetic templates for passive and star-forming galaxies at $z<2$. Following automatic template fitting, individual sources are inspected and classified by two authors (MFu and MFo) in four classes (4, secure redshift with multiple features; 3, good redshift with single but unambiguous feature; 2, possible redshift, based on a single feature; 1, unknown redshift). At $z>4$, spectroscopic identification is based on Ly$\alpha$ emission alone, which often shows an asymmetric profile. Typical redshift uncertanties are $\delta z \approx 0.0003$. The continuum detected $z>4$ objects are listed in Table~\ref{tab:redshifts}, including the quasar itself. Fig.~\ref{fig:contspec} shows the spectra of the three sources (excluding the quasar), focusing on the spectral region where Ly$\alpha$ is detected. 

To assess the completeness of our source catalogue, we perform 10,000 repetitions of the analysis described above on mock images constructed by injecting 80 mock sources at each iteration (to avoid blending issues) in blank sky regions. We repeat this experiment twice, the first time for point sources matched to the image quality of the MUSE data (0.6~arcsec) and the second time considering exponential disks (neglecting inclination) with scale-length of 0.26~arcsec, convolved with the instrument point spread function (PSF).
In Fig.~\ref{fig:contcomp}, we show the fraction of objects recovered compared to the number of injected sources as a function of magnitude, finding that our search is $50\%$ ($90\%$) complete at 26.7 mag (26.4 mag) on the white-light image for point sources and $50\%$ ($90\%$) complete at 25.7 mag (25.3 mag) for extended sources.

\begin{figure*}
    \centering
    \includegraphics[width=\textwidth]{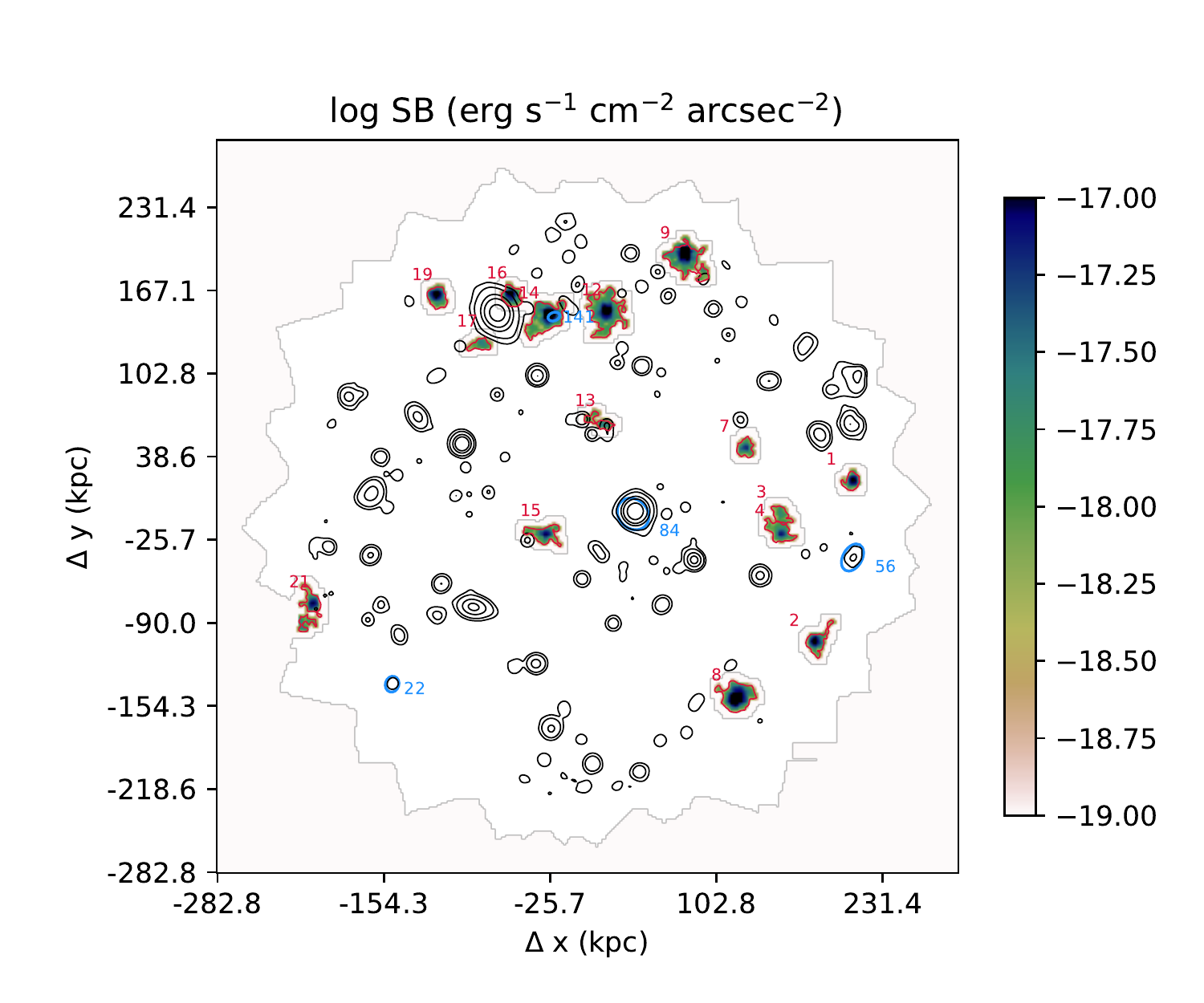}
    \caption{Map of the Ly$\alpha$ emission detected in the MUSE cube at $z>4$ (excluding the quasar itself). Solid contours show continuum emission, whilst the colour scale gives the measured Ly$\alpha$ surface brightness. Objects detected in Ly$\alpha$ emission at $z>4$ are numbered and outlined in red. Continuum sources at $z>4$ are numbered and circled in blue.}
    \label{fig:lyamap}
\end{figure*}

\begin{figure*}
    \centering
    \includegraphics[width=\textwidth]{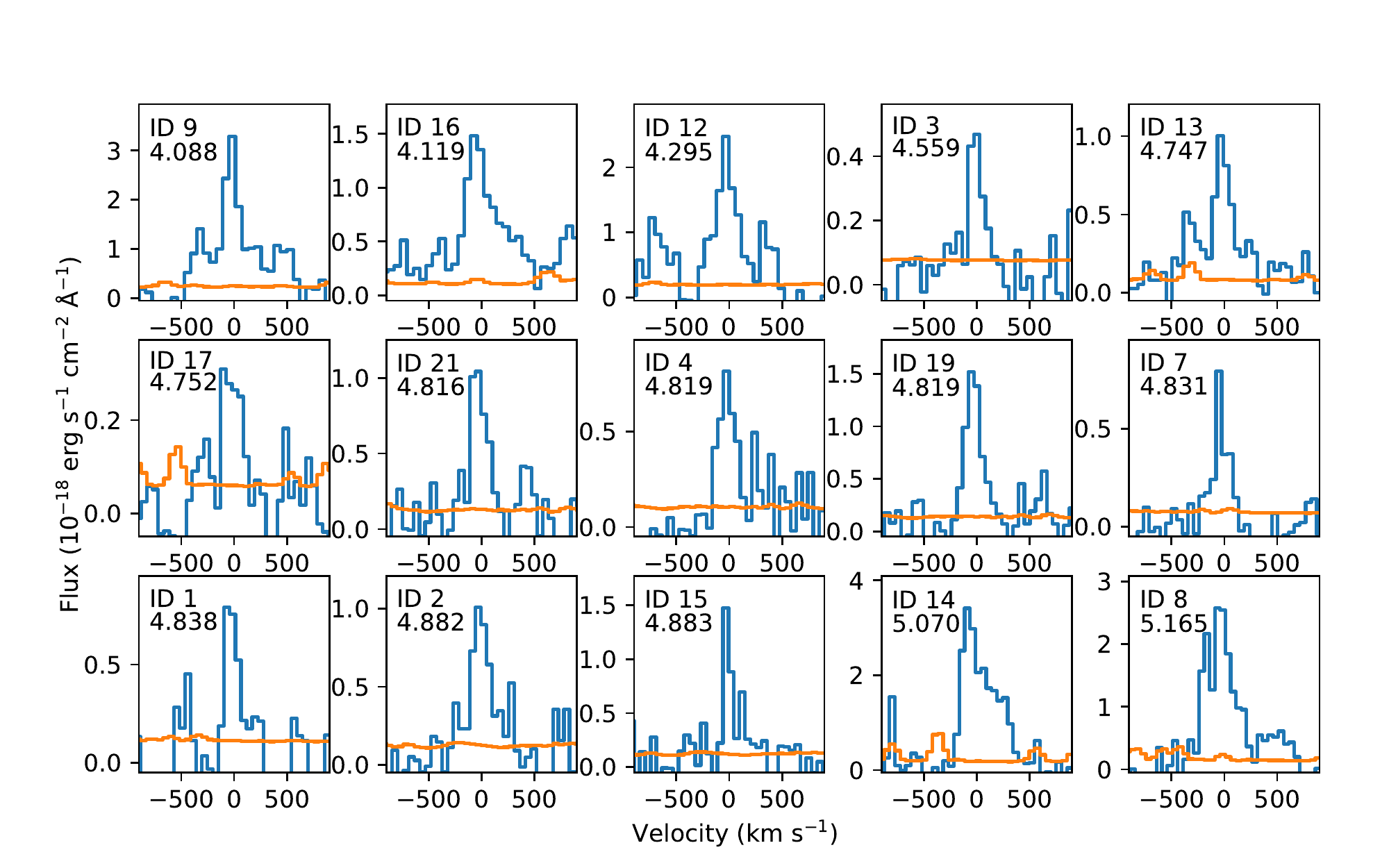}
    \caption{Emission line spectra centred on detected Ly$\alpha$ emission for the 15 $z>4$ line-emission detected sources (in order of increasing redshift, see also Table~\ref{tab:emitters}). The blue histogram shows the detected flux profile and the orange the associated noise spectrum.}
    \label{fig:lyaspec}
\end{figure*}

\begin{figure}
    \centering
    \includegraphics[scale=0.45]{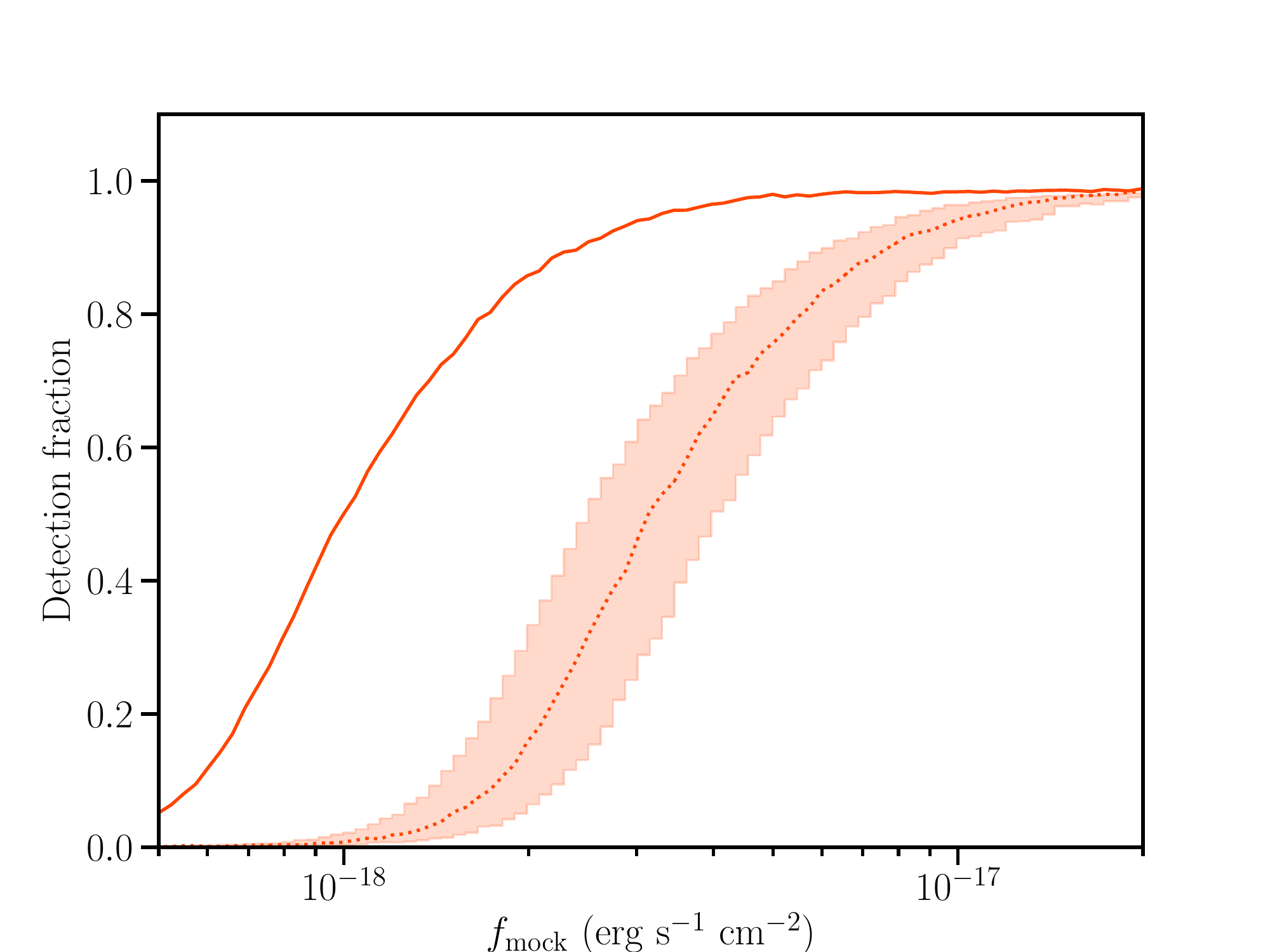}
    \caption{Completeness limits for extraction of line emitters at $S/N \ge 6$. The red solid curve refers to point sources matched to the MUSE PSF. In contrast, the shaded region shows extended sources with exponential profiles of scale-lengths of 2 -- 4 kpc, with the dashed curve showing the 3~kpc scale length result.}
    \label{fig:lyacomp}
\end{figure}

\subsection{Emission line galaxies}
\label{sec:elgdet}

For emission line galaxies we follow the approach described in \citet{2019MNRAS.tmp.2667L}, which we only briefly summarize here. 
After reducing the cube to the wavelength range of interest ($\lambda = 6100-7900~$\AA, covering the redshift interval $z=4-5.5$ for Ly$\alpha$), we subtract the quasar PSF and the continuum of sources using the tools distributed as part of {\sc Cubex} \citep[for details, see e.g.][]{2019MNRAS.483.5188C}. 

We then run {\sc Cubex} to identify groups of at least 27 connected voxels, covering more than 9 pixel$^2$ once projected along the wavelength axis, and 3 channels in wavelength. Objects marked for extraction are then retained if the integrated $S/N$ corrected for correlated noise as discussed above is $S/N > 6.5$. In contrast to previous work conducted mostly at $\lambda < 6000$~\AA, in this analysis we adopt a more conservative cut in $S/N$ to ensure that we minimize the contamination arising from residuals of sky lines.

Following extraction, objects are classified in two confidence classes. Objects in class 1 have integrated $S/N \ge 7$ in the mean coadd, are detected with $S/N > 3$ within independent coadds of only half of the exposures, and the $S/N$ of these two subsets agree within 50\% of their value. This class contains sources with the highest purity at the expense of completeness. 
Objects in class 2 obey a similar classification, but include the remaining objects with $6\le S/N < 7$, which raises the completeness at the possible expense of the purity. For this class, we also monitor the fraction of voxels within the segmentation map that is contained in a $5^3$ voxel volume, which we find to be a good metric to reject spurious identifications such as very extended structures at the edge of the field where the quality of the data is significantly worse. 

Next, we generate optimally-extracted maps \citep[see e.g.][]{Borisova2016}
of the mean, median cubes and two independent-half cubes, and we extract a 1D spectrum by projecting the segmentation map in 2D and summing flux over the spatial direction as a function of wavelength. Using a dedicated GUI, we  inspect these products, including the 3D segmentation map, for all the objects to remove the remaining false-positive (typically objects that present very elongated chains of voxels in their segmentation map that are not well-connected in wavelength). The remaining objects in class 1 are deemed to be real sources, with the objects in class 2 considered only candidate sources as they approach the detection limit. 

At this stage, we also assign a redshift to the sources, according to the following criteria. Sources that present a clear doublet emission can be classified as either [OII]$\lambda\lambda 3726,3729$, CIII]$\lambda\lambda 1907,1909$, or MgII$\lambda\lambda 2796,2803$ emitters. The classification of single line emitters is more ambiguous as multiple rest-frame lines could be in principle detected at any given redshift. For class 1 sources, we can rule out H$\alpha$ trivially for $\lambda < 6563$~\AA\ and can classify H$\alpha$ for the remaining sources that show associated [OIII]$\lambda 5007$ or the [OII] doublet. \ion{C}{IV} can be recognized by the associated strong Ly$\alpha$ emission (unless Lya is absorbed or resonantly scattered) or CIII] doublet emission. Finally, the identification of Ly$\alpha$ is also strengthened by the characteristic shape of the profile (where evident by eye in the spectra). These criteria apply also to class 2 sources, although the varying sensitivity limit across the wavelength range and the different line strengths make the classification more uncertain. 
Following these criteria, we identify 14 additional $z>4$ galaxies, 3 of which are in class 2, and 6 [OII] emitters, one of which is in class 2 and a second one for which the classification is ambiguous (i.e. good signal to noise emission line, but of uncertain redshift). The properties of the sources are detailed in Table~\ref{tab:emitters}, a map of the Ly$\alpha$ emitters is shown in Fig.~\ref{fig:lyamap} and a gallery of the Ly$\alpha$ profiles is shown in Fig.~\ref{fig:lyaspec}.

The three continuum selected galaxies given in Table~\ref{tab:redshifts} are all also detected with the line detection method. One of these (ID 141 in continuum and ID 14 in emission) shows a small offset between the centroid of the continuum emission and the line emission, with the line emission being significantly more extended than the continuum, whilst the remaining two are more consistent in extent and centroid with the continuum detections. As such we include the line detection for this source in Table~\ref{tab:emitters} for reference, but not the other two continuum sources. All three are included and treated as single objects in the analysis that follows, giving a total of 17 $z>4$ galaxies exhibiting Ly$\alpha$ emission.

Similarly to the analysis of continuum sources, we quantify the completeness of our search by analyzing 5,000 mock cubes constructed by injecting 500 mock line emitters in empty regions of each mock cube (to avoid blending). Two types of source are considered: compact emitters with size matched to the MUSE PSF and line spread function with FWHM of 2.5\AA, and extended sources with exponential profiles of scale-length of 2, 3, and 4 kpc convolved with PSF and line spread function with FWHM of 2.5\AA.  By analyzing the mock cubes with {\sc CubEx} as done for the real data, we find the completeness function shown in Fig.~\ref{fig:lyacomp} for $S/N>6$.
Our analysis is $50\%$ complete at $\approx 10^{-18}~\rm erg~s^{-1}~cm^{-2}$ for point sources and $\approx 3.1\times 10^{-18}~\rm erg~s^{-1}~cm^{-2}$ for extended sources with 3~kpc scale-length (at $90\%$ completeness, the limits for point and extended sources become $\approx 2.4\times 10^{-18}~\rm erg~s^{-1}~cm^{-2}$  and $\approx 7.7\times 10^{-18}~\rm erg~s^{-1}~cm^{-2}$ for a 3~kpc scale-length).

\section{Properties of strong absorption lines}
\label{sec:strabs}
\subsection{Identification and measurement of column densities}

We perform a continuum fit of the X-Shooter spectrum using a combination of the {\sc lt\_continuumfit} code, contained within the {\sc linetools} package\footnote{\url{https://github.com/linetools}}, and a template quasar continuum. The {\sc lt\_continuumfit} code follows the steps outlined in \citet{2011MNRAS.414...28C,2017MNRAS.471.2174B}, fitting an initial cubic spline form to the data. However, due to the frequency of Ly$\alpha$ absorbers at $z\sim4-5$, the procedure underestimates the unabsorbed intrinsic continuum below $\lambda_{\rm rest}=1215.67$~\AA\ and so we use the SDSS $0.04\lesssim z\lesssim4.78$ quasar composite spectrum calculated by \citet{vandenberk01} as a guide to correcting the initial cubic-spline fit.

We first normalize and redshift the \citet{vandenberk01} composite to fit the observed quasar continuum at $\lambda_{\rm rest}\approx1200-1400$~\AA. To allow for a tilt in the template, we apply a power-law factor to the composite at $\lambda_{\rm rest}<1215$~\AA\ of the form $(1215.67-\lambda_{\rm rest})^{\alpha}$, finding a slope of $\alpha=0.045$ by normalising to the peak flux at the Ly$\beta$ emission wavelength in the quasar spectrum. We then use this template as a guide when fitting the continuum at $\lambda_{\rm rest}<1215.67$~\AA\ using {\sc lt\_continuumfit}.

\begin{figure}
    \centering
    \includegraphics[width=\columnwidth]{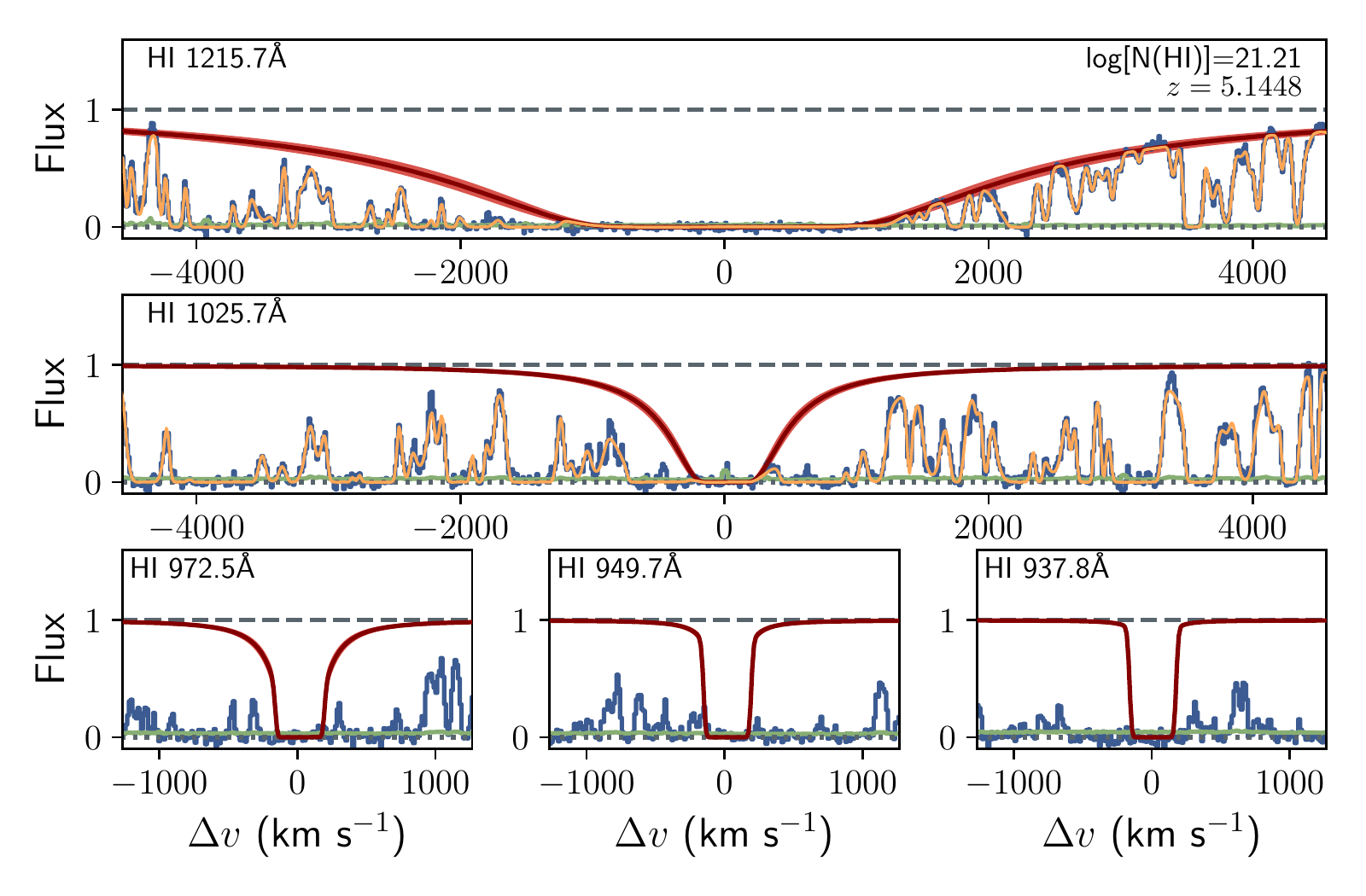}
    \includegraphics[width=\columnwidth]{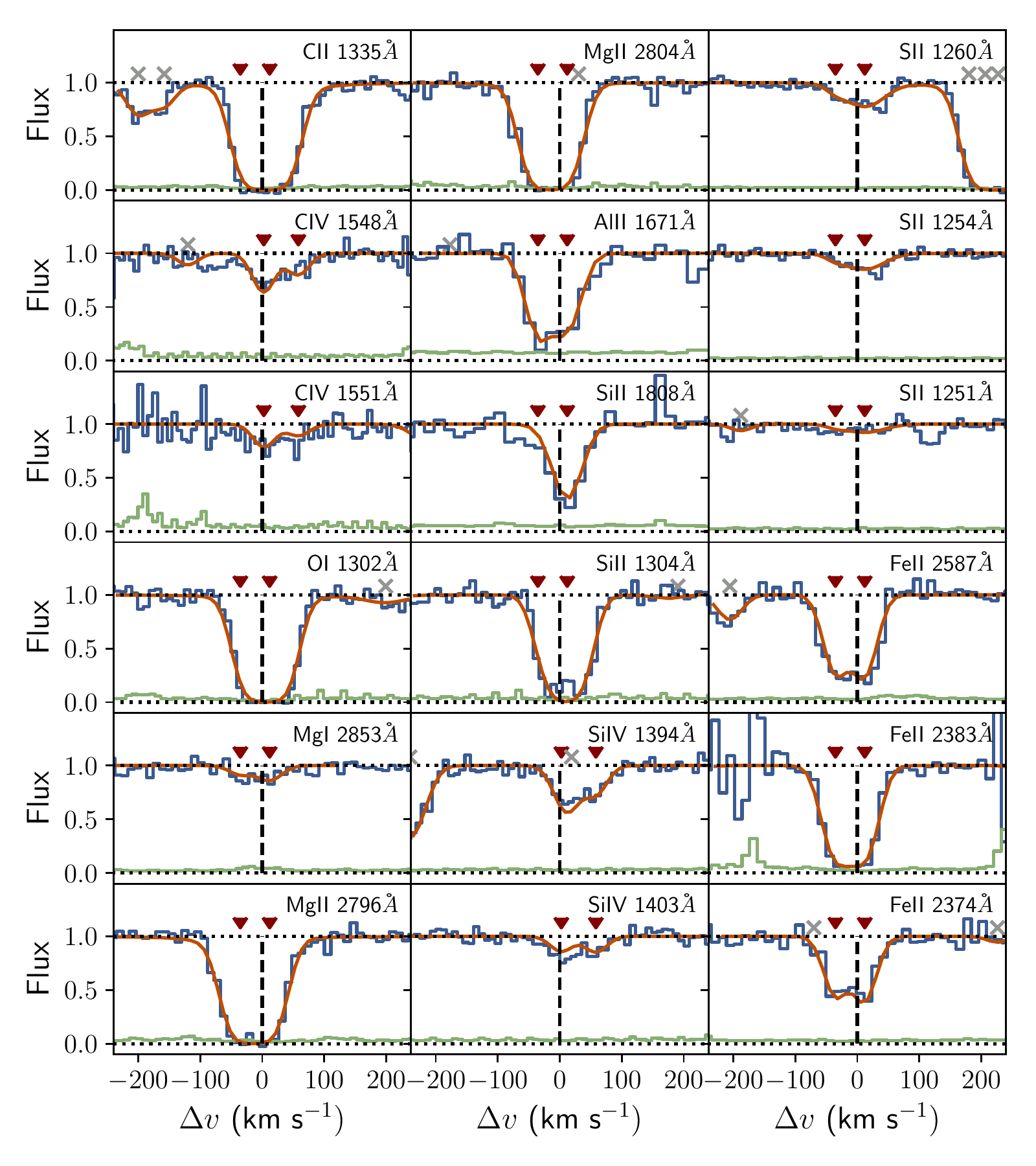}
    \caption{Model fits to the H~{\sc i} absorption and associated metal lines in the X-Shooter spectrum for the DLA system at $z=5.1448$. The data is shown by the blue stepped line in each panel, with the orange curve showing the overall fit within the wavelength range and, in the case of the Lyman series panels, the red curves show the individual strong H~{\sc i} system (with the width of the curves showing the extent given by the best fitting ranges quoted in Table~\ref{tab:als_params}). The upturned triangles in each metal absorption panel identifies centroids of individual absorption components with the given velocity range of the DLA. The grey crosses indicate absorption features identified as not being at the same redshift (within the given velocity range) of the DLA.}
    \label{fig:DLA-ALS_a}
\end{figure}

Using this estimated continuum model, we make an initial census of strong Ly$\alpha$ absorption (in the range bounded by the intrinsic quasar Ly$\alpha$ and Ly$\beta$ emission) and metal absorption systems along the sightline using {\sc pyigm\_guesses} (part of the {\sc pyigm} suite of codes\footnote{\url{https://github.com/pyigm}}). We include all systems where we observe either damping wings in the \ion{H}{i} absorption or detected metal line absorption at a given redshift. This results in a sample of strong absorption systems, encompassing primarily LLSs and DLAs, all of which show associated \ion{C}{iv} absorption features. The \qso\ sightline has a high incidence of metal and strong Ly$\alpha$ absorption system lines in the redshift range $4<z<z_{\rm qso}$, which we list in Table~\ref{tab:als_params} (note that below $z\sim4$, \ion{C}{iv} begins to fall in the Ly$\alpha$ forest for our quasar sample, whilst $z\sim4$ Ly$\alpha$ is obscured by $z\sim5$ Ly$\beta$ absorption). The strongest of these is a clear DLA at $z\approx 5.14$. All but one of the other metal-line detected systems lie at lower redshifts than the DLA, meaning that for these we have no measurement of their Lyman limit strengths, or in general any useful measurement of the Lyman series beyond that of Ly$\beta$.

For each strong absorption system, we perform a fit to the region of the Ly$\alpha$ forest over a velocity range depending on the strength of the absorber, using the {\sc alis} line fitting software package\footnote{\url{https://github.com/rcooke-ast/ALIS}}. In the first iteration, we keep the redshift, column density and velocity broadening of each Ly$\alpha$ absorption line as free parameters, whilst keeping the continuum fixed to that estimated previously using {\sc lt\_continuumfit}. Taking this initial fit to the Ly$\alpha$ regime, we then produce the resultant Ly$\beta$ profile over the same velocity range. Comparing this Ly$\beta$ profile to the data, we add appropriate lower-redshift Ly$\alpha$ absorption systems to complete the model spectrum in the primary Ly$\beta$ regime. We then perform a second iteration of the fitting using {\sc alis} to fit the primary Ly$\alpha$ and Ly$\beta$ absorption as well as the secondary Ly$\alpha$ absorption simultaneously.

As a final step, we perform a further iteration, allowing for addition of further absorbers in both regimes where necessary. Taking this configuration, we then estimate the effect of uncertainties on the continuum level on the fitting profiles. In order to do so, we re-run {\sc alis} using the final model fit, but allowing the continuum level to be a free parameter, re-calculating the parameters for the absorption systems. Any change in the parameters for the strong absorber of interest is then folded into the {\sc alis} fitting estimated uncertainties on each parameter.  

The two lowest redshift ($z\approx 4.03$ and $z\approx 4.12$) absorbers have no associated coverage of Ly$\beta$ (or any higher Lyman series orders), due to the presence, and absorption below the Lyman limit, of the $z\approx 5.14$ DLA. These are therefore purely constrained in H~{\sc i} by Ly$\alpha$ and thus have large associated uncertainties on the H~{\sc i} column density measurements. In contrast to these, the one absorber at $z\approx5.16$ has available coverage up to Ly$\eta$, with all but the Ly$\eta$ line being saturated. From the detected flux at Ly$\eta$ observed wavelengths, we constrain this absorption line to most likely be a partial LLS, with $N({\rm HI})=10^{16.25\pm0.25}~\rm cm^{-2}$. This is the lowest column density absorber in our sample, with the highest being the $z=5.14$ absorber with $N({\rm HI})=10^{21.21\pm0.05}~\rm cm^{-2}$, which is relatively well constrained via the damping wings of the absorber.

Our fits to the H~{\sc i} and metal line absorption profiles for the DLA system at $z=5.1448$ are shown in Fig.~\ref{fig:DLA-ALS_a}, whilst we provide plots of the rest of the $z>4$ absorption systems in the Appendix (Figs.~\ref{fig:LLS-ALS_b}, \ref{fig:LLS-ALS_c}, \ref{fig:LLS-ALS_d} and \ref{fig:LLS-ALS_e}). The fitted parameters and the estimated uncertainties are instead given in Table~\ref{tab:als_params}, where we report only the integrated column densities (i.e. across all components) for each reported species.

\begin{table*}
\centering
\caption{Properties of absorption line systems identified in the sightline of \qso.}
\begin{tabular}{lcccccccccc}
\hline
Redshift & $\log$N(\ion{H}{I})  & $\log$N(\ion{C}{II})  & $\log$N(\ion{C}{IV})  & $\log$N(\ion{Mg}{I})  & $\log$N(\ion{Mg}{II})  & $\log$N(\ion{Al}{II})  & $\log$N(\ion{Al}{III})  & $\log$N(\ion{Si}{II})  & $\log$N(\ion{Si}{IV})  & $\log$N(\ion{Fe}{II})  \\
        & (cm$^{-2}$)  & (cm$^{-2}$)  & (cm$^{-2}$)  & (cm$^{-2}$)  & (cm$^{-2}$)  & (cm$^{-2}$)  & (cm$^{-2}$)  & (cm$^{-2}$)  & (cm$^{-2}$)  \\
\hline 
 5.1612 & 16.25$\pm$0.25 & ---& 13.76$\pm$0.02 & ---& $<$12.91 & ---& $<$12.92 & ---& $<$12.60 & $<$12.81  \\ 
 5.1448 & 21.21$\pm$0.05 & $>$14.94 & 13.63$\pm$0.03 & 11.95$\pm$0.06 & $>$14.00 & $>$13.00 & ---& 15.87$\pm$0.08 & 13.18$\pm$0.04 & 14.77$\pm$0.03  \\ 
 4.8804 & 18.70$\pm$0.70 & $<$13.40 & 14.14$\pm$0.07 & ---& $<$12.78 & ---& $<$12.84 & $<$13.52 & 13.18$\pm$0.04 & --- \\ 
 4.8004 & 19.00$\pm$0.80 & ---& 13.54$\pm$0.02 & ---& $<$12.46 & $<$11.98 & $<$12.91 & $<$14.64 & $<$12.94 & $<$12.94  \\ 
 4.6917 & 17.40$\pm$1.00 & ---& 13.18$\pm$0.04 & ---& $<$12.33 & $<$12.24 & ---& ---& 12.62$\pm$0.05 & $<$12.98  \\ 
 4.6651 & 19.03$\pm$0.40 & ---& 13.49$\pm$0.03 & ---& 12.76$\pm$0.03 & $<$12.59 & $<$12.77 & $<$13.19 & $<$12.80 & $<$12.96  \\ 
 4.6169 & 19.92$\pm$0.17 & ---& 13.60$\pm$0.04 & ---& 13.02$\pm$0.04 & $<$16.08 & $<$12.85 & $<$13.18 & 12.98$\pm$0.04 & $<$13.11  \\ 
 4.1220 & 17.00$\pm$1.80 & ---& 13.74$\pm$0.02 & ---& $>$13.81 & 13.25$\pm$0.05 & $<$12.62 & 14.20$\pm$0.01 & ---& 13.44$\pm$0.03  \\ 
 4.0302 & 17.90$\pm$1.10 & ---& $>$14.06 & ---& 13.47$\pm$0.08 & $<$12.38 & $<$12.94 & $<$13.15 & ---& 12.73$\pm$0.22  \\ 
\hline
\end{tabular}
\label{tab:als_params}
\end{table*}

{bf Whilst the above provides a sample of absorbers that is effectively agnostic of the galaxy positions, we are interested in the absorption properties of the whole galaxy sample. We therefore make a secondary galaxy guided identification of absorption systems. We therefore identify the most proximate candidate strong \ion{H}{i} absorption feature within $\pm1000$~km~s$^{-1}$ of each galaxy with no already identified strong absorption feature and no other galaxy within 1000~km~s$^{-1}$ at a closer impact parameter (i.e. where we have multiple galaxies at a given redshift, we only search for absorption within $\pm1000$~km~s$^{-1}$ of the closest to the sightline). To identify absorption systems, we use the same method as for the primary set of strong systems. We first identify absorption systems within $\pm1000$~km~s$^{-1}$ of the redshift of interest using {\sc pyigm\_guesses} and then perform an iterative fitting process using {\sc alis} (adding lines where appropriate to improve the fit). We then identify the nearest strong absorber (i.e. $N(HI)\gtrsim17$) to the galaxy redshift. The resulting column density estimates and absorber redshifts are listed below our primary sample in Table~\ref{tab:als_params}. Given these systems were not identified in the primary sample (i.e. either through metal lines or clear strong absorption), these secondary systems are all at the lower column density end of our strong absorber range and are poorly constrained, lying as they do in the flat section of the curve of growth.}

\subsection{Absorption line system metallicities}

We now estimate the metallicities of the detected absorption line systems (including only those systems with at least one clearly detected metal line). Whilst the DLA is likely dense enough for partial ionization not to be an issue in determining metallicities, the LLSs are at least partially photoionized \citep[e.g.][]{mf2013}, complicating the estimation of metallicities from the observed ion column densities. Using the identified H~{\sc i} and metal line features, we therefore constrain the metallicity of each LLS using a grid of photo-ionization models created with the {\sc cloudy} code \citep{2017RMxAA..53..385F}. We follow the method outlined in \citet{2016MNRAS.455.4100F}, creating a grid of models given by the parameter ranges provided in Table~\ref{tab:cloudy_grid} using {\sc cloudy}. Using the affine invariant MCMC ensemble sampler {\sc emcee} \citep{2018zndo...1436565F}, we sample the full parameter space, constraining the posterior probability distribution function (PDF) for the metallicity of each absorber taking the column density constraints given in Table~\ref{tab:als_params} as priors \citep[see also][]{2015MNRAS.446...18C}.

For the DLA, we calculate the metallicity directly from the measured column densities. Following \citet{2012ApJ...755...89R}, we use sulphur as first preference for calculating the DLA metallicity finding $[{\rm S/H}]=-1.70\pm0.11$. We also compute the measured Fe abundance, finding $[{\rm Fe/H}]=-1.91\pm0.10$. This translates to a metallicity of $[{\rm M/H}]=-1.61\pm0.19$ if we apply a correction of $+0.3$~dex to correct from Fe abundance to $\alpha$-element metallicity, consistent with the estimate from sulphur (with the caveat that this conversion is estimated from lower redshift samples than our own, e.g. \citealt{2008A&A...480..349P, 2012ApJ...755...89R}). Whilst the estimates are consistent, we use the Sulphur derived metallicity as the total DLA metallicity ([M/H]) in the analysis that follows. 

\begin{table}
    \caption{Parameter ranges and steps adopted to create the photo-ionization model grid used to constrain the metallicities of the strong absorbers.}
    \centering
    \begin{tabular}{lccc}
         \hline
         Parameter & Min & Max & Step  \\
         \hline
         Redshift           & 4.1   & 5.3  & 0.3  \\ 
         $N$(H{\sc i})      & 16.0  & 21.0 & 0.25 \\ 
         $\rm \left[M/H\right]$ & -4.5  & 0.00 & 0.25 \\ 
         $n(H)$             & -4.25 & 0.00 & 0.25 \\ 
         \hline
    \end{tabular}
    \label{tab:cloudy_grid}
\end{table}

\begin{figure}
    \centering
    \includegraphics[width=\columnwidth]{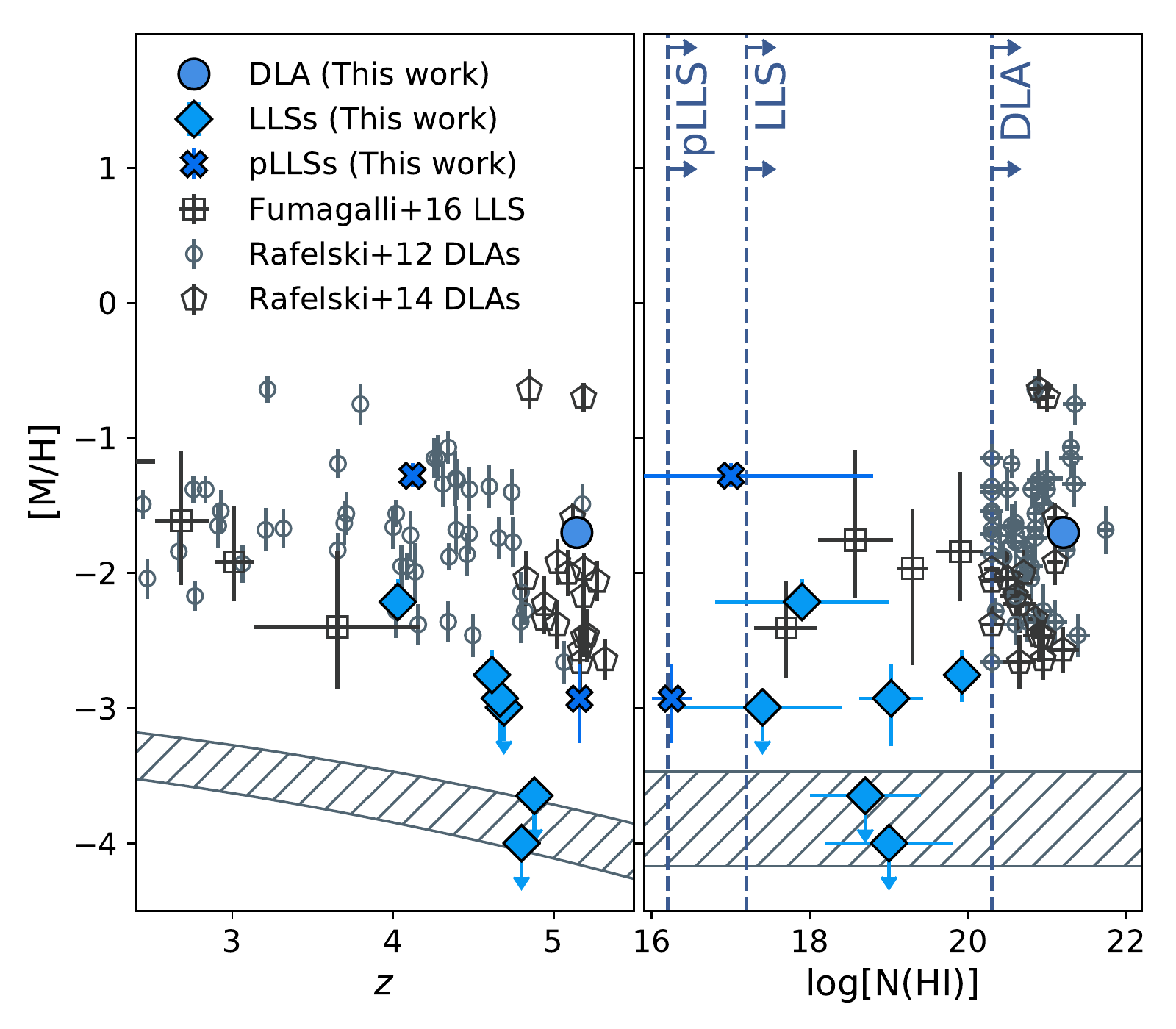}
\caption{Inferred absorber metallicities as a function of redshift (left) and H~{\sc i} column density (right). Blue circles, triangles and squares show our own sample split into DLA, LLS and pLLS subsets. Literature measurements are shown from \citet{2012ApJ...755...89R} and and \citet{2014ApJ...782L..29R} for DLAs (open circles and triangles respectively). The hatched region shows the mean metallicity of the IGM based on the \ion{C}{iv} measurements by \citet{2011ApJ...743...21S}.}
    \label{fig:ALS_props}
\end{figure}

The resultant best estimates of the metallicities, plotted versus redshift and $N$(H~{\sc i}), are shown in Fig.~\ref{fig:ALS_props}, with the error bars showing the 10th and 90th percentile range in the MCMC-derived PDF. The circle, diamonds and $\times$ symbols identify absorbers as DLAs, LLSs and pLLSs (partial Lyman Limit System) respectively (although we note the uncertainties on the pLLS and LLS column density estimates mean the classification are themselves uncertain in some cases). For reference, we also plot the properties of the DLA samples presented in \citet[][open circles]{2012ApJ...755...89R} and \citet[][open pentagons]{2014ApJ...782L..29R}, and the mean results from the LLS sample of \citet[][open squares]{Fumagalli2016}. In addition, the hatched region in the left hand panel shows the mean metallicity as a function of redshift estimated from the \ion{C}{iv} mean density as reported in \citet{2011ApJ...743...21S}.  The 9 absorbers identified in the \qso\ line of sight cover a wide range of metallicity ($\approx 3~\rm dex$), enabling us to study a diverse range of enrichment histories.  

\begin{table*}
    \caption{Galaxy-absorber associations and derived absorber metallicities. The first sample comprises absorbers purely detected via the presence of absorption by hydrogen and metal species in the quasar sightline. The second sample lists the absorbers identified via a search for saturated \ion{H}{i} absorption features within $|\Delta v|<1000$~km~s$^{-1}$.}
    \centering
    \begin{tabular}{ccccccc}
         \hline
         Absorber & $N({\rm HI})$ & [M/H] & Closest galaxy & $\rho$ & $\Delta v$ & $N_{\rm gal}$\\
         redshift & log[cm$^{-2}$] &  &  & (kpc) & (km~s$^{-1}$) & ($<1000~{\rm km~ s^{-1}})$\\
         \hline
         5.1612   & $16.25\pm0.25$  & $-2.93^{+0.25}_{-0.33}$ & Em-8 & 162.4 & $+$209.3 & 1 \\ 
         5.1448   & $21.21\pm0.05$  & $-1.70\pm0.11$  & Em-8 & 162.4 & $+$1009.4 & 1 \\ 
         4.8804   & $18.70\pm0.70$  & $<-3.65$         & Em-15 &  74.2 & $+$134.1 & 2\\ 
         4.8004   & $19.0\pm0.80$  & $<-3.99$         & Em-4 &  118.0 & $+$963.4 & 3 \\ 
         4.6917   & $17.40\pm1.00$  & $<-2.99$         & --- & --- & --- & 0 \\ 
         4.6651   & $19.03\pm0.40$  & $-2.93^{+0.26}_{-0.35}$  & --- & --- & --- & 0 \\ 
         4.6169   & $19.92\pm0.17$  & $-2.75^{+0.18}_{-0.20}$  & Cn-22 & 241.8 & $+$430.5 & 1 \\ 
         4.1220   & $17.00\pm1.80$  & $-1.28^{+0.08}_{-0.09}$ & Em-16 & 210.9 & $-$152.1 & 1 \\ 
         4.0302   & $17.90\pm1.10$  & $-2.21^{+0.16}_{-0.14}$  & --- & --- & --- & 0 \\ 
        \hline
         5.0672   & $16.9\pm1.1$  & ---      & Em-14/Cn-141 & 165.6 & $+$107.5 & 1 \\ 
         4.7437   & $16.9\pm1.4$  & ---      & Em-13 & 74.4 & $+$190.1 & 2 \\ 
         4.5474   & $17.1\pm1.2$  & ---  & Em-3 & 120.2 & $+$649.1 & 1 \\ 
         4.2884   & $17.2\pm1.1$  & ---  & Em-12 & 166.8 & $+$352.7 & 1 \\ 
         4.0862   & $16.9\pm2.1$  & --- & Cn-56 &  193.5 & $-$186.6  & 2 \\ 
         \hline
    \end{tabular}
    \label{tab:metallicity}
\end{table*}

\section{Simulation data}
\label{sec:eagle}

To provide additional context to our observations, we incorporate predictions for absorption line column densities from the {\sc eagle} suite  of cosmological  hydro-dynamical $\Lambda$CDM simulations \citep{2015MNRAS.450.1937C, 2015MNRAS.446..521S}. The simulations were run with a modified version of the smoothed particle hydrodynamics (SPH) code GADGET3, incorporating state-of-the-art numerical techniques and subgrid models used to capture physical processes important to galaxy formation and evolution. These include radiative gas cooling, star formation, mass loss from stars, metal enrichment, energy feedback from star formation and AGN  and  gas  accretion  onto,  and  mergers  of,  super-massive black holes. The efficiency of stellar feedback and the mass accretion onto black holes is calibrated to match the present-day stellar mass function of galaxies (subject to the additional constraint that the galaxy sizes need to be realistic), and the efficiency of AGN feedback is calibrated to match the observed relation between stellar mass and black hole mass.

To compare with observations, we select a representative sample of simulated galaxies from the L100N1504 simulation cube, which consists of a cube of side length 100 (comoving) Mpc and resolution $1504^3$ dark matter particles. Based on LAE clustering analyses \citep{2010ApJ...723..869O}, we select central galaxies of halos in the mass range $10^{11.2-11.8}$~M$_{\odot}$ and with star formation rates of $>0.3$~M$_{\odot}$~yr$^{-1}$. This selection, effectively matching to observed clustering of LAEs, gives galaxy space densities from EAGLE of $\rho=1.04\times10^{-3}$~Mpc$^{-3}$. This space density is consistent with observed number densities of LAEs at $z\sim4-5$ \citep[e.g.][]{2017MNRAS.471..267D}, given the flux limits of our data (see Sec.~\ref{sec:elgdet}).

Calculating accurate simulated \ion{H}{i} column densities from the simulations requires that the main ionizing processes that shape the distribution of neutral hydrogen are accounted for. After the collisional ionization (which is dominant at high temperatures), photoionization by the meta-galactic UVB radiation is the main contributor to the bulk of hydrogen ionization on cosmic scales, particularly at $z\gtrsim1$ \citep[e.g.][]{2013MNRAS.430.2427R}. \citet{2015MNRAS.452.2034R} show that radiation from local sources is important at small scales and high column densities, reducing the covering factors of LLSs and DLAs by only $\approx10\%$ at $\approx R_{\rm vir}$, but by up to $\approx60\%$ at $\approx0.1R_{\rm vir}$ for high ($N(HI)\gtrsim10^{22}$~cm$^{-2}$) column density systems \citep[see also][]{2013MNRAS.431.2261R, 2013ApJ...765...89S, 2014MNRAS.438..529R}. The computationally expensive treatment of ionisation by local sources is not included in the simulation volume used in this work and so we note that covering fractions of \ion{H}{i} may be over-predicted in the simulated volumes by $\approx10-20\%$ for LLSs in our sample and $\approx20-50\%$ for DLAs (albeit within $\lesssim R_{\rm vir}$, equivalent to $\approx30-40$~kpc as discussed later). On the other hand, the HI covering fraction may be underrepresented based on resolution effects of the simulations themselves  as recently demonstrated in zoom-in simulations \citep[][]{ 2019ApJ...873..129P, 2019MNRAS.482L..85V, 2019MNRAS.488.3634R}.

Following \citet{2013MNRAS.431.2261R, 2015MNRAS.452.2034R, 2016MNRAS.460.2157O}, we calculate column densities using SPH interpolation and projecting the ion content of desired regions onto a 2D grid (with 1 pkpc resolution). This projection into 2D was performed through a box of $\pm600$~kpc around each galaxy, equivalent to $\approx20~R_{\rm vir}$ from each galaxy along the line of sight. We project maps for each galaxy in the $x$, $y$, and $z$ directions that are 600 pkpc across with the depth of 1200 kpc. SPH particles are smoothed onto a grid using the SPH kernel function. We then calculate the median column density profiles and percentile ranges from the resultant sample, using the three projections for each galaxy selected.

\section{The circumgalactic medium at \MakeLowercase{$z\approx 5$}}
\label{sec:assoc}

\subsection{Galaxy-absorber associations}

\subsubsection{Distribution of neutral hydrogen around $z=4-5$ LAEs}

\begin{figure*}
    \centering
    \includegraphics[width=\textwidth]{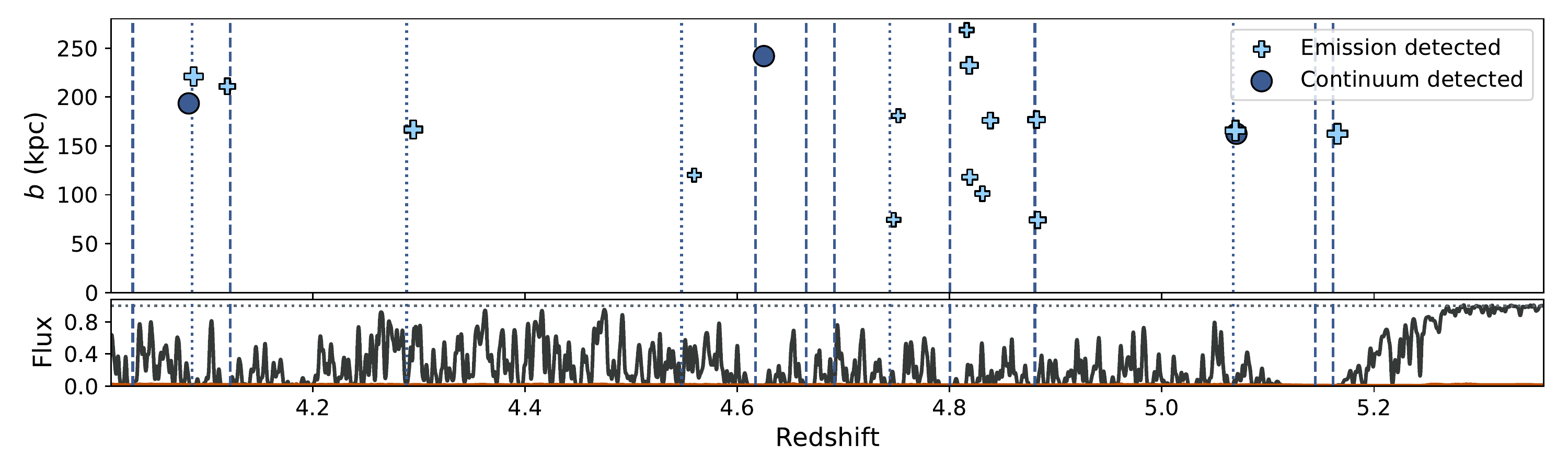}
    \caption{Redshift distribution of galaxies and absorbers. The top panel shows emission (plus symbols) and continuum (circle symbols) detected galaxies as a function of redshift and impact parameter (note the galaxy at $z\sim5.07$ is detected by both the emission and continuum algorithms). The lower panel shows the flux normalised quasar spectrum. Dashed lines covering both panels highlight redshift positions of strong absorption lines with associated \ion{C}{iv} absorption. Dotted lines indicate the positions of the strongest identified absorption features in the proximity of galaxies with no strong absorbers identified in the first pass of the quasar spectrum.}
    \label{fig:als_velimp}
\end{figure*}

\begin{figure}
    \centering
    \includegraphics[width=\columnwidth]{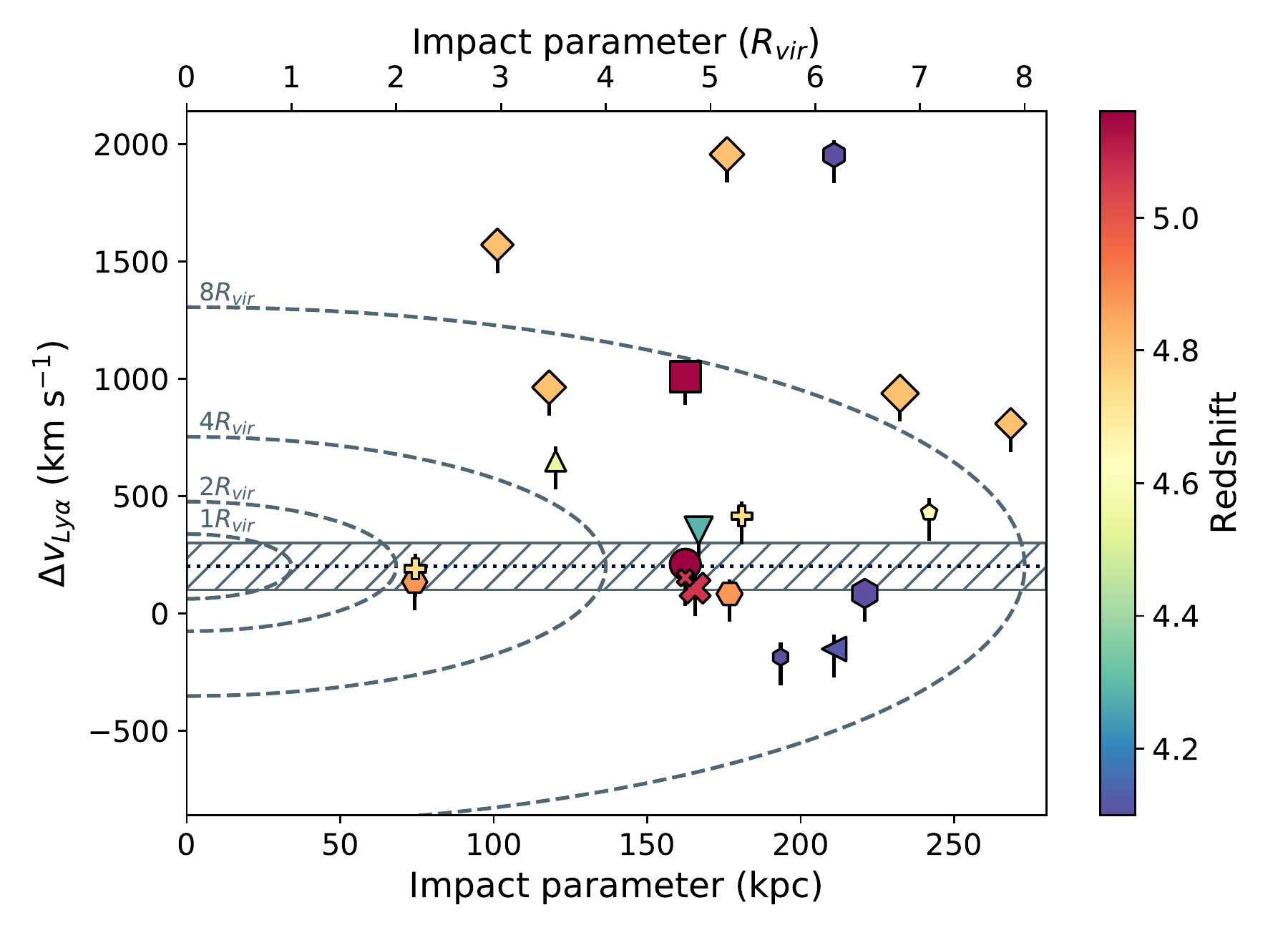}
    \caption{Associations between LAEs and strong absorbers along the sightline to \qso. Points are colour coded by redshift with sizes proportional to the ${\rm log}(L_{\rm line})$ of each galaxy. 
    Galaxies clustered around the same absorber share the same symbol.
    The dotted line and shaded region show $\Delta v=200\pm100$~km~s$^{-1}$, i.e. the typical offset of Ly$\alpha$ emission from intrinsic galaxy redshift. The top axis scale is in units of $R_{\rm vir}$ is calculated assuming a $M_{\rm h}=10^{11.5}$~M$_\odot$ halo at $z=4.8$.}
    \label{fig:als_velimp}
\end{figure}

The observations reveal a wide range of galaxy environments around $z>4$ strong absorption systems. Two of the strong absorbers show galaxies with Ly$\alpha$ emission detected within the field of view at $|\Delta v_{-200}|<100$~km~s$^{-1}$, with one of these showing two LAEs within this velocity range. Three more of the absorption systems are seen to lie within $|\Delta v_{-200}|\approx1000$~km~s$^{-1}$ of single LAEs, whilst a further absorption system lies at $|\Delta v_{-200}|\approx1000-2000$~km~s$^{-1}$ from 5 LAEs tracing a galaxy over-density at $z\approx4.83$. The remaining 3 systems show no detected galaxies within $|\Delta v_{-200}|\approx2000$~km~s$^{-1}$. In summary, 6 of the 9 strong \ion{H}{I} absorption systems are found to lie within $|\Delta v_{-200}|\approx1000$~km~s$^{-1}$ and $\approx250$~kpc of at least one detected Ly$\alpha$ emitter at the depth of our search.

Comparing to targeted and small-scale IFU searches for galaxies around such absorption systems, where detection rates of $\approx10\%$ are common at $z\gtrsim1$ on scales of $\lesssim 100$~kpc \citep[e.g.][]{2012MNRAS.419.3060P}, the MUSE observations offer a more comprehensive overall picture, with at least one galaxy detected for 66\% of the absorbers (up to scales of $\approx300$~kpc). Such high incidence of galaxies identifications per absorber are comparable with similar MUSE searches at $z\approx 3$ \citep{2019MNRAS.487.5070M}.

From clustering analyses \citep[e.g.][]{2007ApJ...671..278G, 2010ApJ...723..869O, 2016MNRAS.456.4061B}, LAE samples across redshifts of $3\lesssim z\lesssim7$ are inferred to inhabit dark matter halos with masses of $M_h\approx10^{11.5\pm0.3}~\rm M_\odot$. We derive an estimated virial radius from this halo mass as:

\begin{equation}
    R_{\rm vir}=\left(\frac{3M_{\rm h}}{4\pi\rho_{\rm crit}(z)\Delta_{\rm c}}\right)^{1/3}
    \label{eq:vir_rad}
\end{equation}

\noindent where $\rho_{\rm crit}$ is the critical density and $\Delta_{\rm c}\approx18\pi^2$ \citep{1998ApJ...495...80B}. This gives  $R_{\rm vir}\approx30-50$~kpc at $z=4-5$. Translating this into a Hubble flow velocity, such a separation would equate to $\Delta v\approx20$~km~s$^{-1}$. Indeed, as the Hubble parameter at $z\approx4-5$ is $H\approx500$~Mpc/(km s$^{-1}$), the vertical axis in Fig.~\ref{fig:als_velimp} extends to $\approx 2$~Mpc. Following \citet{1998ApJ...495...80B}, a halo mass of $\approx10^{11.5\pm0.3}~\rm M_\odot$ at $z\approx4-5$ corresponds to a velocity dispersion of $\approx140$~km~s$^{-1}$. The concentric dashed ellipses in Fig.~\ref{fig:als_velimp} show scales of 1, 2, 4 and 8$\times$ the typical virial radius and velocity dispersion for a $M_{h}=10^{11.5}$~M$_\odot$ halo. Given the assumed halo mass and associated virial radius, all of the proximate galaxies lie $\gtrsim2R_{\rm vir}$ from the sightline absorbers.

Despite the high number of sources identified near strong absorbers, it is important to maintain the awareness that the detected LAE galaxies are not a comprehensive census of $z\sim4-5$ galaxies within the field. Beside the incompleteness related to the sensitivity of our data, MUSE observations have also limited sensitivity to sources with moderate dust obscuration or where the galaxy Ly$\alpha$ emission is significantly absorbed by the ambient gas. For example, \citet{2003ApJ...588...65S} report 60\% of LBGs at $z\sim3$ to show Ly$\alpha$ primarily in absorption, which are more difficult to identify in fainter objects due to limited $S/N$ in the continuum. 
It is therefore possible that undetected sources lie in closer proximity to the sightline for some absorbers, and likewise that the 3 absorbers with no galaxy counterpart may in fact still arise from objects that are undetected. 
Nevertheless, the current observations provide a valuable census of  relatively-bright LAEs up to scales of $\approx 6-8~R_{\rm vir}$.


\begin{figure}
    \centering
    \includegraphics[width=\columnwidth]{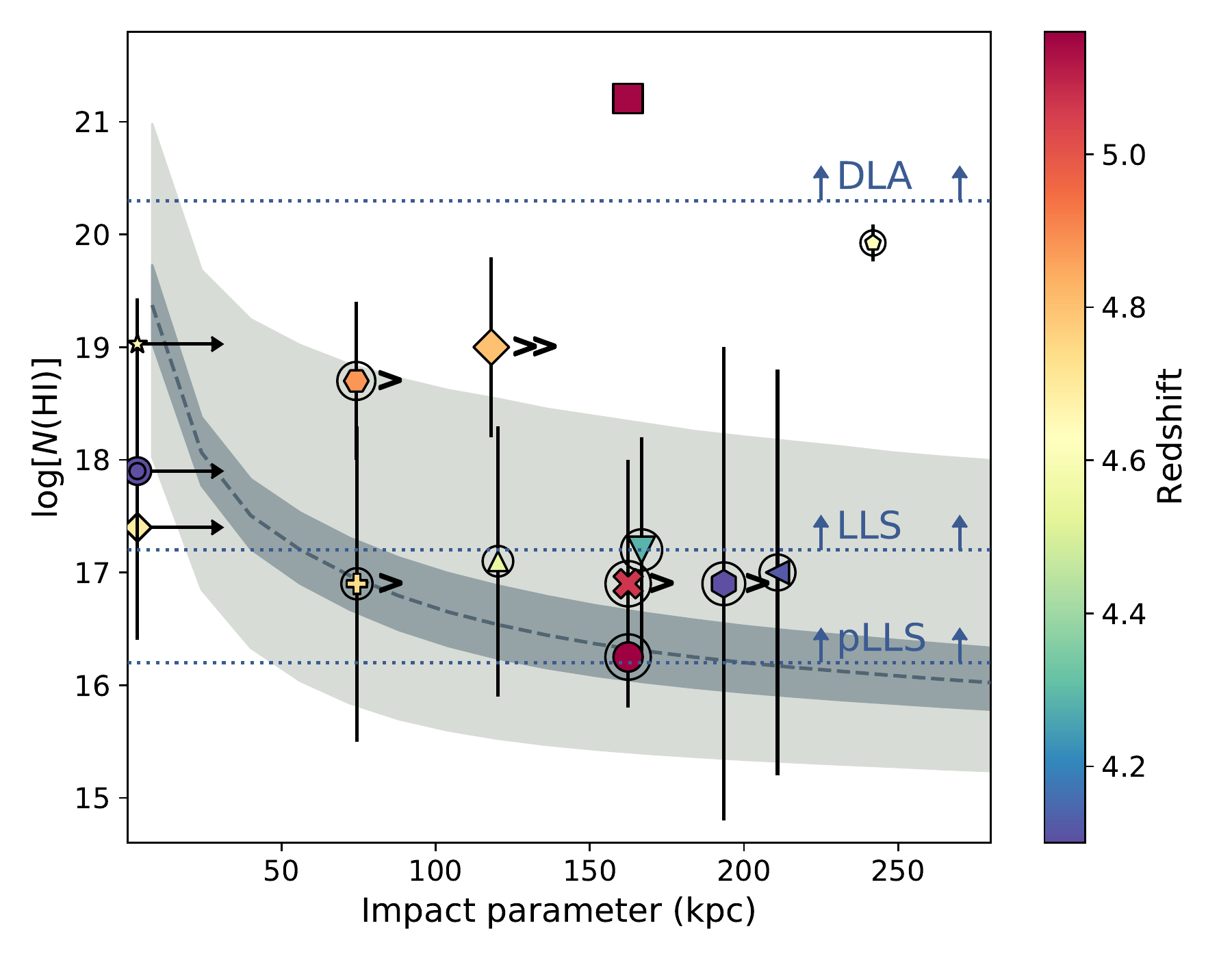}
    \caption{Neutral hydrogen column densities as a function of impact parameter from the nearest detected galaxy within $|\Delta v_{-200}|<1000$~km~s$^{-1}$. Symbols are consistent with those given in Fig.~\ref{fig:als_velimp}. Absorber-galaxy pairs lying within $|\Delta v_{-200}|<500$~km~s$^{-1}$ are circled. Where multiple galaxies are detected at the same redshift as a given absorber, chevrons are added to the symbol denoting the number of additional galaxies. Limits to the right mark the column density of the absorbers for which we do not identify galaxies. The dashed curve and shaded regions illustrate the median and [32\%,68\%] and [10\%,90\%] column density ranges predicted by the {\sc eagle} simulations.}
    \label{fig:als_nhimet}
\end{figure}

In Fig.~\ref{fig:als_nhimet}, we show the \ion{H}{I} column density ranges for the absorbers as a function of galaxy impact parameters for all galaxies within $|\Delta v_{-200}|<1000$~km~s$^{-1}$ of an absorber. The differing symbols again represent the different absorbers, are colour coded by redshift and are assigned consistently with Fig.~\ref{fig:als_velimp}. Galaxies lying within $|\Delta v_{-200}|<500~{\rm km~s}^{-1}$ of an absorber are plotted with a ring around their primary marker, whilst those with additional galaxies at the same redshift are given chevrons (with the number of chevrons denoting the number of additional galaxies within $1000$~km~s$^{-1}$ of the absorber). The dotted horizontal lines denote the observational criteria for the different categories of strong \ion{H}{I} absorbers. The dashed curve and shaded regions show the median, [32\%,68\%] and [10\%,90\%] column density ranges predicted as a function of impact parameter around the galaxy sample extracted from the {\sc eagle} simulation volume.

As would be expected given the galaxy incompleteness and the patchy nature of the CGM, this analysis shows a large scatter in column density versus impact parameter. Four of the detected galaxies show column densities broadly consistent with the range predicted by the simulation, with the weakest two absorbers aligning with galaxies at large impact parameters ($b\gtrsim150$~kpc). The comparison with simulations further reinforce the idea that galaxies may exist at smaller impact parameters in any of these 4 cases, as well as the 2 cases where we see enhanced column densities at large impact parameters compared to the {\sc eagle} predictions. As a slight aside, it is also interesting to note that the {\sc eagle} predicted distribution appears to trace lower redshift results well \citep[e.g.][]{2001ApJ...559..654C, 2018ApJS..237...11K}. 

\begin{figure}
    \centering
    \includegraphics[width=\columnwidth]{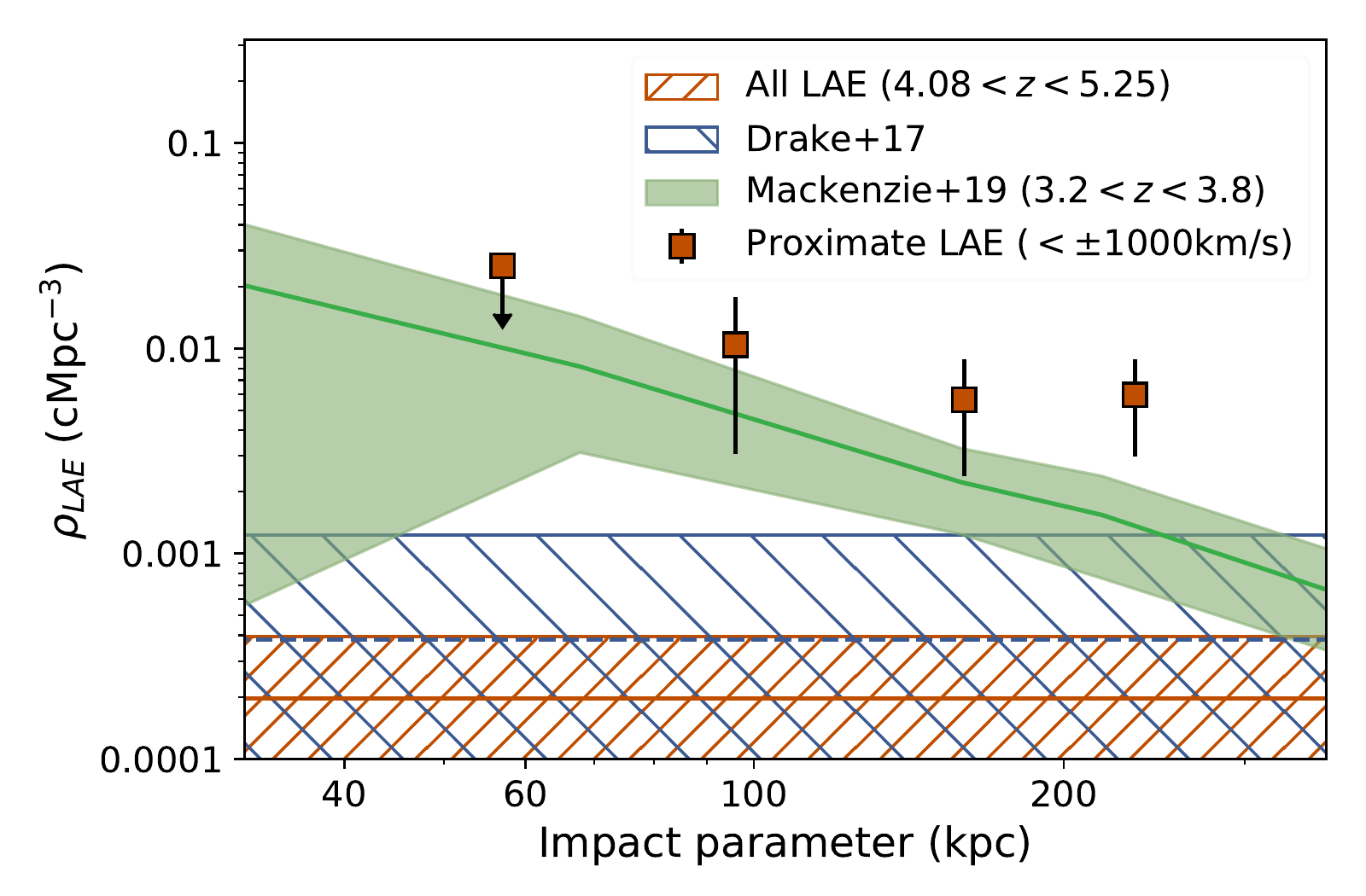}
    \caption{LAE space density within $|\Delta v_{-200}|=\pm1000$~km~s$^{-1}$ of identified strong absorption systems, as a function of impact parameter (square points). The light green band shows the overdensity measured around lower-redshift DLAs by \citet{2019MNRAS.487.5070M}.
    The hatched regions show the mean spatial density of all galaxies detected over the redshift range $4.08<z<5.16$ in our MUSE field and the space density of LAEs at $z=4.5$ from the luminosity function of \citet{2017A&A...608A...6D}, given our observational Ly$\alpha$ flux limits.}
    \label{fig:LAErhoVimp}
\end{figure}

Whilst inevitably there will be galaxies at the studied redshifts that lie below both our continuum and line emission detection thresholds, the MUSE detected galaxies still provide us with a valuable tracer population for the over-density of objects in the proximity of strong absorbers. To this end, we now estimate the average galaxy environment around our sample of absorption systems with $N({\rm HI})>10^{16.2}~\rm cm^{-2}$, and compare it with previous studies and the general field population. 
Specifically, we compute the space density of galaxies detected in our sample within $|\Delta v_{-200}|<1000$~km~s$^{-1}$ of the identified absorption line systems as a function of impact parameter (i.e. within cylindrical volumes assuming the velocity offsets transform to a distance via the Hubble flow), treating the upper end of this velocity range as a physical spatial separation along the line of sight. Note that for this analysis, we only use the 9 absorption systems that were detected without prior knowledge of the galaxies identified in the MUSE data. The result is shown in Fig.~\ref{fig:LAErhoVimp} and is given by the square points. The hatched regions show the mean spatial density of galaxies over the redshift range $4.08<z<5.16$ in our MUSE field and the space density of LAEs at $z=4.5$, given our observational Ly$\alpha$ flux limits, calculated from the luminosity functions of \citet{2017MNRAS.471..267D, 2017A&A...608A...6D}. A clear and significant over-density of galaxies is evident on average around the absorption systems in our field, establishing a physical connection between the gas probed in absorption and the distribution of galaxies within the adopted window. \citet{2019MNRAS.487.5070M} presented a similar analysis showing the over-density of LAEs around DLAs at $z\approx3.2-3.8$, shown here as the filled green region. Our own result at this higher redshift shows a comparable clustering of LAEs (in agreement within $1\sigma$ level) around absorption systems, including lower column density ones. 

\subsubsection{Metal absorption around $z\sim4-5$ LAEs}

\ion{C}{iv} absorption in quasar sightlines has been shown to have a significant correlation with galaxies via clustering analyses at $z\sim3$ \citep[e.g.][]{2003ApJ...584...45A, 2011MNRAS.414...28C, 2014MNRAS.445..794T}, whilst lower redshift studies (i.e. $z\sim0.01$) have shown that \ion{C}{iv} systems are more common in low galaxy-density regions \citep[e.g.][]{2016ApJ...832..124B}.

\begin{figure}
    \centering
    \includegraphics[width=\columnwidth]{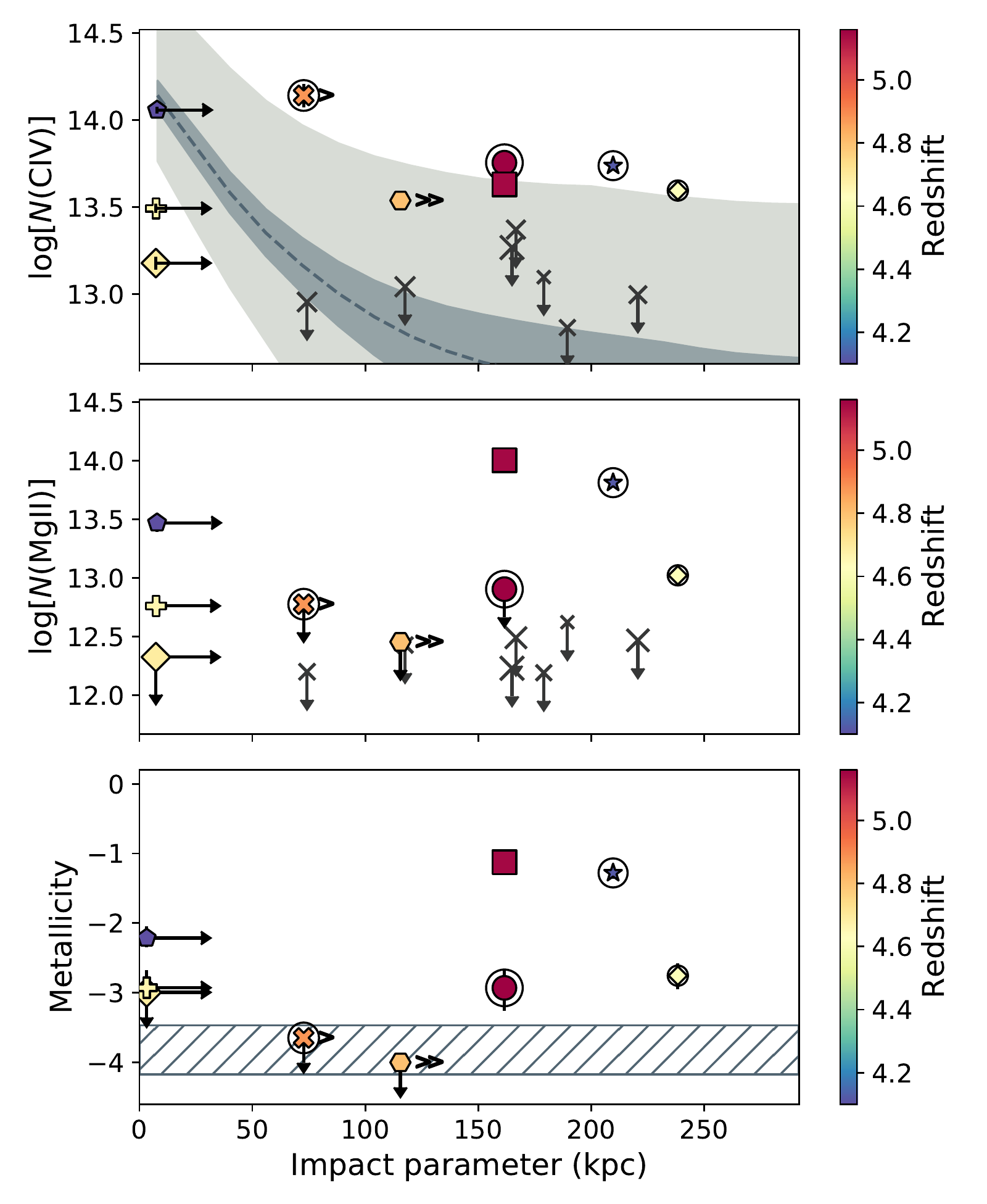}
    \caption{Column density of \ion{C}{iv} and Mg~{\sc ii} and inferred metallicity as a function of impact parameter to galaxies identified within $|\Delta v_{-200}|<1000$~km~s$^{-1}$. Galaxies lying within $\Delta v=\pm500$~km~s$^{-1}$ of absorbers are circled.
    Each detection is colour coded by redshift as given by the colour bar, with the symbols assigned as in Fig.~\ref{fig:als_velimp}. Where multiple galaxies are
    detected at the same redshift as a given absorber, chevrons are added to
the symbol denoting the number of additional galaxies.
Upper limits (grey $\times$ symbols, 3$\sigma$ upper limits) are estimated within $\pm250$~km~s$^{-2}$ of high-confidence galaxies at $4<z<5.2$ with no detected \ion{C}{iv}. Absorbers with no detected proximate galaxy are instead shown by lower limits in impact parameter.}
    \label{fig:als_nciv}
\end{figure}

With these new observations, we can now investigate the link between  \ion{C}{iv} and galaxies at $z>4$. In the top panel of Fig.~\ref{fig:als_nciv}, we show \ion{C}{iv} column density versus galaxy impact parameters. Where a \ion{C}{iv} detection is made, we show all galaxies within $|\Delta v_{-200}|=\pm1000$~km~s$^{-1}$. In this figure, we also include upper limits on the column density (grey crosses) computed with $|\Delta v_{-200}|=\pm250$~km~s$^{-1}$ from galaxies within the field of view with no detected absorber. Finally, we mark the three absorbers where we fail to detect any galaxies by lower limits with respect to the impact parameter. As in Fig.~\ref{fig:als_nhimet}, the dashed and shaded regions show the median, [32\%,68\%] and [10\%,90\%] column density ranges predicted around the galaxy sample extracted from the {\sc eagle} simulation volume. The central panel shows our equivalent to the top panel, but for Mg~{\sc ii}, whilst the lower panel shows the estimated metallicities derived from the MCMC analysis described previously. The hatched region in the lower panel marks the range in metallicities of the CGM at $4<z<5$ estimated from \citet{2011ApJ...743...21S} as in Fig.~\ref{fig:ALS_props}.

We find that of the 17 galaxies detected in the MUSE field of view between $4 \lesssim z \lesssim 5.3$, 8 lie within 1000 km~s$^{-1}$ of a detected C~{\sc iv} absorber (with a further 2 galaxies within 2000 km~s$^{-1}$ of C~{\sc iv} absorption). The remaining galaxies show upper limits on C~{\sc iv} absorption. Fig.~\ref{fig:als_nciv} highlights that the 3 systems with the highest \ion{C}{iv} column densities are all within 500~km~s$^{-1}$ of a detected galaxy. Mindful of the small sample size and selection biases described above, this finding is at least consistent with the conclusion of previous work \citep{meyer2019,keating2019}, according to which galaxies at these redshifts are expected to reside within ionized regions, and thus cluster more with high-ionisation lines rather than neutral species. We note, however, that this correspondence is not unique, and that gas with different ionization stages is still found close to galaxies at these redshifts. Indeed, our sample contains also galaxies in proximity to absorbers showing high column density of both  C~{\sc iv} and  Mg~{\sc ii}, but with low column density of  H~{\sc i}. We note how there seems to be a lack of clear correlation between metallicity and proximity to galaxies both in projection and in velocity space. Evidence of inhomogeneous enrichment is indeed building up at lower redshifts \citep[e.g.][]{2011Sci...334.1245F, 2019MNRAS.487.5070M, 2019MNRAS.tmp.2667L, 2019MNRAS.490.1451F}, and it is thus not surprising to find hints of similar inhomogeneity at these higher redshifts, where the time available for enriched pockets to grow is limited. 

\begin{figure}
    \centering
    \includegraphics[width=\columnwidth]{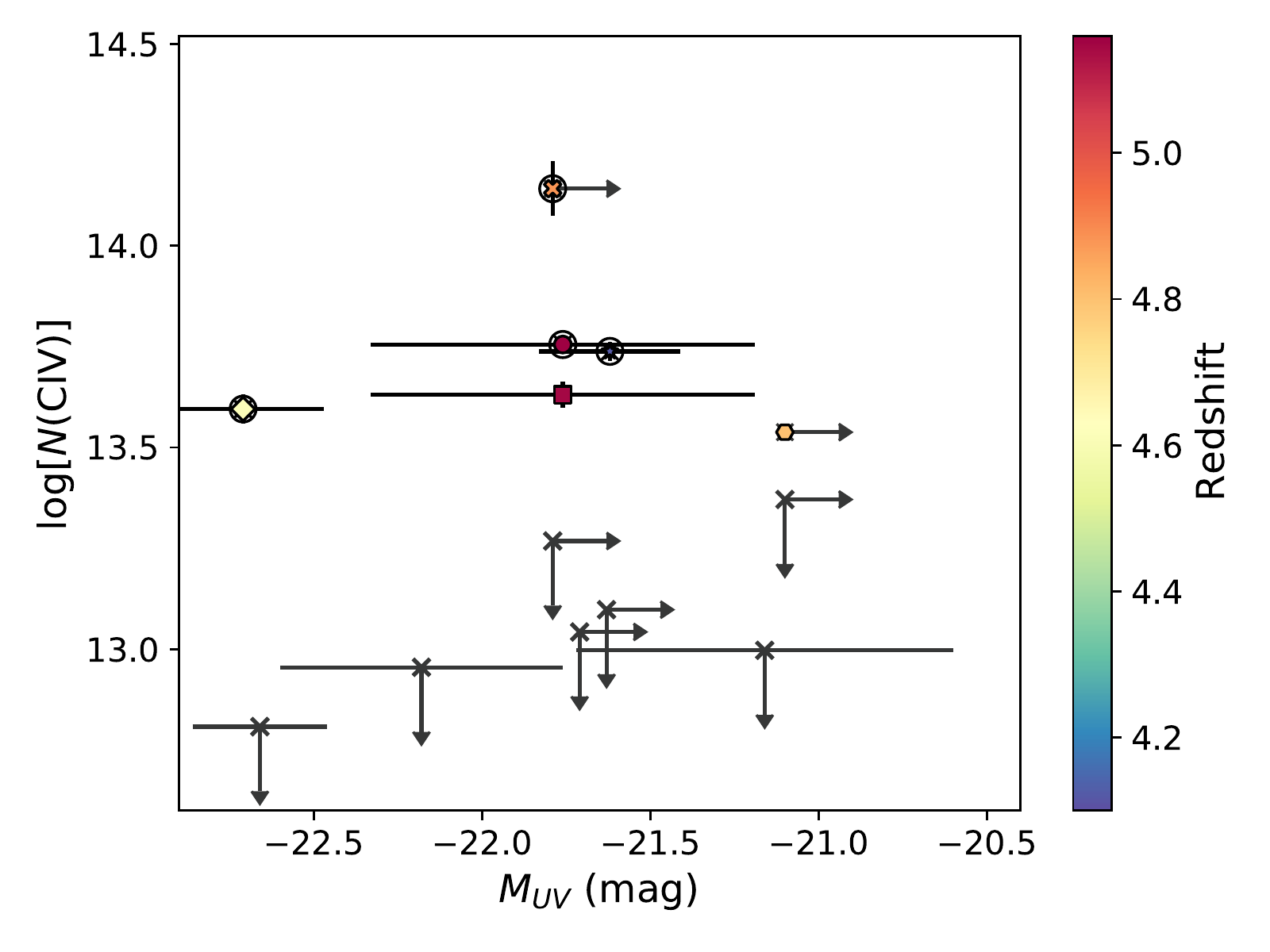}
    \caption{Column density of \ion{C}{iv} as a function of the absolute UV magnitude of the nearest galaxy identified within $|\Delta v_{-200}|<1000$~km~s$^{-1}$. Galaxies lying within $\Delta v=\pm500$~km~s$^{-1}$ of absorbers are circled.
    Each detection is colour coded by redshift as given by the colour bar, with the symbols assigned as in Fig.~\ref{fig:als_velimp}. Where multiple galaxies are detected at the same redshift as a given absorber, chevrons are added to the symbol denoting the number of additional galaxies. Upper limits (grey $\times$ symbols, 3$\sigma$ upper limits) are estimated within $\pm250$~km~s$^{-2}$ of high-confidence galaxies at $4<z<5.2$ with no detected \ion{C}{iv}.}
    \label{fig:civ_muv}
\end{figure}

When compared to the {\sc eagle} simulated data, we find in our observations a significant fraction of upper limits at impact parameters of $> 100~\rm kpc$, consistent with the predictions from simulation. However, we also find several absorbers with high column density of C~{\sc iv}, in excess to the {\sc eagle} predictions for a given impact parameter. While this may reflect again incompleteness in our search towards low impact parameters, a deficit of strong C~{\sc iv} near simulated galaxies has been documented in the literature before \citep{2016MNRAS.459.2299F}, and may be common feature of simulations. Adding to this line of enquiry, we show in Fig.~\ref{fig:civ_muv} the \ion{C}{iv} absorber column densities as a function of the UV absolute magnitude of the nearest galaxy. There is no clear correlation between the UV brightness of the galaxies and the \ion{C}{iv} column density, although the strongest absorber does lie in the proximity of one of the faintest galaxies (which also happens to be the closest galaxy to the quasar sightline).

\begin{figure*}
    \centering
    \includegraphics[width=0.8\textwidth]{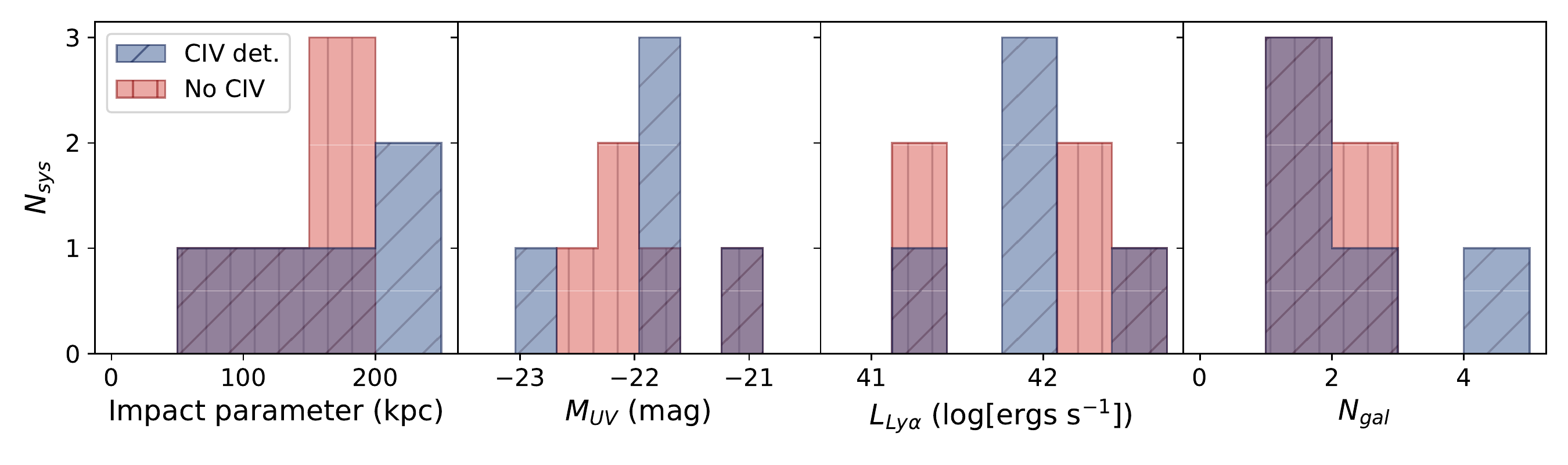}
    \caption{Properties of closest detected galaxies (including both emission and contfinuum detected) split by whether proximate \ion{C}{iv} absorption is detected (blue diagonal hatched histograms) or not detected (red vertical hatched histograms).}
    \label{fig:civ_gals}
\end{figure*}

Further, we plot in Fig.~\ref{fig:civ_gals} the distributions of closest galaxy properties with the blue diagonal hatched histograms showing galaxies with proximate \ion{C}{iv} absorption  and the red vertical hatched histograms showing galaxies with no proximate \ion{C}{iv} absorption. We see no clear difference between the two samples for impact parameter, UV brightness $M_{\rm UV}$, Ly$\alpha$ flux and galaxy environment ($N_{\rm gal}$).

\begin{figure}
    \centering
    \includegraphics[width=0.8\columnwidth]{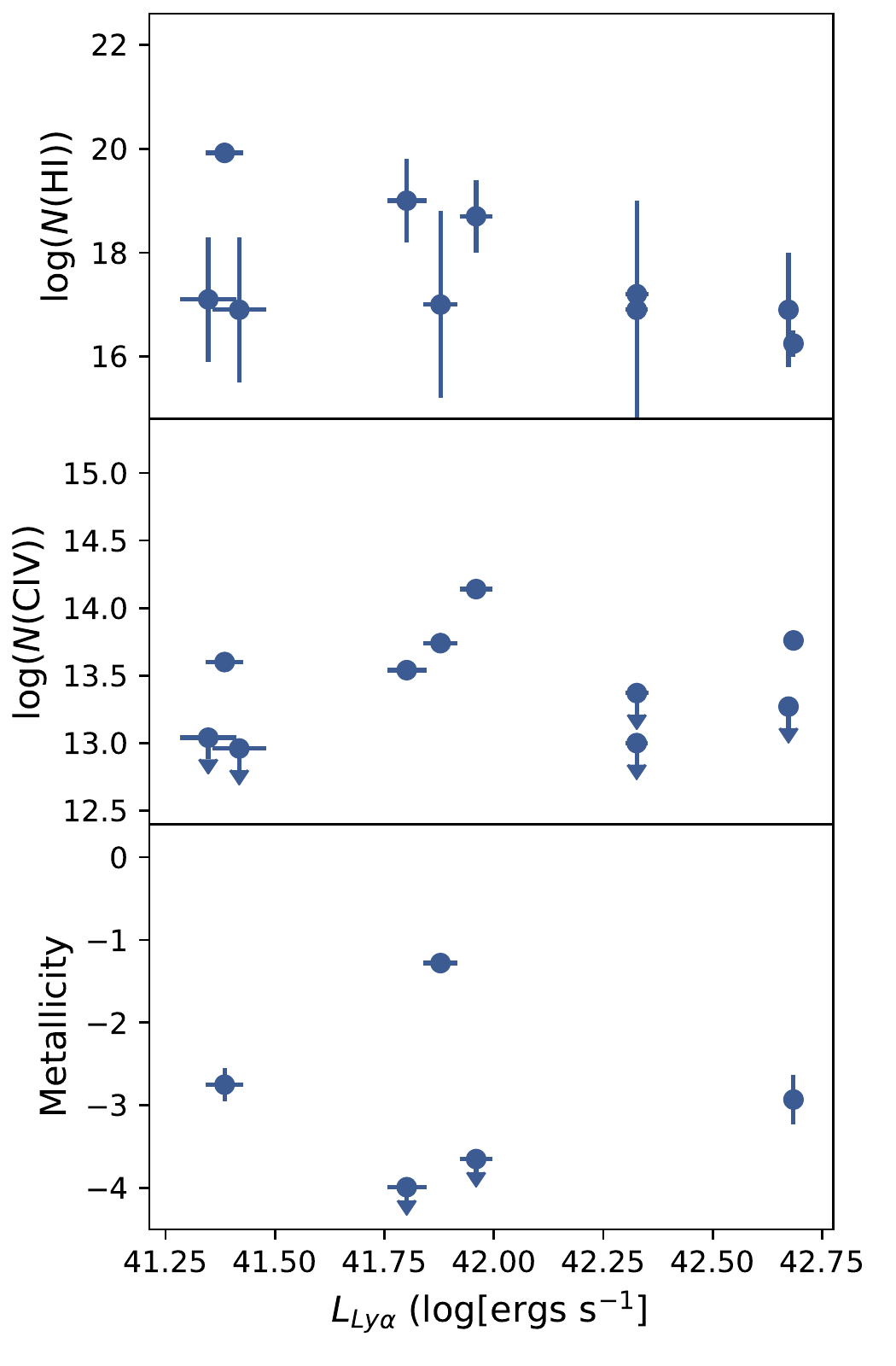}
    \caption{Absorption line properties as a function of nearest galaxy Ly$\alpha$ luminosity, showing the \ion{H}{i} column density (top panel), the \ion{C}{iv} column density (middle panel) and the absorber metallicity (bottom panel).}
    \label{fig:gLya_aprops}
\end{figure}

Inverting the question, we now investigate the absorption line system properties as a function of the nearest galaxy Ly$\alpha$ luminosity. We show the proximate absorber \ion{H}{i} column density, \ion{C}{iv} column density, and metallicity as a function of galaxy Ly$\alpha$ luminosity in the top, middle and bottom panels of Fig.~\ref{fig:gLya_aprops} respectively. 

We see little indication of any clear correlations between the absorber system properties and the proximate galaxy Ly$\alpha$ luminosity. The galaxies are found in the proximity of a range of absorbers at $N({\rm HI})\gtrsim10^{16}$~cm$^{-2}$, covering a wide range in metallicities with no clear correlation with galaxy Ly$\alpha$ emission. Of course the sample probes a wide range of impact parameter between galaxy and absorber, which as suggested by the EAGLE predictions in Fig.~\ref{fig:als_nhimet} and Fig.~\ref{fig:als_nciv} will contribute to scatter in the results. In addition, the simulations predict a significant scatter in the absorber column densities for a given galaxy halo, such that the column density of gas any given sightline may pass through is modulated by the patchy covering factor at any specific location. Our observations seem to support this in so much as the absorber properties show little correlation with the measured galaxy properties in this small sample. 

\subsection{The gas environment of $z>4$ groups}

Looking in more detail at the two incidences of multiple galaxies at the same redshift as a given absorption line system (those highlighted by $\times$ and hexagonal symbols), we find in both cases prominent \ion{H}{I} absorption giving rise to LLSs, which themselves are found to have low metallicity ($\lesssim10^{-3}$). 

Considering the group of five emitters detected at $z\approx4.82$, these lie along an axis on the plane of the sky from East to West. This can be seen in Fig.~\ref{fig:lyamap}, where the $z\sim4.81$ structure is traced by LAE sources 1, 4 and 7 clustered together closely in the East to 19 and 21 scattered in the West of the image. The galaxy pair at $z=4.88$ (LAEs 2 and 15) lies along a similarly projected axis on the sky. Should the five $z\approx4.82$ galaxies lie in a galaxy group or proto-cluster environment, a halo mass of $M_{\rm h}\approx10^{12.5-13.5}$~M$_\odot$ would perhaps be reasonable \citep[e.g.][]{2001MNRAS.321..372J, 2005Natur.435..629S}, which would give a virial radius of $r_{\rm vir}\approx80-160$~kpc. In projection then at least, the strong absorber may well lie within $\approx1-2R_{\rm vir}$ of a group environment traced by these galaxies.


Similarly to the C~{\sc iv} case  of \citet{2013ApJ...779L..17B} at low redshift, for example, the coincidence of multiple galaxies with metal poor gas may be a case of intergalactic gas accreting onto the galaxy group, or it may be tracing the CGM of an undetected galaxy in the observed group environment. Indeed, that these two relatively pristine absorption systems align with apparently dense galaxy environments also finds commonality with recent studies of the galaxy environment of metal poor LLSs at $z\sim3$ \citep{Fumagalli2016, 2019MNRAS.tmp.2667L}. As in the two cases here, the $z\sim3$ metal-poor environments show galaxy over-densities at $80~{\rm kpc}\lesssim b\lesssim 300$~kpc, within $|\Delta v_{-200} |\lesssim1000$~km~s$^{-1}$. Similarly to these previous works, we argue that the low metallicities that we measure for these systems suggests they are not outflows in nature, but more likely tracing cold gas within the cosmic web of gas surrounding the galaxy over-density. Indeed, as with the previous examples at $z\approx3.1$, the presence of an over-density strongly favours this gas ultimately being accreted onto the galaxy population with time, further fuelling star formation in this region.

Our own results resolving LAEs around strong absorption systems coincident with \ion{C}{iv} absorption are complemented by two concurrent studies: \citet{2020arXiv200103498F} and \citet{2020arXiv200104453G}. Both studies see a clear trend for \ion{C}{iv} absorption to be coincident with the galaxy population at scales of $\sim100-200$~kpc, as we have also shown here. Further \citet{2020arXiv200104453G} also report a preference for the galaxies in the proximity of \ion{C}{iv} absorption to be towards the faint end of the Ly$\alpha$ luminosity function, something that we do not observe in our own sample.

\subsection{Extended emission associated to the quasar}

Complementing the investigation of the CGM in absorption, the MUSE data also provides a probe of the CGM via Ly$\alpha$ emission around the background quasar, exploiting the presence of the bright central active galactic nucleus which illuminates the surrounding gas \citep[e.g.][]{2006A&A...459..717C, 2009MNRAS.400..843G, Cantalupo+14, Martin+14, Hennawi+15, Fumagalli2016, Borisova2016, ArrigoniBattaia2019, 2019ApJ...887..196F}. Whilst the LAEs and LBGs predominantly probe halo masses of $\sim10^{11}$~M$_\odot$ \citep[e.g.][]{2003ApJ...584...45A, 2010ApJ...723..869O, 2016MNRAS.456.4061B, 2013MNRAS.430..425B}, quasars are more often found in higher mass halos \citep[e.g.][]{2009ApJ...697.1634R, 
2016MNRAS.459.1179C, stottinprep} offering a probe of the gas in halos at the higher end of the halo mass function.

For this analysis, we first prepare the MUSE data cube by subtracting the quasar PSF and the continuum of other sources using the {\sc CubEx} tools, as described in detail in \citet{Borisova2016} and \citet{ArrigoniBattaia2019}. After smoothing the cube with a Gaussian filter of 3 pixels in radius, we search for extended Ly$\alpha$ emission at the quasar redshift, running {\rm CubEx} to detect connected pixels with $S/N \ge 3$. We repeat this procedure on the mean- and median-combined cube, finding consistent results.  

\begin{figure}
    \centering
    \includegraphics[scale=0.56]{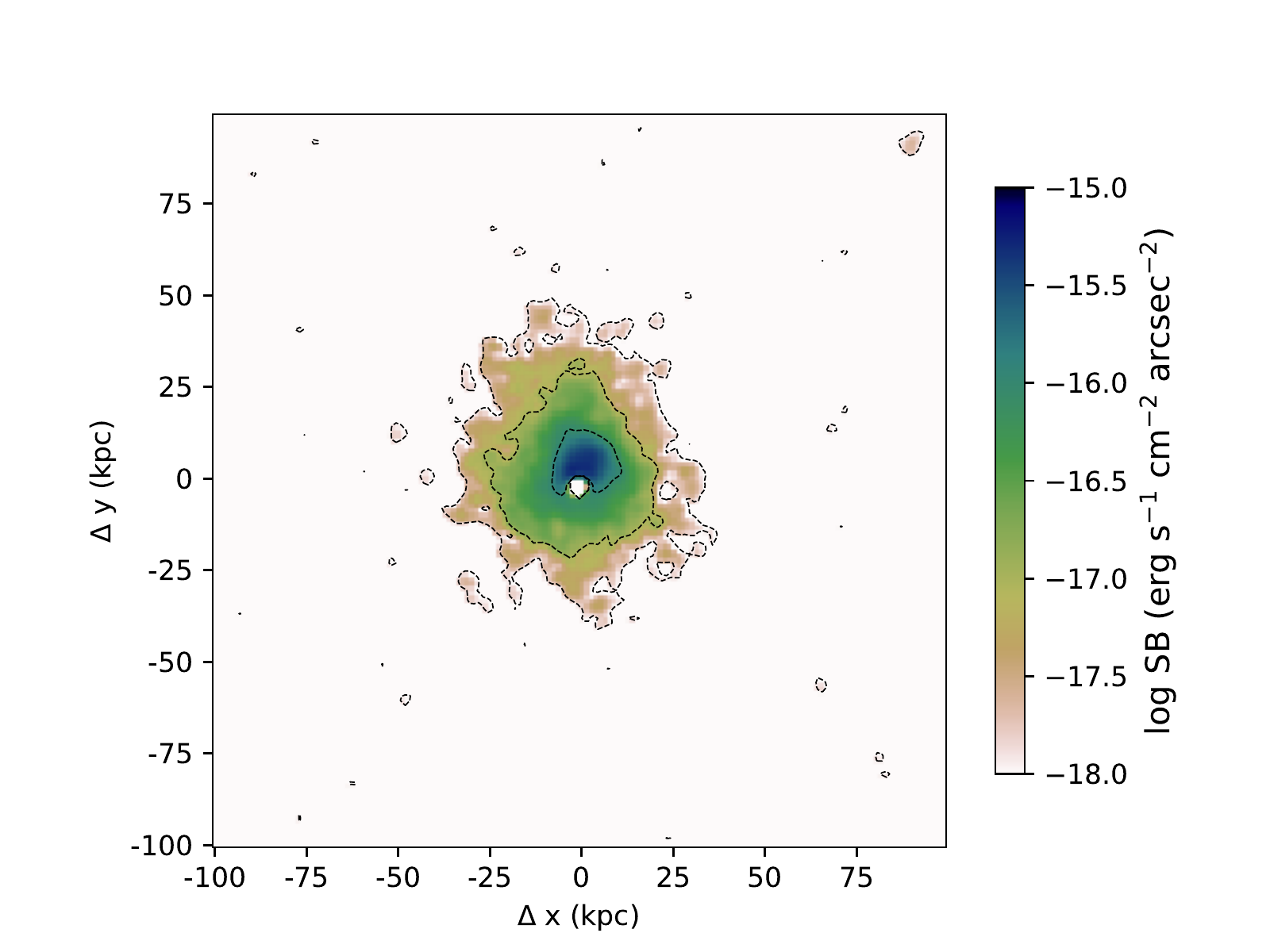}
    \caption{Optimally-extracted narrow-band image of the diffuse Ly$\alpha$ emission at the redshift of the quasar \qso. The dotted contours are at surface brightness levels of $10^{-18,-17,-16}$~\sbline.}\label{fig:qsoneb}
\end{figure}

\begin{figure*}
    \centering
    \begin{tabular}{c|c}
    \includegraphics[width=\textwidth]{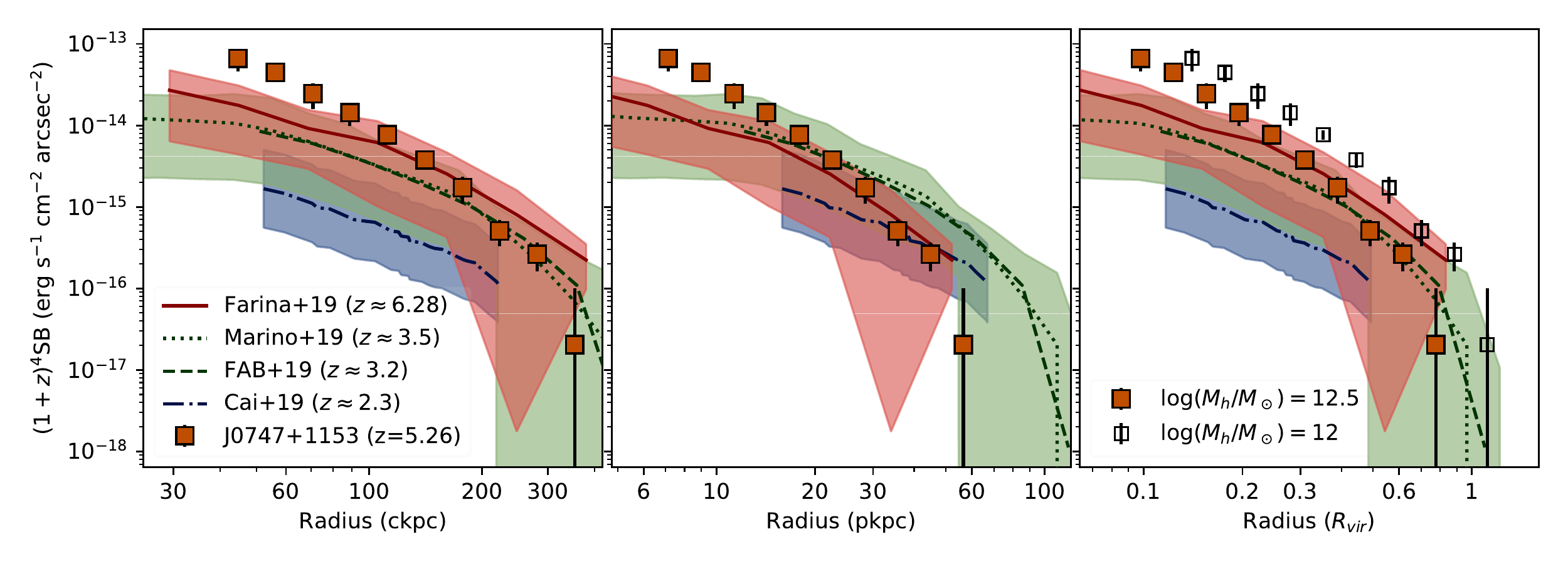}
     \end{tabular}
    \caption{Surface brightness profiles of Ly$\alpha$ nebulae, as a function of comoving radius (left), proper radius (centre) and units of the virial radius of the inferred host halo mass (assuming $M_{\rm h}=10^{12.5}$~M$_\odot$, right). The results based on \qso\ in this work are shown by the filled black sqaures. Also shown are the average profiles from \citet{2019ApJS..245...23C} at $z\approx 2.3$ (blue dot-dashed line and shaded region - which shows the 10-90th percentile range), \citet{ArrigoniBattaia2019} at $z\approx 3.17$ (dashed line), \citet{Borisova2016} at $z\approx 3.5$ (green dotted line and shaded region - which shows the 10-90th percentile range taken from \citealt{2019ApJ...880...47M}); and \citet{2019ApJ...887..196F} at $z\approx6.3$ (solid red curve and shaded region). The open black squares in the right panel show the profile for \qso\ renormalized assuming a virial radius of $50~\rm kpc$ (i.e. $M_{\rm h}=10^{12}$~M$_\odot$) instead of $80~\rm kpc$. The profile for \qso\ appears to be brighter and steeper compared to lower redshifts systems.}\label{fig:qsoprofile}
\end{figure*}

With this analysis, we clearly identify an extended nebula
with size of $\approx 50~\rm kpc$ at a Ly$\alpha$ surface brightness level of $\approx 10^{-18}$~\sbline\  and a peak surface brightness of $> 10^{-16}$~\sbline. The nebula, shown in Fig.~\ref{fig:qsoneb} in an optimally-extracted narrow-band image, appears roughly symmetric at the current surface brightness limit of $3.8\times 10^{-18}$~\sbline\ integrated in a 25\AA\ window, with a hint of elongation in the North direction. In Fig.~\ref{fig:qsoprofile}, we compare the nebula's surface brightness profile (extracted from a narrow-band image of 25~\AA\ in width and centred at the peak line emission of the nebula) with the average profiles of other nebulae detected with MUSE at $z>3$ \citep{Borisova2016,ArrigoniBattaia2019,2019ApJ...887..196F} and with the Keck Cosmic Web Imager (KCWI) at $z\sim2$ \citep{2019ApJS..245...23C}. Once corrected for the cosmological surface brightness dimming, the nebula around \qso\ is found to be broadly consistent with the ranges measured for such nebulae at $z>3$, but with somewhat enhanced surface brightnesses at scales of $r\lesssim80$ comoving kpc. The surface brightness profile of \qso\ is also comparable to the profiles reported around $z\approx 6$ quasars in \citet{drake2019} and \citet{2019ApJ...887..196F}.

Although only a single object, \qso\ aligns with the mild redshift evolution noted in \citet{2019ApJS..245...23C} between $z\lesssim3$ and $z\gtrsim3$. Based on the argument in \citet{ArrigoniBattaia2019}, we could be witnessing an increase in the amount of cold gas at $z\approx 5$ in this host galaxy possibly due to intense gas accretion. It should be noted, however, that \qso\ is among the brightest quasars known at $z\approx 5$ \citep{Wang2016}.  Indeed, after re-normalising the X-Shooter spectrum using a curve-of-growth analysis of the quasar photometry in the reconstructed $i-$band from MUSE data, we measure a quasar luminosity of $M_{1450} = -28.069 \pm 0.003$ mag. This is in excellent agreement with the value reported by \citet{Wang2016}, $M_{1450} = -28.04$ mag, thus ruling out significant variability over the time-scale of our observations. Compared to the characteristic luminosity of $z\approx 5$ quasars, $M^*_{1459} = -26.98 \pm 0.23$ \citep{Yang2016}, \qso\ is clearly among the brightest quasars at these redshifts, and is at the bright end of the luminosity distribution probed at $z\approx 3$ (e.g. $-28.29 \le  M_{1450} \le -25.65$ in \citealt{ArrigoniBattaia2019}). Therefore, while no significant trend has been observed thus far between the quasar luminosity and the brightness of the Ly$\alpha$ nebulae \citep[e.g.][]{2006A&A...459..717C, ArrigoniBattaia2019}, it remains possible that the surface brightness enhancement may be linked to the quasar luminosity.

The apparent steeper surface brightness profile (compared to the average nebulae profiles) shown in Fig.~\ref{fig:qsoprofile} would appear suggestive of a more compact nebula. In the right panel of Fig.~\ref{fig:qsoprofile}, we investigate whether this difference can be attributed to a scaling proportional to the typical size of the host halo. For this, we re-scale the profiles assuming the virial radius at a characteristic halo mass of $\log M_{\rm halo}/M_\odot \approx 12.5$, independent of redshift \citep[see discussion in][]{ArrigoniBattaia2019}, which is $R_{\rm vir}\approx 80~\rm kpc$ at $z\approx 5.26$. Admittedly, there is very large uncertainty on the evolution of the typical halo mass of quasar hosts beyond $z>3$ \citep{Timlin2018}, with uncertainties over a factor of $2-3$. However, unless the virial radius is $\lesssim 50~\rm kpc$, corresponding to halos with $\log M_{\rm halo}/M_\odot \lesssim 12$ at this redshift (open squares in Fig.~\ref{fig:qsoprofile}), the nebula around \qso\ suggests an intrinsic difference, e.g. due to the underlying density profile. Again, we note the quasar's high intrinsic luminosity, which would suggest a larger, rather than smaller, host halo. Indeed, given the high luminosity of \qso\, a virial radius of $R_{\rm vir}\approx100$~kpc would not be unreasonable.

\section{Summary and Conclusions}
\label{sec:summary}

We have presented a survey of galaxies in a single field observed with the VLT MUSE IFU, in conjunction with VLT X-Shooter moderate resolution spectroscopy of a background quasar  at $z\approx 5.26$ exhibiting multiple strong absorption lines. Our analysis shows the following key results.

\begin{enumerate}
\item Based on 9 C~{\sc iv} detected, $N({\rm HI})\gtrsim10^{16}$~cm$^{-2}$, absorbers we find a galaxy detection fraction of 67\% (6/9) for galaxies within $b<250$~kpc and $|\Delta v_{-200} |<1000$~km~s$^{-1}$. Applying a more stringent velocity offset cut of $|\Delta v_{-200} |<500$~km~s$^{-1}$, we find a detection fraction of 44\% (4/9). This is based on a Ly$\alpha$ flux limit of $3.1\times10^{18}$~erg~s$^{-1}$~cm$^{-2}$ (50\% completeness for extended sources).
\item Taking the systems together as a whole, we find that galaxies are correlated with the strong absorption systems, showing a statistically significant over-density of galaxy numbers when compared to the background population at $z\sim4-5$ and given the flux limit of our observations.
\item We detect two low metallicity absorption systems apparently associated (at $\approx100$~kpc separation) with candidate group environments, with 2 or more detected LAE galaxies within the MUSE field of view.
\item We detect extended Ly$\alpha$ emission around the $z=5.26$ quasar, reaching extents of $\approx50$~kpc at a surface brightness limit of $3.8\times10^{-18}$~erg~s$^{-1}$~cm$^{-2}$~arcsec$^{-2}$.  After scaling for surface brightness dimming, we find that this nebula is centrally brighter, with a steeper radial profile, than the average for nebulae studied at $z\sim3$, hinting at a mild redshift evolution in such nebulae.
\end{enumerate}

Overall, we find a picture that, whilst strong \ion{H}{i} systems are correlated with LAE galaxies at $z\sim5$, this coverage is patchy. Our results have shown that, on average, the galaxy environment is over-dense around LLSs and DLAs at $z\sim4-5$, similarly to \citet{2019MNRAS.487.5070M} and \citet{2019MNRAS.tmp.2667L} at $z\approx3-4$. The individual systems lie across a wide range of galaxy environments however, from groups of LAEs to no detected LAEs at all (although of course there will inevitably be fainter galaxies below our detection limits in most cases, c.f. \citealt{2014MNRAS.438..529R}). Uttimately, we find that \ion{H}{i} gas is often associated with complex structures as opposed to, or as well as, single galaxies. 

All the detected LAEs in our sample are at impact parameters of $b\gtrsim2R_{\rm vir}$, lending support to the analysis of \citet{keating2019}, suggestive of Ly$\alpha$ emission from galaxies in close proximity to strong absorption being absorbed, potentially due to patchy re-ionization. Whilst \citet{keating2019} and \citet{2018ApJ...863...92B} measure under-densities of LAEs at impact parameters of up to $\approx3$~Mpc in simulations and observations respectively, we find a lack of LAEs only at $\lesssim60$~kpc around the LLS and DLA systems presented here. This smaller on-sky scale may indeed be expected if the strong absorbers in our sample are tracing multiple smaller islands of non-reionized material compared to the extended trough observed by \citet{2015MNRAS.447.3402B} and \citet{2018ApJ...863...92B}. The statistical uncertainties on our sample however can not rule out a continuation of the over-density of LAEs detected at $60~{\rm kpc}\lesssim b\lesssim300$~kpc to these smaller scales at $b\lesssim60$~kpc from strong \ion{H}{i} absorbers. Complementing the analyses of the neutral \ion{H}{i} gas distribution, our analysis of metals and system metallicity suggest an inhomogeneously enriched medium. Both observations and simulation predictions show a large scatter in \ion{C}{iv} column density around $z\sim5$ galaxies, whilst no correlation is evident between metallicity and galaxy proximity.

These data, alongside similar recent studies at lower redshift \citep{2017MNRAS.468.1373B, 2019MNRAS.487.5070M, 2019ApJ...878L..33C, 2019MNRAS.tmp.2667L, 2019MNRAS.490.1451F}, highlight the value of deep IFU observations in performing blind identification of galaxies in the proximity of quasar sightline data. Going forward the available VLT X-Shooter sample of bright $z\sim5$ quasars present an indispensable base for developing these studies with further VLT/MUSE IFU observations. In conjunction with this, the simulations have shown that if we are to properly understand the galaxy-absorber connection, higher signal to noise sightline spectra are required beyond the existing data in order to probe to lower column density metal absorption features. In summary, these data lay the ground for a survey comprising deep sightline spectra combined with extensive MUSE IFU data and ultimately supported by NIR spectroscopic observations, either with JWST or the E-ELT, to probe more fully the component of the galaxy population not obscured in Ly$\alpha$ emission.  

\section*{Acknowledgements}

Based on observations made with ESO Telescopes at the La Silla Paranal Observatory under programme ID 0102.A$-$0261. RMB and MF acknowledge support by the Science and Technology Facilities Council [grant number ST/P000541/1]. This project has received funding from the European Research Council (ERC) under the European Union's Horizon 2020 research and innovation programme (grant agreement No. 757535). MR acknowledges support by Space Telescope Science Institute's Director's Research Funds. SC gratefully acknowledges support from Swiss National Science Foundation grant PP00P2-163824. The Cosmic Dawn center is funded by the Danish National Research Foundation (DNRF). SL was funded by project  FONDECYT 1191232. For access to the codes and advanced data products used in this work, please contact the authors or visit \url{http://www.michelefumagalli.com/codes.html}. Raw data are publicly available via the ESO Science Archive Facility.




\bibliographystyle{mnras}
\bibliography{bibref} 

\begin{thebibliography}{}
\makeatletter
\relax
\def\mn@urlcharsother{\let\do\@makeother \do\$\do\&\do\#\do\^\do\_\do\%\do\~}
\def\mn@doi{\begingroup\mn@urlcharsother \@ifnextchar [ {\mn@doi@}
  {\mn@doi@[]}}
\def\mn@doi@[#1]#2{\def\@tempa{#1}\ifx\@tempa\@empty \href
  {http://dx.doi.org/#2} {doi:#2}\else \href {http://dx.doi.org/#2} {#1}\fi
  \endgroup}
\def\mn@eprint#1#2{\mn@eprint@#1:#2::\@nil}
\def\mn@eprint@arXiv#1{\href {http://arxiv.org/abs/#1} {{\tt arXiv:#1}}}
\def\mn@eprint@dblp#1{\href {http://dblp.uni-trier.de/rec/bibtex/#1.xml}
  {dblp:#1}}
\def\mn@eprint@#1:#2:#3:#4\@nil{\def\@tempa {#1}\def\@tempb {#2}\def\@tempc
  {#3}\ifx \@tempc \@empty \let \@tempc \@tempb \let \@tempb \@tempa \fi \ifx
  \@tempb \@empty \def\@tempb {arXiv}\fi \@ifundefined
  {mn@eprint@\@tempb}{\@tempb:\@tempc}{\expandafter \expandafter \csname
  mn@eprint@\@tempb\endcsname \expandafter{\@tempc}}}

\bibitem[\protect\citeauthoryear{{Adelberger}, {Steidel}, {Shapley}  \&
  {Pettini}}{{Adelberger} et~al.}{2003}]{2003ApJ...584...45A}
{Adelberger} K.~L.,  {Steidel} C.~C.,  {Shapley} A.~E.,   {Pettini} M.,  2003,
  \mn@doi [\apj] {10.1086/345660}, \href
  {https://ui.adsabs.harvard.edu/abs/2003ApJ...584...45A} {584, 45}

\bibitem[\protect\citeauthoryear{{Arrigoni Battaia}, {Hennawi}, {Prochaska},
  {O{\~n}orbe}, {Farina}, {Cantalupo}  \& {Lusso}}{{Arrigoni Battaia}
  et~al.}{2019}]{ArrigoniBattaia2019}
{Arrigoni Battaia} F.,  {Hennawi} J.~F.,  {Prochaska} J.~X.,  {O{\~n}orbe} J.,
  {Farina} E.~P.,  {Cantalupo} S.,   {Lusso} E.,  2019, \mn@doi [\mnras]
  {10.1093/mnras/sty2827}, \href
  {https://ui.adsabs.harvard.edu/abs/2019MNRAS.482.3162A} {482, 3162}

\bibitem[\protect\citeauthoryear{{Bacon} et~al.,}{{Bacon}
  et~al.}{2010}]{Bacon2010MUSE}
{Bacon} R.,  et~al., 2010, in \procspie. p. 773508, \mn@doi{10.1117/12.856027}

\bibitem[\protect\citeauthoryear{{Barnes} \& {Haehnelt}}{{Barnes} \&
  {Haehnelt}}{2014}]{2014MNRAS.440.2313B}
{Barnes} L.~A.,  {Haehnelt} M.~G.,  2014, \mn@doi [\mnras]
  {10.1093/mnras/stu445}, \href
  {https://ui.adsabs.harvard.edu/abs/2014MNRAS.440.2313B} {440, 2313}

\bibitem[\protect\citeauthoryear{{Barnes}, {Haehnelt}, {Tescari}  \&
  {Viel}}{{Barnes} et~al.}{2011}]{2011MNRAS.416.1723B}
{Barnes} L.~A.,  {Haehnelt} M.~G.,  {Tescari} E.,   {Viel} M.,  2011, \mn@doi
  [\mnras] {10.1111/j.1365-2966.2011.18789.x}, \href
  {https://ui.adsabs.harvard.edu/abs/2011MNRAS.416.1723B} {416, 1723}

\bibitem[\protect\citeauthoryear{{Becker}, {Sargent}, {Rauch}  \&
  {Carswell}}{{Becker} et~al.}{2012}]{Becker2012}
{Becker} G.~D.,  {Sargent} W. L.~W.,  {Rauch} M.,   {Carswell} R.~F.,  2012,
  \mn@doi [\apj] {10.1088/0004-637X/744/2/91}, \href
  {https://ui.adsabs.harvard.edu/abs/2012ApJ...744...91B} {744, 91}

\bibitem[\protect\citeauthoryear{{Becker}, {Bolton}, {Madau}, {Pettini},
  {Ryan-Weber}  \& {Venemans}}{{Becker} et~al.}{2015}]{2015MNRAS.447.3402B}
{Becker} G.~D.,  {Bolton} J.~S.,  {Madau} P.,  {Pettini} M.,  {Ryan-Weber}
  E.~V.,   {Venemans} B.~P.,  2015, \mn@doi [\mnras] {10.1093/mnras/stu2646},
  \href {https://ui.adsabs.harvard.edu/abs/2015MNRAS.447.3402B} {447, 3402}

\bibitem[\protect\citeauthoryear{{Becker}, {Davies}, {Furlanetto}, {Malkan},
  {Boera}  \& {Douglass}}{{Becker} et~al.}{2018}]{2018ApJ...863...92B}
{Becker} G.~D.,  {Davies} F.~B.,  {Furlanetto} S.~R.,  {Malkan} M.~A.,  {Boera}
  E.,   {Douglass} C.,  2018, \mn@doi [\apj] {10.3847/1538-4357/aacc73}, \href
  {https://ui.adsabs.harvard.edu/abs/2018ApJ...863...92B} {863, 92}

\bibitem[\protect\citeauthoryear{{Becker} et~al.,}{{Becker}
  et~al.}{2019}]{2019ApJ...883..163B}
{Becker} G.~D.,  et~al., 2019, \mn@doi [\apj] {10.3847/1538-4357/ab3eb5}, \href
  {https://ui.adsabs.harvard.edu/abs/2019ApJ...883..163B} {883, 163}

\bibitem[\protect\citeauthoryear{{Bergeron} \& {Boisse}}{{Bergeron} \&
  {Boisse}}{1986}]{1986A&A...168....6B}
{Bergeron} J.,  {Boisse} P.,  1986, \aap, \href
  {https://ui.adsabs.harvard.edu/abs/1986A&A...168....6B} {168, 6}

\bibitem[\protect\citeauthoryear{{Bertin} \& {Arnouts}}{{Bertin} \&
  {Arnouts}}{1996}]{sextractor}
{Bertin} E.,  {Arnouts} S.,  1996, \mn@doi [Astronomy and Astrophysics
  Supplement Series] {10.1051/aas:1996164}, \href
  {https://ui.adsabs.harvard.edu/\#abs/1996A&AS..117..393B} {117, 393}

\bibitem[\protect\citeauthoryear{{Bielby} et~al.,}{{Bielby}
  et~al.}{2013}]{2013MNRAS.430..425B}
{Bielby} R.,  et~al., 2013, \mn@doi [\mnras] {10.1093/mnras/sts639}, \href
  {https://ui.adsabs.harvard.edu/abs/2013MNRAS.430..425B} {430, 425}

\bibitem[\protect\citeauthoryear{{Bielby} et~al.,}{{Bielby}
  et~al.}{2016}]{2016MNRAS.456.4061B}
{Bielby} R.~M.,  et~al., 2016, \mn@doi [\mnras] {10.1093/mnras/stv2914}, \href
  {https://ui.adsabs.harvard.edu/abs/2016MNRAS.456.4061B} {456, 4061}

\bibitem[\protect\citeauthoryear{{Bielby}, {Crighton}, {Fumagalli}, {Morris},
  {Stott}, {Tejos}  \& {Cantalupo}}{{Bielby}
  et~al.}{2017a}]{2017MNRAS.468.1373B}
{Bielby} R.,  {Crighton} N.~H.~M.,  {Fumagalli} M.,  {Morris} S.~L.,  {Stott}
  J.~P.,  {Tejos} N.,   {Cantalupo} S.,  2017a, \mn@doi [\mnras]
  {10.1093/mnras/stx528}, \href
  {https://ui.adsabs.harvard.edu/abs/2017MNRAS.468.1373B} {468, 1373}

\bibitem[\protect\citeauthoryear{{Bielby} et~al.,}{{Bielby}
  et~al.}{2017b}]{2017MNRAS.471.2174B}
{Bielby} R.~M.,  et~al., 2017b, \mn@doi [\mnras] {10.1093/mnras/stx1772}, \href
  {https://ui.adsabs.harvard.edu/abs/2017MNRAS.471.2174B} {471, 2174}

\bibitem[\protect\citeauthoryear{{Bielby} et~al.,}{{Bielby}
  et~al.}{2019}]{2019MNRAS.486...21B}
{Bielby} R.~M.,  et~al., 2019, \mn@doi [\mnras] {10.1093/mnras/stz774}, \href
  {https://ui.adsabs.harvard.edu/abs/2019MNRAS.486...21B} {486, 21}

\bibitem[\protect\citeauthoryear{{Bigiel}, {Leroy}, {Walter}, {Brinks}, {de
  Blok}, {Madore}  \& {Thornley}}{{Bigiel} et~al.}{2008}]{2008AJ....136.2846B}
{Bigiel} F.,  {Leroy} A.,  {Walter} F.,  {Brinks} E.,  {de Blok} W.~J.~G.,
  {Madore} B.,   {Thornley} M.~D.,  2008, \mn@doi [\aj]
  {10.1088/0004-6256/136/6/2846}, \href
  {https://ui.adsabs.harvard.edu/abs/2008AJ....136.2846B} {136, 2846}

\bibitem[\protect\citeauthoryear{{Bigiel} et~al.,}{{Bigiel}
  et~al.}{2011}]{2011ApJ...730L..13B}
{Bigiel} F.,  et~al., 2011, \mn@doi [\apjl] {10.1088/2041-8205/730/2/L13},
  \href {https://ui.adsabs.harvard.edu/abs/2011ApJ...730L..13B} {730, L13}

\bibitem[\protect\citeauthoryear{{Bird}, {Haehnelt}, {Neeleman}, {Genel},
  {Vogelsberger}  \& {Hernquist}}{{Bird} et~al.}{2015}]{2015MNRAS.447.1834B}
{Bird} S.,  {Haehnelt} M.,  {Neeleman} M.,  {Genel} S.,  {Vogelsberger} M.,
  {Hernquist} L.,  2015, \mn@doi [\mnras] {10.1093/mnras/stu2542}, \href
  {https://ui.adsabs.harvard.edu/abs/2015MNRAS.447.1834B} {447, 1834}

\bibitem[\protect\citeauthoryear{{Borisova} et~al.,}{{Borisova}
  et~al.}{2016}]{Borisova2016}
{Borisova} E.,  et~al., 2016, \mn@doi [\apj] {10.3847/0004-637X/831/1/39},
  \href {https://ui.adsabs.harvard.edu/abs/2016ApJ...831...39B} {831, 39}

\bibitem[\protect\citeauthoryear{{Bouch{\'e}}, {Lowenthal}, {Charlton},
  {Bershady}, {Churchill}  \& {Steidel}}{{Bouch{\'e}}
  et~al.}{2001}]{2001ApJ...550..585B}
{Bouch{\'e}} N.,  {Lowenthal} J.~D.,  {Charlton} J.~C.,  {Bershady} M.~A.,
  {Churchill} C.~W.,   {Steidel} C.~C.,  2001, \mn@doi [\apj] {10.1086/319805},
  \href {https://ui.adsabs.harvard.edu/abs/2001ApJ...550..585B} {550, 585}

\bibitem[\protect\citeauthoryear{{Bouch{\'e}}, {Murphy}, {P{\'e}roux},
  {Davies}, {Eisenhauer}, {F{\"o}rster Schreiber}  \& {Tacconi}}{{Bouch{\'e}}
  et~al.}{2007}]{2007ApJ...669L...5B}
{Bouch{\'e}} N.,  {Murphy} M.~T.,  {P{\'e}roux} C.,  {Davies} R.,  {Eisenhauer}
  F.,  {F{\"o}rster Schreiber} N.~M.,   {Tacconi} L.,  2007, \mn@doi [\apjl]
  {10.1086/523594}, \href
  {https://ui.adsabs.harvard.edu/abs/2007ApJ...669L...5B} {669, L5}

\bibitem[\protect\citeauthoryear{{Bouch{\'e}} et~al.,}{{Bouch{\'e}}
  et~al.}{2010}]{2010ApJ...718.1001B}
{Bouch{\'e}} N.,  et~al., 2010, \mn@doi [\apj] {10.1088/0004-637X/718/2/1001},
  \href {https://ui.adsabs.harvard.edu/abs/2010ApJ...718.1001B} {718, 1001}

\bibitem[\protect\citeauthoryear{{Brinchmann}, {Charlot}, {White}, {Tremonti},
  {Kauffmann}, {Heckman}  \& {Brinkmann}}{{Brinchmann}
  et~al.}{2004}]{2004MNRAS.351.1151B}
{Brinchmann} J.,  {Charlot} S.,  {White} S.~D.~M.,  {Tremonti} C.,  {Kauffmann}
  G.,  {Heckman} T.,   {Brinkmann} J.,  2004, \mn@doi [\mnras]
  {10.1111/j.1365-2966.2004.07881.x}, \href
  {https://ui.adsabs.harvard.edu/abs/2004MNRAS.351.1151B} {351, 1151}

\bibitem[\protect\citeauthoryear{{Bryan} \& {Norman}}{{Bryan} \&
  {Norman}}{1998}]{1998ApJ...495...80B}
{Bryan} G.~L.,  {Norman} M.~L.,  1998, \mn@doi [\apj] {10.1086/305262}, \href
  {https://ui.adsabs.harvard.edu/abs/1998ApJ...495...80B} {495, 80}

\bibitem[\protect\citeauthoryear{{Bunker}, {Warren}, {Clements}, {Williger}  \&
  {Hewett}}{{Bunker} et~al.}{1999}]{1999MNRAS.309..875B}
{Bunker} A.~J.,  {Warren} S.~J.,  {Clements} D.~L.,  {Williger} G.~M.,
  {Hewett} P.~C.,  1999, \mn@doi [\mnras] {10.1046/j.1365-8711.1999.02876.x},
  \href {https://ui.adsabs.harvard.edu/abs/1999MNRAS.309..875B} {309, 875}

\bibitem[\protect\citeauthoryear{{Burchett}, {Tripp}, {Werk}, {Howk},
  {Prochaska}, {Ford}  \& {Dav{\'e}}}{{Burchett}
  et~al.}{2013}]{2013ApJ...779L..17B}
{Burchett} J.~N.,  {Tripp} T.~M.,  {Werk} J.~K.,  {Howk} J.~C.,  {Prochaska}
  J.~X.,  {Ford} A. a.~B.,   {Dav{\'e}} R.,  2013, \mn@doi [\apjl]
  {10.1088/2041-8205/779/2/L17}, \href
  {https://ui.adsabs.harvard.edu/abs/2013ApJ...779L..17B} {779, L17}

\bibitem[\protect\citeauthoryear{{Burchett} et~al.,}{{Burchett}
  et~al.}{2016}]{2016ApJ...832..124B}
{Burchett} J.~N.,  et~al., 2016, \mn@doi [\apj] {10.3847/0004-637X/832/2/124},
  \href {https://ui.adsabs.harvard.edu/abs/2016ApJ...832..124B} {832, 124}

\bibitem[\protect\citeauthoryear{{Cai}, {Fan}, {Dave}, {Finlator}  \&
  {Oppenheimer}}{{Cai} et~al.}{2017}]{cai2017}
{Cai} Z.,  {Fan} X.,  {Dave} R.,  {Finlator} K.,   {Oppenheimer} B.,  2017,
  \mn@doi [\apj] {10.3847/2041-8213/aa8fc6}, \href
  {https://ui.adsabs.harvard.edu/abs/2017ApJ...849L..18C} {849, L18}

\bibitem[\protect\citeauthoryear{{Cai} et~al.,}{{Cai}
  et~al.}{2019}]{2019ApJS..245...23C}
{Cai} Z.,  et~al., 2019, \mn@doi [\apjs] {10.3847/1538-4365/ab4796}, \href
  {https://ui.adsabs.harvard.edu/abs/2019ApJS..245...23C} {245, 23}

\bibitem[\protect\citeauthoryear{{Cantalupo}, {Arrigoni-Battaia}, {Prochaska},
  {Hennawi}  \& {Madau}}{{Cantalupo} et~al.}{2014}]{Cantalupo+14}
{Cantalupo} S.,  {Arrigoni-Battaia} F.,  {Prochaska} J.~X.,  {Hennawi} J.~F.,
  {Madau} P.,  2014, \mn@doi [\nat] {10.1038/nature12898}, \href
  {https://ui.adsabs.harvard.edu/abs/2014Natur.506...63C} {506, 63}

\bibitem[\protect\citeauthoryear{{Cantalupo} et~al.,}{{Cantalupo}
  et~al.}{2019}]{2019MNRAS.483.5188C}
{Cantalupo} S.,  et~al., 2019, \mn@doi [\mnras] {10.1093/mnras/sty3481}, \href
  {https://ui.adsabs.harvard.edu/abs/2019MNRAS.483.5188C} {483, 5188}

\bibitem[\protect\citeauthoryear{{Chehade} et~al.,}{{Chehade}
  et~al.}{2016}]{2016MNRAS.459.1179C}
{Chehade} B.,  et~al., 2016, \mn@doi [\mnras] {10.1093/mnras/stw616}, \href
  {https://ui.adsabs.harvard.edu/abs/2016MNRAS.459.1179C} {459, 1179}

\bibitem[\protect\citeauthoryear{{Chen}, {Lanzetta}, {Webb}  \&
  {Barcons}}{{Chen} et~al.}{2001}]{2001ApJ...559..654C}
{Chen} H.-W.,  {Lanzetta} K.~M.,  {Webb} J.~K.,   {Barcons} X.,  2001, \mn@doi
  [\apj] {10.1086/322414}, \href
  {https://ui.adsabs.harvard.edu/abs/2001ApJ...559..654C} {559, 654}

\bibitem[\protect\citeauthoryear{{Chen}, {Boettcher}, {Johnson}, {Zahedy},
  {Rudie}, {Cooksey}, {Rauch}  \& {Mulchaey}}{{Chen}
  et~al.}{2019}]{2019ApJ...878L..33C}
{Chen} H.-W.,  {Boettcher} E.,  {Johnson} S.~D.,  {Zahedy} F.~S.,  {Rudie}
  G.~C.,  {Cooksey} K.~L.,  {Rauch} M.,   {Mulchaey} J.~S.,  2019, \mn@doi
  [\apjl] {10.3847/2041-8213/ab25ec}, \href
  {https://ui.adsabs.harvard.edu/abs/2019ApJ...878L..33C} {878, L33}

\bibitem[\protect\citeauthoryear{{Christensen}, {Jahnke}, {Wisotzki}  \&
  {S{\'a}nchez}}{{Christensen} et~al.}{2006}]{2006A&A...459..717C}
{Christensen} L.,  {Jahnke} K.,  {Wisotzki} L.,   {S{\'a}nchez} S.~F.,  2006,
  \mn@doi [\aap] {10.1051/0004-6361:20065318}, \href
  {https://ui.adsabs.harvard.edu/abs/2006A&A...459..717C} {459, 717}

\bibitem[\protect\citeauthoryear{{Crain} et~al.,}{{Crain}
  et~al.}{2015}]{2015MNRAS.450.1937C}
{Crain} R.~A.,  et~al., 2015, \mn@doi [\mnras] {10.1093/mnras/stv725}, \href
  {https://ui.adsabs.harvard.edu/abs/2015MNRAS.450.1937C} {450, 1937}

\bibitem[\protect\citeauthoryear{{Crighton} et~al.,}{{Crighton}
  et~al.}{2011}]{2011MNRAS.414...28C}
{Crighton} N.~H.~M.,  et~al., 2011, \mn@doi [\mnras]
  {10.1111/j.1365-2966.2011.17247.x}, \href
  {https://ui.adsabs.harvard.edu/abs/2011MNRAS.414...28C} {414, 28}

\bibitem[\protect\citeauthoryear{{Crighton}, {Hennawi}, {Simcoe}, {Cooksey},
  {Murphy}, {Fumagalli}, {Prochaska}  \& {Shanks}}{{Crighton}
  et~al.}{2015}]{2015MNRAS.446...18C}
{Crighton} N. H.~M.,  {Hennawi} J.~F.,  {Simcoe} R.~A.,  {Cooksey} K.~L.,
  {Murphy} M.~T.,  {Fumagalli} M.,  {Prochaska} J.~X.,   {Shanks} T.,  2015,
  \mn@doi [\mnras] {10.1093/mnras/stu2088}, \href
  {https://ui.adsabs.harvard.edu/abs/2015MNRAS.446...18C} {446, 18}

\bibitem[\protect\citeauthoryear{{D'Odorico} et~al.,}{{D'Odorico}
  et~al.}{2018}]{dodorico2018}
{D'Odorico} V.,  et~al., 2018, \mn@doi [\apj] {10.3847/2041-8213/aad7b7}, \href
  {https://ui.adsabs.harvard.edu/abs/2018ApJ...863L..29D} {863, L29}

\bibitem[\protect\citeauthoryear{{Daddi} et~al.,}{{Daddi}
  et~al.}{2007}]{2007ApJ...670..156D}
{Daddi} E.,  et~al., 2007, \mn@doi [\apj] {10.1086/521818}, \href
  {https://ui.adsabs.harvard.edu/abs/2007ApJ...670..156D} {670, 156}

\bibitem[\protect\citeauthoryear{{Dav{\'e}}, {Finlator}  \&
  {Oppenheimer}}{{Dav{\'e}} et~al.}{2012}]{2012MNRAS.421...98D}
{Dav{\'e}} R.,  {Finlator} K.,   {Oppenheimer} B.~D.,  2012, \mn@doi [\mnras]
  {10.1111/j.1365-2966.2011.20148.x}, \href
  {https://ui.adsabs.harvard.edu/abs/2012MNRAS.421...98D} {421, 98}

\bibitem[\protect\citeauthoryear{{Dekel} \& {Mandelker}}{{Dekel} \&
  {Mandelker}}{2014}]{2014MNRAS.444.2071D}
{Dekel} A.,  {Mandelker} N.,  2014, \mn@doi [\mnras] {10.1093/mnras/stu1427},
  \href {https://ui.adsabs.harvard.edu/abs/2014MNRAS.444.2071D} {444, 2071}

\bibitem[\protect\citeauthoryear{{Dekel}, {Sari}  \& {Ceverino}}{{Dekel}
  et~al.}{2009}]{2009ApJ...703..785D}
{Dekel} A.,  {Sari} R.,   {Ceverino} D.,  2009, \mn@doi [\apj]
  {10.1088/0004-637X/703/1/785}, \href
  {https://ui.adsabs.harvard.edu/abs/2009ApJ...703..785D} {703, 785}

\bibitem[\protect\citeauthoryear{{D{\'\i}az}, {Koyama}, {Ryan-Weber}, {Cooke},
  {Ouchi}, {Shimasaku}  \& {Nakata}}{{D{\'\i}az}
  et~al.}{2014}]{2014MNRAS.442..946D}
{D{\'\i}az} C.~G.,  {Koyama} Y.,  {Ryan-Weber} E.~V.,  {Cooke} J.,  {Ouchi} M.,
   {Shimasaku} K.,   {Nakata} F.,  2014, \mn@doi [\mnras]
  {10.1093/mnras/stu914}, \href
  {https://ui.adsabs.harvard.edu/abs/2014MNRAS.442..946D} {442, 946}

\bibitem[\protect\citeauthoryear{{D{\'\i}az}, {Ryan-Weber}, {Cooke}, {Koyama}
  \& {Ouchi}}{{D{\'\i}az} et~al.}{2015}]{2015MNRAS.448.1240D}
{D{\'\i}az} C.~G.,  {Ryan-Weber} E.~V.,  {Cooke} J.,  {Koyama} Y.,   {Ouchi}
  M.,  2015, \mn@doi [\mnras] {10.1093/mnras/stu2738}, \href
  {https://ui.adsabs.harvard.edu/abs/2015MNRAS.448.1240D} {448, 1240}

\bibitem[\protect\citeauthoryear{{D{\'\i}az}, {Ryan-Weber}, {Karman}, {Caputi},
  {Salvadori}, {Crighton}, {Ouchi}  \& {Vanzella}}{{D{\'\i}az}
  et~al.}{2020}]{2020arXiv200104453G}
{D{\'\i}az} C.~G.,  {Ryan-Weber} E.,  {Karman} W.,  {Caputi} K.,  {Salvadori}
  S.,  {Crighton} N.,  {Ouchi} M.,   {Vanzella} E.,  2020, arXiv e-prints,
  \href {https://ui.adsabs.harvard.edu/abs/2020arXiv200104453G} {p.
  arXiv:2001.04453}

\bibitem[\protect\citeauthoryear{{Djorgovski}, {Pahre}, {Bechtold}  \&
  {Elston}}{{Djorgovski} et~al.}{1996}]{1996Natur.382..234D}
{Djorgovski} S.~G.,  {Pahre} M.~A.,  {Bechtold} J.,   {Elston} R.,  1996,
  \mn@doi [\nat] {10.1038/382234a0}, \href
  {https://ui.adsabs.harvard.edu/abs/1996Natur.382..234D} {382, 234}

\bibitem[\protect\citeauthoryear{{Drake} et~al.,}{{Drake}
  et~al.}{2017a}]{2017MNRAS.471..267D}
{Drake} A.~B.,  et~al., 2017a, \mn@doi [\mnras] {10.1093/mnras/stx1515}, \href
  {https://ui.adsabs.harvard.edu/abs/2017MNRAS.471..267D} {471, 267}

\bibitem[\protect\citeauthoryear{{Drake} et~al.,}{{Drake}
  et~al.}{2017b}]{2017A&A...608A...6D}
{Drake} A.~B.,  et~al., 2017b, \mn@doi [\aap] {10.1051/0004-6361/201731431},
  \href {https://ui.adsabs.harvard.edu/abs/2017A&A...608A...6D} {608, A6}

\bibitem[\protect\citeauthoryear{{Drake}, {Farina}, {Neeleman}, {Walter},
  {Venemans}, {Banados}, {Mazzucchelli}  \& {Decarli}}{{Drake}
  et~al.}{2019}]{drake2019}
{Drake} A.~B.,  {Farina} E.~P.,  {Neeleman} M.,  {Walter} F.,  {Venemans} B.,
  {Banados} E.,  {Mazzucchelli} C.,   {Decarli} R.,  2019, arXiv e-prints,
  \href {https://ui.adsabs.harvard.edu/abs/2019arXiv190607197D} {p.
  arXiv:1906.07197}

\bibitem[\protect\citeauthoryear{{Elbaz} et~al.,}{{Elbaz}
  et~al.}{2007}]{2007A&A...468...33E}
{Elbaz} D.,  et~al., 2007, \mn@doi [\aap] {10.1051/0004-6361:20077525}, \href
  {https://ui.adsabs.harvard.edu/abs/2007A&A...468...33E} {468, 33}

\bibitem[\protect\citeauthoryear{{Farina} et~al.,}{{Farina}
  et~al.}{2019}]{2019ApJ...887..196F}
{Farina} E.~P.,  et~al., 2019, \mn@doi [\apj] {10.3847/1538-4357/ab5847}, \href
  {https://ui.adsabs.harvard.edu/abs/2019ApJ...887..196F} {887, 196}

\bibitem[\protect\citeauthoryear{{Faucher-Gigu{\`e}re} \&
  {Kere{\v{s}}}}{{Faucher-Gigu{\`e}re} \& {Kere{\v{s}}}}{2011}]{FG2011}
{Faucher-Gigu{\`e}re} C.-A.,  {Kere{\v{s}}} D.,  2011, \mn@doi [\mnras]
  {10.1111/j.1745-3933.2011.01018.x}, \href
  {https://ui.adsabs.harvard.edu/abs/2011MNRAS.412L.118F} {412, L118}

\bibitem[\protect\citeauthoryear{{Faucher-Gigu{\`e}re}, {Feldmann}, {Quataert},
  {Kere{\v{s}}}, {Hopkins}  \& {Murray}}{{Faucher-Gigu{\`e}re}
  et~al.}{2016}]{FG2016}
{Faucher-Gigu{\`e}re} C.-A.,  {Feldmann} R.,  {Quataert} E.,  {Kere{\v{s}}} D.,
   {Hopkins} P.~F.,   {Murray} N.,  2016, \mn@doi [\mnras]
  {10.1093/mnrasl/slw091}, \href
  {https://ui.adsabs.harvard.edu/abs/2016MNRAS.461L..32F} {461, L32}

\bibitem[\protect\citeauthoryear{{Ferland} et~al.,}{{Ferland}
  et~al.}{2017}]{2017RMxAA..53..385F}
{Ferland} G.~J.,  et~al., 2017, \rmxaa, \href
  {https://ui.adsabs.harvard.edu/abs/2017RMxAA..53..385F} {53, 385}

\bibitem[\protect\citeauthoryear{{Finlator}, {Oppenheimer}, {Dav{\'e}},
  {Zackrisson}, {Thompson}  \& {Huang}}{{Finlator}
  et~al.}{2016}]{2016MNRAS.459.2299F}
{Finlator} K.,  {Oppenheimer} B.~D.,  {Dav{\'e}} R.,  {Zackrisson} E.,
  {Thompson} R.,   {Huang} S.,  2016, \mn@doi [\mnras] {10.1093/mnras/stw805},
  \href {https://ui.adsabs.harvard.edu/abs/2016MNRAS.459.2299F} {459, 2299}

\bibitem[\protect\citeauthoryear{{Finlator}, {Doughty}, {Cai}  \&
  {D{\'\i}az}}{{Finlator} et~al.}{2020}]{2020arXiv200103498F}
{Finlator} K.,  {Doughty} C.,  {Cai} Z.,   {D{\'\i}az} G.,  2020, arXiv
  e-prints, \href {https://ui.adsabs.harvard.edu/abs/2020arXiv200103498F} {p.
  arXiv:2001.03498}

\bibitem[\protect\citeauthoryear{{Foreman-Mackey} et~al.,}{{Foreman-Mackey}
  et~al.}{2018}]{2018zndo...1436565F}
{Foreman-Mackey} D.,  et~al., 2018, {Dfm/Emcee: Emcee V3.0Rc2},
  \mn@doi{10.5281/zenodo.1436565}

\bibitem[\protect\citeauthoryear{{Fossati} et~al.,}{{Fossati}
  et~al.}{2019}]{2019MNRAS.490.1451F}
{Fossati} M.,  et~al., 2019, \mn@doi [\mnras] {10.1093/mnras/stz2693}, \href
  {https://ui.adsabs.harvard.edu/abs/2019MNRAS.490.1451F} {490, 1451}

\bibitem[\protect\citeauthoryear{{Fumagalli}, {O'Meara}  \&
  {Prochaska}}{{Fumagalli} et~al.}{2011a}]{2011Sci...334.1245F}
{Fumagalli} M.,  {O'Meara} J.~M.,   {Prochaska} J.~X.,  2011a, \mn@doi
  [Science] {10.1126/science.1213581}, \href
  {https://ui.adsabs.harvard.edu/abs/2011Sci...334.1245F} {334, 1245}

\bibitem[\protect\citeauthoryear{{Fumagalli}, {Prochaska}, {Kasen}, {Dekel},
  {Ceverino}  \& {Primack}}{{Fumagalli} et~al.}{2011b}]{Fumagalli2011}
{Fumagalli} M.,  {Prochaska} J.~X.,  {Kasen} D.,  {Dekel} A.,  {Ceverino} D.,
  {Primack} J.~R.,  2011b, \mn@doi [\mnras] {10.1111/j.1365-2966.2011.19599.x},
  \href {https://ui.adsabs.harvard.edu/abs/2011MNRAS.418.1796F} {418, 1796}

\bibitem[\protect\citeauthoryear{{Fumagalli}, {O'Meara}, {Prochaska}  \&
  {Worseck}}{{Fumagalli} et~al.}{2013}]{mf2013}
{Fumagalli} M.,  {O'Meara} J.~M.,  {Prochaska} J.~X.,   {Worseck} G.,  2013,
  \mn@doi [\apj] {10.1088/0004-637X/775/1/78}, \href
  {https://ui.adsabs.harvard.edu/abs/2013ApJ...775...78F} {775, 78}

\bibitem[\protect\citeauthoryear{{Fumagalli}, {O'Meara}  \&
  {Prochaska}}{{Fumagalli} et~al.}{2016a}]{2016MNRAS.455.4100F}
{Fumagalli} M.,  {O'Meara} J.~M.,   {Prochaska} J.~X.,  2016a, \mn@doi [\mnras]
  {10.1093/mnras/stv2616}, \href
  {https://ui.adsabs.harvard.edu/abs/2016MNRAS.455.4100F} {455, 4100}

\bibitem[\protect\citeauthoryear{{Fumagalli}, {Cantalupo}, {Dekel}, {Morris},
  {O'Meara}, {Prochaska}  \& {Theuns}}{{Fumagalli}
  et~al.}{2016b}]{Fumagalli2016}
{Fumagalli} M.,  {Cantalupo} S.,  {Dekel} A.,  {Morris} S.~L.,  {O'Meara}
  J.~M.,  {Prochaska} J.~X.,   {Theuns} T.,  2016b, \mn@doi [\mnras]
  {10.1093/mnras/stw1782}, \href
  {https://ui.adsabs.harvard.edu/\#abs/2016MNRAS.462.1978F} {462, 1978}

\bibitem[\protect\citeauthoryear{{Fumagalli}, {Haardt}, {Theuns}, {Morris},
  {Cantalupo}, {Madau}  \& {Fossati}}{{Fumagalli} et~al.}{2017a}]{MF2017UVB}
{Fumagalli} M.,  {Haardt} F.,  {Theuns} T.,  {Morris} S.~L.,  {Cantalupo} S.,
  {Madau} P.,   {Fossati} M.,  2017a, \mn@doi [\mnras] {10.1093/mnras/stx398},
  \href {https://ui.adsabs.harvard.edu/abs/2017MNRAS.467.4802F} {467, 4802}

\bibitem[\protect\citeauthoryear{{Fumagalli} et~al.,}{{Fumagalli}
  et~al.}{2017b}]{Fumagalli2017}
{Fumagalli} M.,  et~al., 2017b, \mn@doi [\mnras] {10.1093/mnras/stx1896}, \href
  {https://ui.adsabs.harvard.edu/\#abs/2017MNRAS.471.3686F} {471, 3686}

\bibitem[\protect\citeauthoryear{{Fynbo}, {Burud}  \& {M{\o}ller}}{{Fynbo}
  et~al.}{2000}]{2000A&A...358...88F}
{Fynbo} J.~U.,  {Burud} I.,   {M{\o}ller} P.,  2000, \aap, \href
  {https://ui.adsabs.harvard.edu/abs/2000A&A...358...88F} {358, 88}

\bibitem[\protect\citeauthoryear{{Fynbo} et~al.,}{{Fynbo}
  et~al.}{2013}]{2013MNRAS.436..361F}
{Fynbo} J.~P.~U.,  et~al., 2013, \mn@doi [\mnras] {10.1093/mnras/stt1579},
  \href {https://ui.adsabs.harvard.edu/abs/2013MNRAS.436..361F} {436, 361}

\bibitem[\protect\citeauthoryear{{Gawiser} et~al.,}{{Gawiser}
  et~al.}{2007}]{2007ApJ...671..278G}
{Gawiser} E.,  et~al., 2007, \mn@doi [\apj] {10.1086/522955}, \href
  {https://ui.adsabs.harvard.edu/abs/2007ApJ...671..278G} {671, 278}

\bibitem[\protect\citeauthoryear{{Goto}, {Utsumi}, {Furusawa}, {Miyazaki}  \&
  {Komiyama}}{{Goto} et~al.}{2009}]{2009MNRAS.400..843G}
{Goto} T.,  {Utsumi} Y.,  {Furusawa} H.,  {Miyazaki} S.,   {Komiyama} Y.,
  2009, \mn@doi [\mnras] {10.1111/j.1365-2966.2009.15486.x}, \href
  {https://ui.adsabs.harvard.edu/abs/2009MNRAS.400..843G} {400, 843}

\bibitem[\protect\citeauthoryear{{Grove}, {Fynbo}, {Ledoux}, {Limousin},
  {M{\o}ller}, {Nilsson}  \& {Thomsen}}{{Grove}
  et~al.}{2009}]{2009A&A...497..689G}
{Grove} L.~F.,  {Fynbo} J.~P.~U.,  {Ledoux} C.,  {Limousin} M.,  {M{\o}ller}
  P.,  {Nilsson} K.~K.,   {Thomsen} B.,  2009, \mn@doi [\aap]
  {10.1051/0004-6361/200811429}, \href
  {https://ui.adsabs.harvard.edu/abs/2009A&A...497..689G} {497, 689}

\bibitem[\protect\citeauthoryear{{Hartoog} et~al.,}{{Hartoog}
  et~al.}{2015}]{2015A&A...580A.139H}
{Hartoog} O.~E.,  et~al., 2015, \mn@doi [\aap] {10.1051/0004-6361/201425001},
  \href {https://ui.adsabs.harvard.edu/abs/2015A&A...580A.139H} {580, A139}

\bibitem[\protect\citeauthoryear{{Heckman}, {Armus}  \& {Miley}}{{Heckman}
  et~al.}{1990}]{1990ApJS...74..833H}
{Heckman} T.~M.,  {Armus} L.,   {Miley} G.~K.,  1990, \mn@doi [\apjs]
  {10.1086/191522}, \href
  {https://ui.adsabs.harvard.edu/abs/1990ApJS...74..833H} {74, 833}

\bibitem[\protect\citeauthoryear{{Hennawi}, {Prochaska}, {Cantalupo}  \&
  {Arrigoni-Battaia}}{{Hennawi} et~al.}{2015}]{Hennawi+15}
{Hennawi} J.~F.,  {Prochaska} J.~X.,  {Cantalupo} S.,   {Arrigoni-Battaia} F.,
  2015, \mn@doi [Science] {10.1126/science.aaa5397}, \href
  {https://ui.adsabs.harvard.edu/abs/2015Sci...348..779H} {348, 779}

\bibitem[\protect\citeauthoryear{{Hinton}, {Davis}, {Lidman}, {Glazebrook}  \&
  {Lewis}}{{Hinton} et~al.}{2016}]{Marz}
{Hinton} S.~R.,  {Davis} T.~M.,  {Lidman} C.,  {Glazebrook} K.,   {Lewis}
  G.~F.,  2016, \mn@doi [Astronomy and Computing]
  {10.1016/j.ascom.2016.03.001}, \href
  {http://adsabs.harvard.edu/abs/2016A%26C....15...61H} {15, 61}

\bibitem[\protect\citeauthoryear{{Jenkins}, {Frenk}, {White}, {Colberg},
  {Cole}, {Evrard}, {Couchman}  \& {Yoshida}}{{Jenkins}
  et~al.}{2001}]{2001MNRAS.321..372J}
{Jenkins} A.,  {Frenk} C.~S.,  {White} S.~D.~M.,  {Colberg} J.~M.,  {Cole} S.,
  {Evrard} A.~E.,  {Couchman} H.~M.~P.,   {Yoshida} N.,  2001, \mn@doi [\mnras]
  {10.1046/j.1365-8711.2001.04029.x}, \href
  {https://ui.adsabs.harvard.edu/abs/2001MNRAS.321..372J} {321, 372}

\bibitem[\protect\citeauthoryear{{Kacprzak}, {Murphy}  \&
  {Churchill}}{{Kacprzak} et~al.}{2010}]{2010MNRAS.406..445K}
{Kacprzak} G.~G.,  {Murphy} M.~T.,   {Churchill} C.~W.,  2010, \mn@doi [\mnras]
  {10.1111/j.1365-2966.2010.16667.x}, \href
  {https://ui.adsabs.harvard.edu/abs/2010MNRAS.406..445K} {406, 445}

\bibitem[\protect\citeauthoryear{{Keating}, {Weinberger}, {Kulkarni},
  {Haehnelt}, {Chardin}  \& {Aubert}}{{Keating} et~al.}{2019}]{keating2019}
{Keating} L.~C.,  {Weinberger} L.~H.,  {Kulkarni} G.,  {Haehnelt} M.~G.,
  {Chardin} J.,   {Aubert} D.,  2019, arXiv e-prints, \href
  {https://ui.adsabs.harvard.edu/abs/2019arXiv190512640K} {p. arXiv:1905.12640}

\bibitem[\protect\citeauthoryear{{Keeney} et~al.,}{{Keeney}
  et~al.}{2018}]{2018ApJS..237...11K}
{Keeney} B.~A.,  et~al., 2018, \mn@doi [\apjs] {10.3847/1538-4365/aac727},
  \href {https://ui.adsabs.harvard.edu/abs/2018ApJS..237...11K} {237, 11}

\bibitem[\protect\citeauthoryear{{Kere{\v{s}}}, {Katz}, {Weinberg}  \&
  {Dav{\'e}}}{{Kere{\v{s}}} et~al.}{2005}]{2005MNRAS.363....2K}
{Kere{\v{s}}} D.,  {Katz} N.,  {Weinberg} D.~H.,   {Dav{\'e}} R.,  2005,
  \mn@doi [\mnras] {10.1111/j.1365-2966.2005.09451.x}, \href
  {https://ui.adsabs.harvard.edu/abs/2005MNRAS.363....2K} {363, 2}

\bibitem[\protect\citeauthoryear{{Krogager}, {M{\o}ller}, {Fynbo}  \&
  {Noterdaeme}}{{Krogager} et~al.}{2017}]{2017MNRAS.469.2959K}
{Krogager} J.~K.,  {M{\o}ller} P.,  {Fynbo} J.~P.~U.,   {Noterdaeme} P.,  2017,
  \mn@doi [\mnras] {10.1093/mnras/stx1011}, \href
  {https://ui.adsabs.harvard.edu/abs/2017MNRAS.469.2959K} {469, 2959}

\bibitem[\protect\citeauthoryear{{Kulkarni}, {Keating}, {Haehnelt}, {Bosman},
  {Puchwein}, {Chardin}  \& {Aubert}}{{Kulkarni}
  et~al.}{2019}]{2019MNRAS.485L..24K}
{Kulkarni} G.,  {Keating} L.~C.,  {Haehnelt} M.~G.,  {Bosman} S. E.~I.,
  {Puchwein} E.,  {Chardin} J.,   {Aubert} D.,  2019, \mn@doi [\mnras]
  {10.1093/mnrasl/slz025}, \href
  {https://ui.adsabs.harvard.edu/abs/2019MNRAS.485L..24K} {485, L24}

\bibitem[\protect\citeauthoryear{{Leroy}, {Walter}, {Brinks}, {Bigiel}, {de
  Blok}, {Madore}  \& {Thornley}}{{Leroy} et~al.}{2008}]{2008AJ....136.2782L}
{Leroy} A.~K.,  {Walter} F.,  {Brinks} E.,  {Bigiel} F.,  {de Blok} W.~J.~G.,
  {Madore} B.,   {Thornley} M.~D.,  2008, \mn@doi [\aj]
  {10.1088/0004-6256/136/6/2782}, \href
  {https://ui.adsabs.harvard.edu/abs/2008AJ....136.2782L} {136, 2782}

\bibitem[\protect\citeauthoryear{{Lilly}, {Carollo}, {Pipino}, {Renzini}  \&
  {Peng}}{{Lilly} et~al.}{2013}]{2013ApJ...772..119L}
{Lilly} S.~J.,  {Carollo} C.~M.,  {Pipino} A.,  {Renzini} A.,   {Peng} Y.,
  2013, \mn@doi [\apj] {10.1088/0004-637X/772/2/119}, \href
  {https://ui.adsabs.harvard.edu/abs/2013ApJ...772..119L} {772, 119}

\bibitem[\protect\citeauthoryear{{Lofthouse} et~al.,}{{Lofthouse}
  et~al.}{2019}]{2019MNRAS.tmp.2667L}
{Lofthouse} E.~K.,  et~al., 2019, \mn@doi [\mnras] {10.1093/mnras/stz3066},
  \href {https://ui.adsabs.harvard.edu/abs/2019MNRAS.tmp.2667L} {p.~2667}

\bibitem[\protect\citeauthoryear{{L{\'o}pez} et~al.,}{{L{\'o}pez}
  et~al.}{2016}]{2016A&A...594A..91L}
{L{\'o}pez} S.,  et~al., 2016, \mn@doi [\aap] {10.1051/0004-6361/201628161},
  \href {https://ui.adsabs.harvard.edu/abs/2016A%26A...594A..91L} {594, A91}

\bibitem[\protect\citeauthoryear{{Lowenthal}, {Hogan}, {Green}, {Woodgate},
  {Caulet}, {Brown}  \& {Bechtold}}{{Lowenthal}
  et~al.}{1995}]{1995ApJ...451..484L}
{Lowenthal} J.~D.,  {Hogan} C.~J.,  {Green} R.~F.,  {Woodgate} B.,  {Caulet}
  A.,  {Brown} L.,   {Bechtold} J.,  1995, \mn@doi [\apj] {10.1086/176237},
  \href {https://ui.adsabs.harvard.edu/abs/1995ApJ...451..484L} {451, 484}

\bibitem[\protect\citeauthoryear{{Mackenzie} et~al.,}{{Mackenzie}
  et~al.}{2019}]{2019MNRAS.487.5070M}
{Mackenzie} R.,  et~al., 2019, \mn@doi [\mnras] {10.1093/mnras/stz1501}, \href
  {https://ui.adsabs.harvard.edu/abs/2019MNRAS.487.5070M} {487, 5070}

\bibitem[\protect\citeauthoryear{{Marino} et~al.,}{{Marino}
  et~al.}{2019}]{2019ApJ...880...47M}
{Marino} R.~A.,  et~al., 2019, \mn@doi [\apj] {10.3847/1538-4357/ab2881}, \href
  {https://ui.adsabs.harvard.edu/abs/2019ApJ...880...47M} {880, 47}

\bibitem[\protect\citeauthoryear{{Martin}}{{Martin}}{2005}]{2005ApJ...621..227M}
{Martin} C.~L.,  2005, \mn@doi [\apj] {10.1086/427277}, \href
  {https://ui.adsabs.harvard.edu/abs/2005ApJ...621..227M} {621, 227}

\bibitem[\protect\citeauthoryear{{Martin}, {Chang}, {Matuszewski}, {Morrissey},
  {Rahman}, {Moore}  \& {Steidel}}{{Martin} et~al.}{2014}]{Martin+14}
{Martin} D.~C.,  {Chang} D.,  {Matuszewski} M.,  {Morrissey} P.,  {Rahman} S.,
  {Moore} A.,   {Steidel} C.~C.,  2014, \mn@doi [\apj]
  {10.1088/0004-637X/786/2/106}, \href
  {https://ui.adsabs.harvard.edu/abs/2014ApJ...786..106M} {786, 106}

\bibitem[\protect\citeauthoryear{{Meyer}, {Bosman}, {Kakiichi}  \&
  {Ellis}}{{Meyer} et~al.}{2019a}]{meyer2019}
{Meyer} R.~A.,  {Bosman} S. E.~I.,  {Kakiichi} K.,   {Ellis} R.~S.,  2019a,
  \mn@doi [\mnras] {10.1093/mnras/sty2954}, \href
  {https://ui.adsabs.harvard.edu/abs/2019MNRAS.483...19M} {483, 19}

\bibitem[\protect\citeauthoryear{{Meyer}, {Bosman}  \& {Ellis}}{{Meyer}
  et~al.}{2019b}]{2019MNRAS.487.3305M}
{Meyer} R.~A.,  {Bosman} S. E.~I.,   {Ellis} R.~S.,  2019b, \mn@doi [\mnras]
  {10.1093/mnras/stz1504}, \href
  {https://ui.adsabs.harvard.edu/abs/2019MNRAS.487.3305M} {487, 3305}

\bibitem[\protect\citeauthoryear{{M{\o}ller} \& {Warren}}{{M{\o}ller} \&
  {Warren}}{1993}]{1993A&A...270...43M}
{M{\o}ller} P.,  {Warren} S.~J.,  1993, \aap, \href
  {https://ui.adsabs.harvard.edu/abs/1993A&A...270...43M} {270, 43}

\bibitem[\protect\citeauthoryear{{M{\o}ller} \& {Warren}}{{M{\o}ller} \&
  {Warren}}{1998}]{1998MNRAS.299..661M}
{M{\o}ller} P.,  {Warren} S.~J.,  1998, \mn@doi [\mnras]
  {10.1046/j.1365-8711.1998.01749.x}, \href
  {https://ui.adsabs.harvard.edu/abs/1998MNRAS.299..661M} {299, 661}

\bibitem[\protect\citeauthoryear{{Nasir} \& {D'Aloisio}}{{Nasir} \&
  {D'Aloisio}}{2019}]{2019arXiv191003570N}
{Nasir} F.,  {D'Aloisio} A.,  2019, arXiv e-prints, \href
  {https://ui.adsabs.harvard.edu/abs/2019arXiv191003570N} {p. arXiv:1910.03570}

\bibitem[\protect\citeauthoryear{{Neeleman}, {Kanekar}, {Prochaska},
  {Rafelski}, {Carilli}  \& {Wolfe}}{{Neeleman}
  et~al.}{2017}]{2017Sci...355.1285N}
{Neeleman} M.,  {Kanekar} N.,  {Prochaska} J.~X.,  {Rafelski} M.,  {Carilli}
  C.~L.,   {Wolfe} A.~M.,  2017, \mn@doi [Science] {10.1126/science.aal1737},
  \href {https://ui.adsabs.harvard.edu/abs/2017Sci...355.1285N} {355, 1285}

\bibitem[\protect\citeauthoryear{{Neeleman}, {Kanekar}, {Prochaska},
  {Christensen}, {Dessauges-Zavadsky}, {Fynbo}, {M{\o}ller}  \&
  {Zwaan}}{{Neeleman} et~al.}{2018}]{2018ApJ...856L..12N}
{Neeleman} M.,  {Kanekar} N.,  {Prochaska} J.~X.,  {Christensen} L.,
  {Dessauges-Zavadsky} M.,  {Fynbo} J. P.~U.,  {M{\o}ller} P.,   {Zwaan} M.~A.,
   2018, \mn@doi [\apjl] {10.3847/2041-8213/aab5b1}, \href
  {https://ui.adsabs.harvard.edu/abs/2018ApJ...856L..12N} {856, L12}

\bibitem[\protect\citeauthoryear{{Neeleman}, {Kanekar}, {Prochaska}, {Rafelski}
   \& {Carilli}}{{Neeleman} et~al.}{2019}]{2019ApJ...870L..19N}
{Neeleman} M.,  {Kanekar} N.,  {Prochaska} J.~X.,  {Rafelski} M.~A.,
  {Carilli} C.~L.,  2019, \mn@doi [\apjl] {10.3847/2041-8213/aaf871}, \href
  {https://ui.adsabs.harvard.edu/abs/2019ApJ...870L..19N} {870, L19}

\bibitem[\protect\citeauthoryear{{Noeske} et~al.,}{{Noeske}
  et~al.}{2007}]{2007ApJ...660L..43N}
{Noeske} K.~G.,  et~al., 2007, \mn@doi [\apjl] {10.1086/517926}, \href
  {https://ui.adsabs.harvard.edu/abs/2007ApJ...660L..43N} {660, L43}

\bibitem[\protect\citeauthoryear{{Oppenheimer} et~al.,}{{Oppenheimer}
  et~al.}{2016}]{2016MNRAS.460.2157O}
{Oppenheimer} B.~D.,  et~al., 2016, \mn@doi [\mnras] {10.1093/mnras/stw1066},
  \href {https://ui.adsabs.harvard.edu/abs/2016MNRAS.460.2157O} {460, 2157}

\bibitem[\protect\citeauthoryear{{Ouchi} et~al.,}{{Ouchi}
  et~al.}{2010}]{2010ApJ...723..869O}
{Ouchi} M.,  et~al., 2010, \mn@doi [\apj] {10.1088/0004-637X/723/1/869}, \href
  {https://ui.adsabs.harvard.edu/abs/2010ApJ...723..869O} {723, 869}

\bibitem[\protect\citeauthoryear{{Peeples} et~al.,}{{Peeples}
  et~al.}{2019}]{2019ApJ...873..129P}
{Peeples} M.~S.,  et~al., 2019, \mn@doi [\apj] {10.3847/1538-4357/ab0654},
  \href {https://ui.adsabs.harvard.edu/abs/2019ApJ...873..129P} {873, 129}

\bibitem[\protect\citeauthoryear{{P{\'e}roux}, {Bouch{\'e}}, {Kulkarni}, {York}
   \& {Vladilo}}{{P{\'e}roux} et~al.}{2011}]{2011MNRAS.410.2251P}
{P{\'e}roux} C.,  {Bouch{\'e}} N.,  {Kulkarni} V.~P.,  {York} D.~G.,
  {Vladilo} G.,  2011, \mn@doi [\mnras] {10.1111/j.1365-2966.2010.17597.x},
  \href {https://ui.adsabs.harvard.edu/abs/2011MNRAS.410.2251P} {410, 2251}

\bibitem[\protect\citeauthoryear{{P{\'e}roux}, {Bouch{\'e}}, {Kulkarni}, {York}
   \& {Vladilo}}{{P{\'e}roux} et~al.}{2012}]{2012MNRAS.419.3060P}
{P{\'e}roux} C.,  {Bouch{\'e}} N.,  {Kulkarni} V.~P.,  {York} D.~G.,
  {Vladilo} G.,  2012, \mn@doi [\mnras] {10.1111/j.1365-2966.2011.19947.x},
  \href {https://ui.adsabs.harvard.edu/abs/2012MNRAS.419.3060P} {419, 3060}

\bibitem[\protect\citeauthoryear{{P{\'e}roux} et~al.,}{{P{\'e}roux}
  et~al.}{2016}]{2016MNRAS.457..903P}
{P{\'e}roux} C.,  et~al., 2016, \mn@doi [\mnras] {10.1093/mnras/stw016}, \href
  {https://ui.adsabs.harvard.edu/abs/2016MNRAS.457..903P} {457, 903}

\bibitem[\protect\citeauthoryear{{P{\'e}roux} et~al.,}{{P{\'e}roux}
  et~al.}{2017}]{2017MNRAS.464.2053P}
{P{\'e}roux} C.,  et~al., 2017, \mn@doi [\mnras] {10.1093/mnras/stw2444}, \href
  {https://ui.adsabs.harvard.edu/abs/2017MNRAS.464.2053P} {464, 2053}

\bibitem[\protect\citeauthoryear{{P{\'e}roux} et~al.,}{{P{\'e}roux}
  et~al.}{2019}]{2019MNRAS.485.1595P}
{P{\'e}roux} C.,  et~al., 2019, \mn@doi [\mnras] {10.1093/mnras/stz202}, \href
  {https://ui.adsabs.harvard.edu/abs/2019MNRAS.485.1595P} {485, 1595}

\bibitem[\protect\citeauthoryear{{Petitjean}, {Ledoux}  \&
  {Srianand}}{{Petitjean} et~al.}{2008}]{2008A&A...480..349P}
{Petitjean} P.,  {Ledoux} C.,   {Srianand} R.,  2008, \mn@doi [\aap]
  {10.1051/0004-6361:20078607}, \href
  {https://ui.adsabs.harvard.edu/abs/2008A&A...480..349P} {480, 349}

\bibitem[\protect\citeauthoryear{{Pettini}, {Shapley}, {Steidel}, {Cuby},
  {Dickinson}, {Moorwood}, {Adelberger}  \& {Giavalisco}}{{Pettini}
  et~al.}{2001}]{2001ApJ...554..981P}
{Pettini} M.,  {Shapley} A.~E.,  {Steidel} C.~C.,  {Cuby} J.-G.,  {Dickinson}
  M.,  {Moorwood} A. F.~M.,  {Adelberger} K.~L.,   {Giavalisco} M.,  2001,
  \mn@doi [\apj] {10.1086/321403}, \href
  {https://ui.adsabs.harvard.edu/abs/2001ApJ...554..981P} {554, 981}

\bibitem[\protect\citeauthoryear{{Planck Collaboration} et~al.,}{{Planck
  Collaboration} et~al.}{2016}]{Planck15}
{Planck Collaboration} et~al., 2016, \mn@doi [\aap]
  {10.1051/0004-6361/201525830}, \href
  {https://ui.adsabs.harvard.edu/\#abs/2016A&A...594A..13P} {594, A13}

\bibitem[\protect\citeauthoryear{{Prochaska} \& {Wolfe}}{{Prochaska} \&
  {Wolfe}}{2009}]{Prochaska2009}
{Prochaska} J.~X.,  {Wolfe} A.~M.,  2009, \mn@doi [\apj]
  {10.1088/0004-637X/696/2/1543}, \href
  {https://ui.adsabs.harvard.edu/abs/2009ApJ...696.1543P} {696, 1543}

\bibitem[\protect\citeauthoryear{{Prochaska}, {O'Meara}  \&
  {Worseck}}{{Prochaska} et~al.}{2010}]{Prochaska2010}
{Prochaska} J.~X.,  {O'Meara} J.~M.,   {Worseck} G.,  2010, \mn@doi [\apj]
  {10.1088/0004-637X/718/1/392}, \href
  {https://ui.adsabs.harvard.edu/abs/2010ApJ...718..392P} {718, 392}

\bibitem[\protect\citeauthoryear{{Rafelski}, {Wolfe}, {Prochaska}, {Neeleman}
  \& {Mendez}}{{Rafelski} et~al.}{2012}]{2012ApJ...755...89R}
{Rafelski} M.,  {Wolfe} A.~M.,  {Prochaska} J.~X.,  {Neeleman} M.,   {Mendez}
  A.~J.,  2012, \mn@doi [\apj] {10.1088/0004-637X/755/2/89}, \href
  {https://ui.adsabs.harvard.edu/abs/2012ApJ...755...89R} {755, 89}

\bibitem[\protect\citeauthoryear{{Rafelski}, {Neeleman}, {Fumagalli}, {Wolfe}
  \& {Prochaska}}{{Rafelski} et~al.}{2014}]{2014ApJ...782L..29R}
{Rafelski} M.,  {Neeleman} M.,  {Fumagalli} M.,  {Wolfe} A.~M.,   {Prochaska}
  J.~X.,  2014, \mn@doi [\apjl] {10.1088/2041-8205/782/2/L29}, \href
  {https://ui.adsabs.harvard.edu/abs/2014ApJ...782L..29R} {782, L29}

\bibitem[\protect\citeauthoryear{{Rahmati} \& {Schaye}}{{Rahmati} \&
  {Schaye}}{2014}]{2014MNRAS.438..529R}
{Rahmati} A.,  {Schaye} J.,  2014, \mn@doi [\mnras] {10.1093/mnras/stt2235},
  \href {https://ui.adsabs.harvard.edu/abs/2014MNRAS.438..529R} {438, 529}

\bibitem[\protect\citeauthoryear{{Rahmati}, {Pawlik}, {Rai{\v{c}}evi{\'c}}  \&
  {Schaye}}{{Rahmati} et~al.}{2013a}]{2013MNRAS.430.2427R}
{Rahmati} A.,  {Pawlik} A.~H.,  {Rai{\v{c}}evi{\'c}} M.,   {Schaye} J.,  2013a,
  \mn@doi [\mnras] {10.1093/mnras/stt066}, \href
  {https://ui.adsabs.harvard.edu/abs/2013MNRAS.430.2427R} {430, 2427}

\bibitem[\protect\citeauthoryear{{Rahmati}, {Schaye}, {Pawlik}  \&
  {Rai{\v{c}}evi{\'c}}}{{Rahmati} et~al.}{2013b}]{2013MNRAS.431.2261R}
{Rahmati} A.,  {Schaye} J.,  {Pawlik} A.~H.,   {Rai{\v{c}}evi{\'c}} M.,  2013b,
  \mn@doi [\mnras] {10.1093/mnras/stt324}, \href
  {https://ui.adsabs.harvard.edu/abs/2013MNRAS.431.2261R} {431, 2261}

\bibitem[\protect\citeauthoryear{{Rahmati}, {Schaye}, {Bower}, {Crain},
  {Furlong}, {Schaller}  \& {Theuns}}{{Rahmati}
  et~al.}{2015}]{2015MNRAS.452.2034R}
{Rahmati} A.,  {Schaye} J.,  {Bower} R.~G.,  {Crain} R.~A.,  {Furlong} M.,
  {Schaller} M.,   {Theuns} T.,  2015, \mn@doi [\mnras]
  {10.1093/mnras/stv1414}, \href
  {https://ui.adsabs.harvard.edu/abs/2015MNRAS.452.2034R} {452, 2034}

\bibitem[\protect\citeauthoryear{{Rhodin}, {Agertz}, {Christensen}, {Renaud}
  \& {Fynbo}}{{Rhodin} et~al.}{2019}]{2019MNRAS.488.3634R}
{Rhodin} N.~H.~P.,  {Agertz} O.,  {Christensen} L.,  {Renaud} F.,   {Fynbo}
  J.~P.~U.,  2019, \mn@doi [\mnras] {10.1093/mnras/stz1479}, \href
  {https://ui.adsabs.harvard.edu/abs/2019MNRAS.488.3634R} {488, 3634}

\bibitem[\protect\citeauthoryear{{Ross} et~al.,}{{Ross}
  et~al.}{2009}]{2009ApJ...697.1634R}
{Ross} N.~P.,  et~al., 2009, \mn@doi [\apj] {10.1088/0004-637X/697/2/1634},
  \href {https://ui.adsabs.harvard.edu/abs/2009ApJ...697.1634R} {697, 1634}

\bibitem[\protect\citeauthoryear{{Salim} et~al.,}{{Salim}
  et~al.}{2007}]{2007ApJS..173..267S}
{Salim} S.,  et~al., 2007, \mn@doi [\apjs] {10.1086/519218}, \href
  {https://ui.adsabs.harvard.edu/abs/2007ApJS..173..267S} {173, 267}

\bibitem[\protect\citeauthoryear{{Schaye} et~al.,}{{Schaye}
  et~al.}{2015}]{2015MNRAS.446..521S}
{Schaye} J.,  et~al., 2015, \mn@doi [\mnras] {10.1093/mnras/stu2058}, \href
  {https://ui.adsabs.harvard.edu/abs/2015MNRAS.446..521S} {446, 521}

\bibitem[\protect\citeauthoryear{{Shapley}, {Steidel}, {Pettini}  \&
  {Adelberger}}{{Shapley} et~al.}{2003}]{2003ApJ...588...65S}
{Shapley} A.~E.,  {Steidel} C.~C.,  {Pettini} M.,   {Adelberger} K.~L.,  2003,
  \mn@doi [\apj] {10.1086/373922}, \href
  {https://ui.adsabs.harvard.edu/abs/2003ApJ...588...65S} {588, 65}

\bibitem[\protect\citeauthoryear{{Shen}, {Madau}, {Guedes}, {Mayer},
  {Prochaska}  \& {Wadsley}}{{Shen} et~al.}{2013}]{2013ApJ...765...89S}
{Shen} S.,  {Madau} P.,  {Guedes} J.,  {Mayer} L.,  {Prochaska} J.~X.,
  {Wadsley} J.,  2013, \mn@doi [\apj] {10.1088/0004-637X/765/2/89}, \href
  {https://ui.adsabs.harvard.edu/abs/2013ApJ...765...89S} {765, 89}

\bibitem[\protect\citeauthoryear{{Simcoe} et~al.,}{{Simcoe}
  et~al.}{2011}]{2011ApJ...743...21S}
{Simcoe} R.~A.,  et~al., 2011, \mn@doi [\apj] {10.1088/0004-637X/743/1/21},
  \href {https://ui.adsabs.harvard.edu/abs/2011ApJ...743...21S} {743, 21}

\bibitem[\protect\citeauthoryear{{Springel} et~al.,}{{Springel}
  et~al.}{2005}]{2005Natur.435..629S}
{Springel} V.,  et~al., 2005, \mn@doi [\nat] {10.1038/nature03597}, \href
  {https://ui.adsabs.harvard.edu/abs/2005Natur.435..629S} {435, 629}

\bibitem[\protect\citeauthoryear{{Steidel}, {Pettini}, {Dickinson}  \&
  {Persson}}{{Steidel} et~al.}{1994}]{1994AJ....108.2046S}
{Steidel} C.~C.,  {Pettini} M.,  {Dickinson} M.,   {Persson} S.~E.,  1994,
  \mn@doi [\aj] {10.1086/117217}, \href
  {https://ui.adsabs.harvard.edu/abs/1994AJ....108.2046S} {108, 2046}

\bibitem[\protect\citeauthoryear{{Stott et al.}}{{Stott et
  al.}}{prep}]{stottinprep}
{Stott et al.} {in prep}, \apj, 000, 000

\bibitem[\protect\citeauthoryear{{Str{\"o}bele} et~al.,}{{Str{\"o}bele}
  et~al.}{2012}]{2012SPIE.8447E..37S}
{Str{\"o}bele} S.,  et~al., 2012, in \procspie. p. 844737,
  \mn@doi{10.1117/12.926110}

\bibitem[\protect\citeauthoryear{{Tacchella}, {Dekel}, {Carollo}, {Ceverino},
  {DeGraf}, {Lapiner}, {Mand elker}  \& {Primack Joel}}{{Tacchella}
  et~al.}{2016}]{2016MNRAS.457.2790T}
{Tacchella} S.,  {Dekel} A.,  {Carollo} C.~M.,  {Ceverino} D.,  {DeGraf} C.,
  {Lapiner} S.,  {Mand elker} N.,   {Primack Joel} R.,  2016, \mn@doi [\mnras]
  {10.1093/mnras/stw131}, \href
  {https://ui.adsabs.harvard.edu/abs/2016MNRAS.457.2790T} {457, 2790}

\bibitem[\protect\citeauthoryear{{Tacconi} et~al.,}{{Tacconi}
  et~al.}{2013}]{2013ApJ...768...74T}
{Tacconi} L.~J.,  et~al., 2013, \mn@doi [\apj] {10.1088/0004-637X/768/1/74},
  \href {https://ui.adsabs.harvard.edu/abs/2013ApJ...768...74T} {768, 74}

\bibitem[\protect\citeauthoryear{{Timlin} et~al.,}{{Timlin}
  et~al.}{2018}]{Timlin2018}
{Timlin} J.~D.,  et~al., 2018, \mn@doi [\apj] {10.3847/1538-4357/aab9ac}, \href
  {https://ui.adsabs.harvard.edu/abs/2018ApJ...859...20T} {859, 20}

\bibitem[\protect\citeauthoryear{{Turner}, {Schaye}, {Steidel}, {Rudie}  \&
  {Strom}}{{Turner} et~al.}{2014}]{2014MNRAS.445..794T}
{Turner} M.~L.,  {Schaye} J.,  {Steidel} C.~C.,  {Rudie} G.~C.,   {Strom}
  A.~L.,  2014, \mn@doi [\mnras] {10.1093/mnras/stu1801}, \href
  {https://ui.adsabs.harvard.edu/abs/2014MNRAS.445..794T} {445, 794}

\bibitem[\protect\citeauthoryear{{Vanden Berk} et~al.,}{{Vanden Berk}
  et~al.}{2001}]{vandenberk01}
{Vanden Berk} D.~E.,  et~al., 2001, \mn@doi [\aj] {10.1086/321167}, \href
  {https://ui.adsabs.harvard.edu/abs/2001AJ....122..549V} {122, 549}

\bibitem[\protect\citeauthoryear{{Vernet} et~al.,}{{Vernet}
  et~al.}{2011}]{2011A&A...536A.105V}
{Vernet} J.,  et~al., 2011, \mn@doi [\aap] {10.1051/0004-6361/201117752}, \href
  {https://ui.adsabs.harvard.edu/abs/2011A&A...536A.105V} {536, A105}

\bibitem[\protect\citeauthoryear{{Wang} et~al.,}{{Wang}
  et~al.}{2016}]{Wang2016}
{Wang} F.,  et~al., 2016, \mn@doi [\apj] {10.3847/0004-637X/819/1/24}, \href
  {https://ui.adsabs.harvard.edu/abs/2016ApJ...819...24W} {819, 24}

\bibitem[\protect\citeauthoryear{{Weilbacher}, {Streicher}, {Urrutia},
  {P{\'e}contal-Rousset}, {Jarno}  \& {Bacon}}{{Weilbacher}
  et~al.}{2014}]{MUSEPipe2014}
{Weilbacher} P.~M.,  {Streicher} O.,  {Urrutia} T.,  {P{\'e}contal-Rousset} A.,
   {Jarno} A.,   {Bacon} R.,  2014, in {Manset} N.,  {Forshay} P.,  eds,
  Astronomical Society of the Pacific Conference Series Vol. 485, Astronomical
  Data Analysis Software and Systems XXIII. p.~451 (\mn@eprint {arXiv}
  {1507.00034})

\bibitem[\protect\citeauthoryear{{Yang} et~al.,}{{Yang}
  et~al.}{2016}]{Yang2016}
{Yang} J.,  et~al., 2016, \mn@doi [\apj] {10.3847/0004-637X/829/1/33}, \href
  {http://adsabs.harvard.edu/abs/2016ApJ...829...33Y} {829, 33}

\bibitem[\protect\citeauthoryear{{Yang} et~al.,}{{Yang}
  et~al.}{2017}]{2017AJ....153..184Y}
{Yang} J.,  et~al., 2017, \mn@doi [\aj] {10.3847/1538-3881/aa6577}, \href
  {https://ui.adsabs.harvard.edu/abs/2017AJ....153..184Y} {153, 184}

\bibitem[\protect\citeauthoryear{{van de Voort}, {Springel}, {Mandelker}, {van
  den Bosch}  \& {Pakmor}}{{van de Voort} et~al.}{2019}]{2019MNRAS.482L..85V}
{van de Voort} F.,  {Springel} V.,  {Mandelker} N.,  {van den Bosch} F.~C.,
  {Pakmor} R.,  2019, \mn@doi [\mnras] {10.1093/mnrasl/sly190}, \href
  {https://ui.adsabs.harvard.edu/abs/2019MNRAS.482L..85V} {482, L85}

\makeatother
\end{thebibliography}




\appendix
\section{Absorption Line Profiles}
\begin{figure*}
    \centering
    \includegraphics[width=0.90\columnwidth]{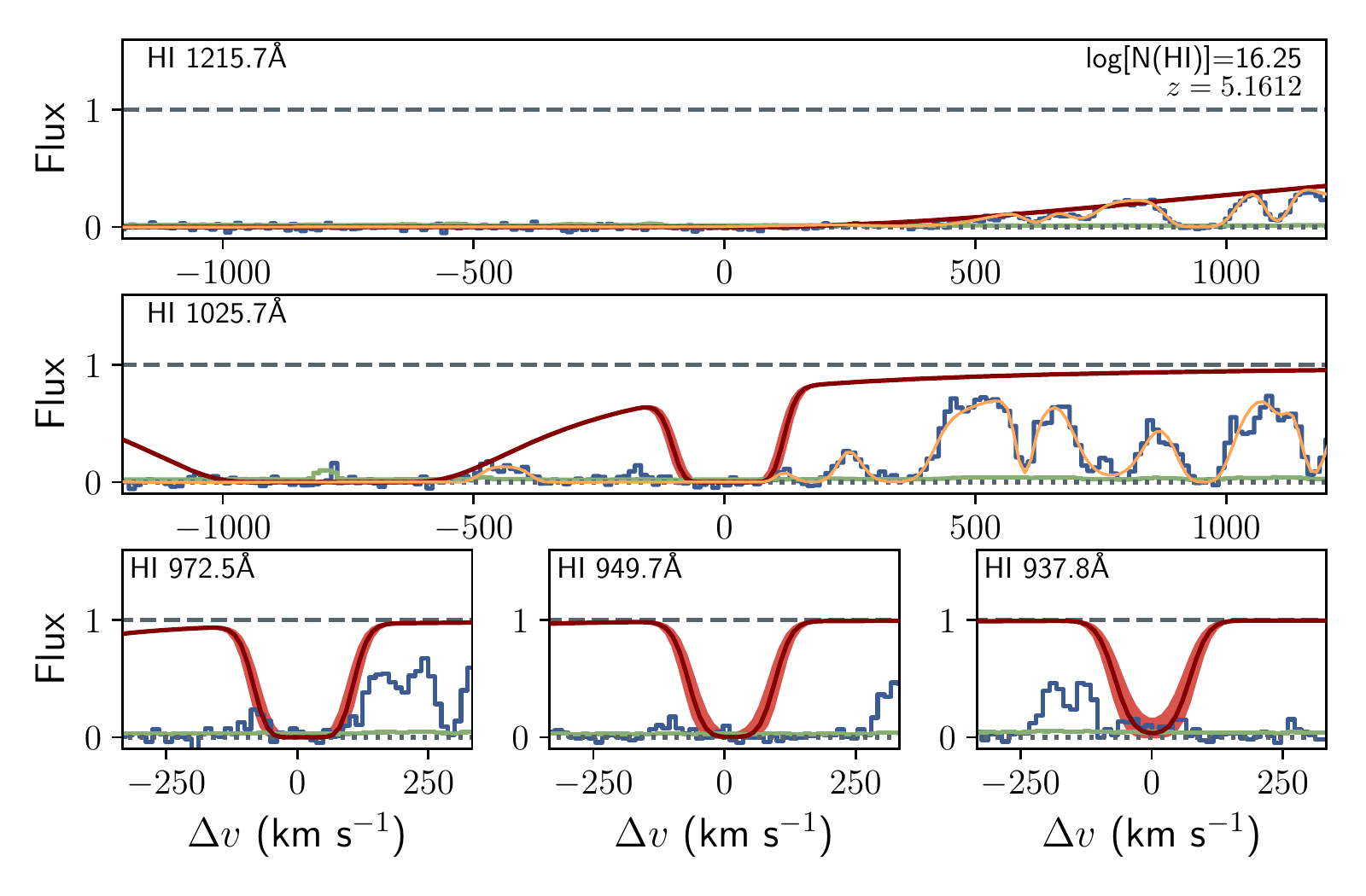}
    \includegraphics[width=0.90\columnwidth]{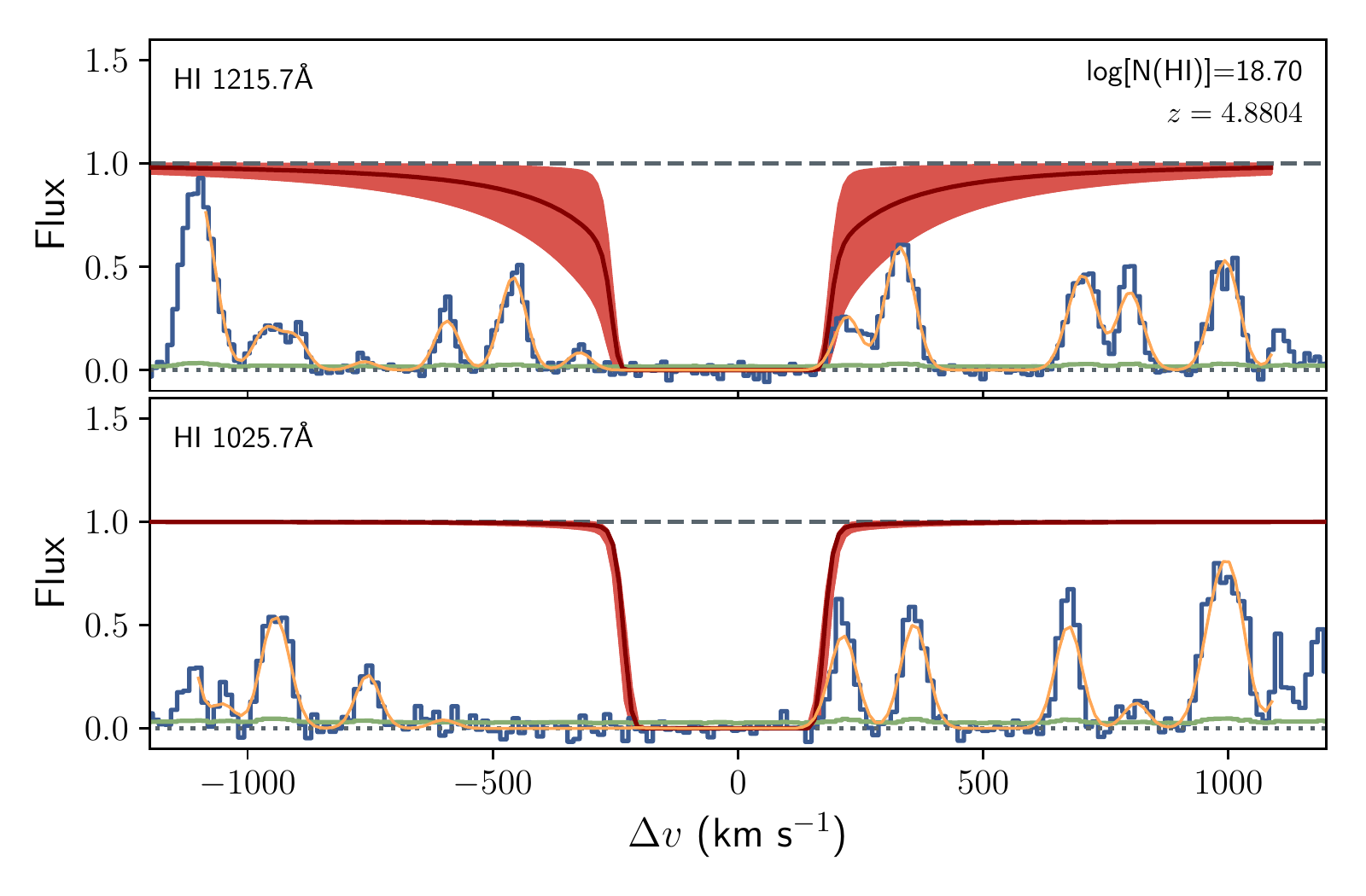}
    \includegraphics[width=0.90\columnwidth]{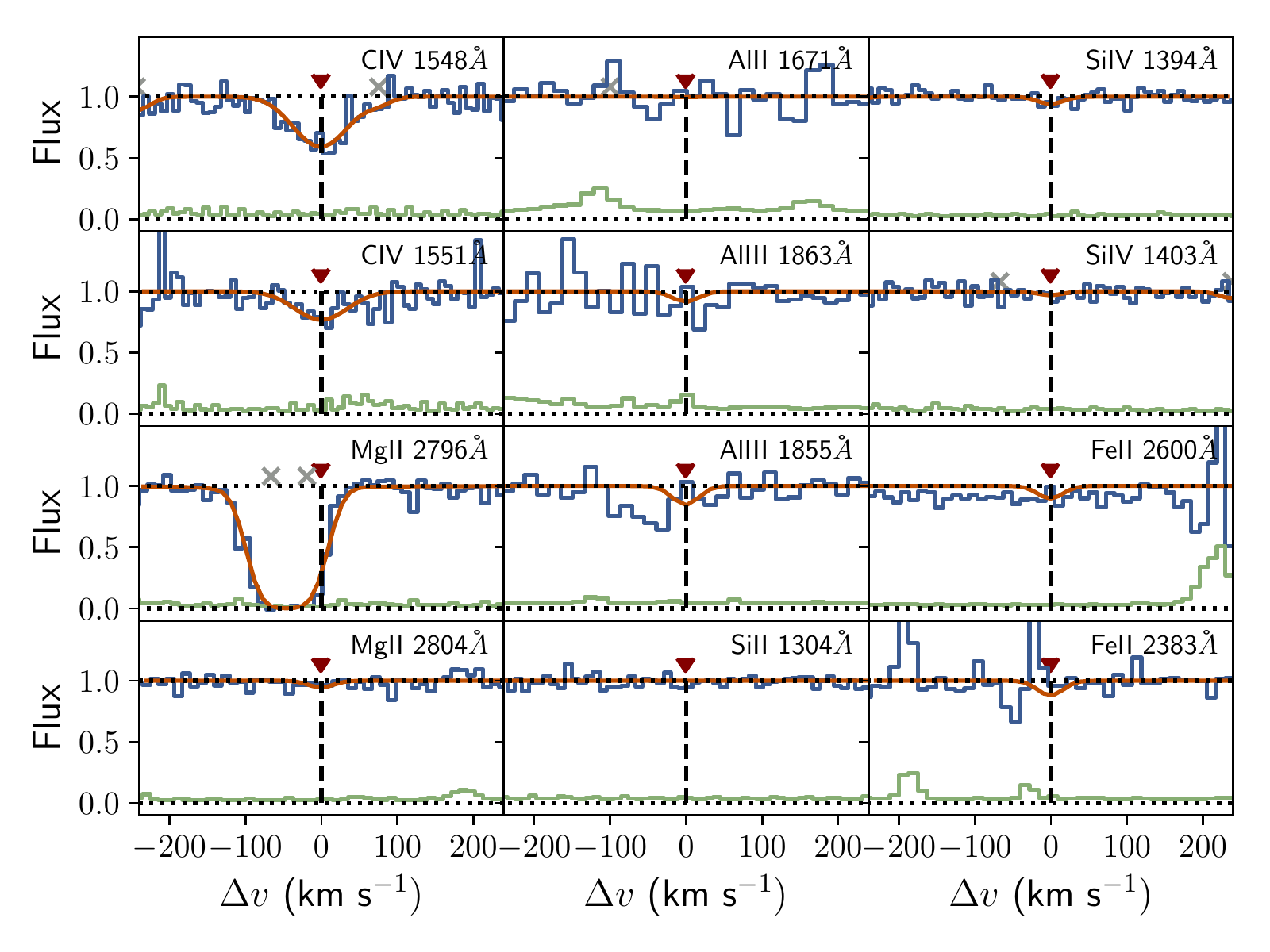}
    \includegraphics[width=0.90\columnwidth]{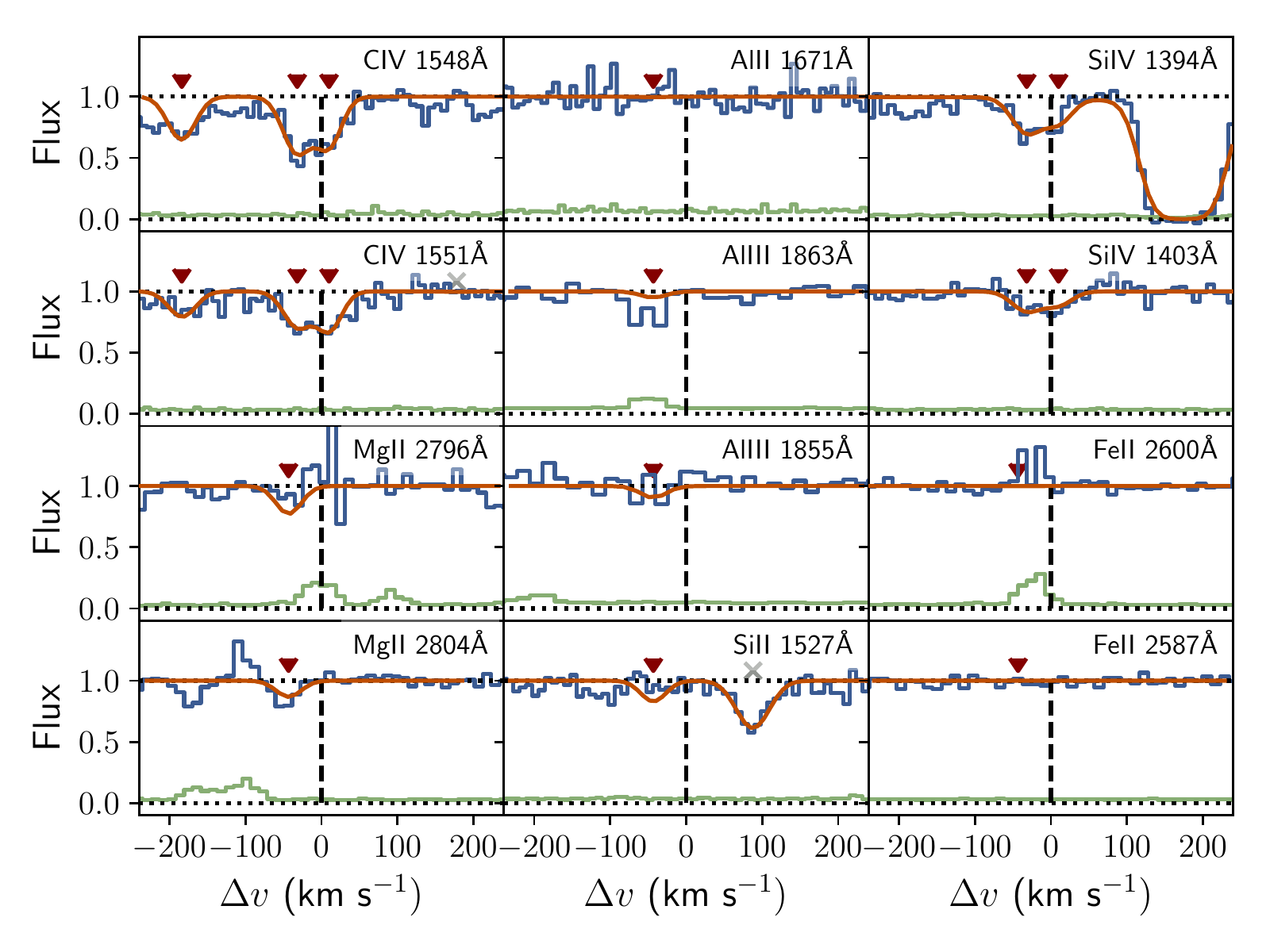}
    \caption{As in Fig.~\ref{fig:DLA-ALS_a}, but for systems at $z=5.16$ (left) and $z=4.88$ (right).}
    \label{fig:LLS-ALS_b}
\end{figure*}

\begin{figure*}
    \centering
    \includegraphics[width=0.90\columnwidth]{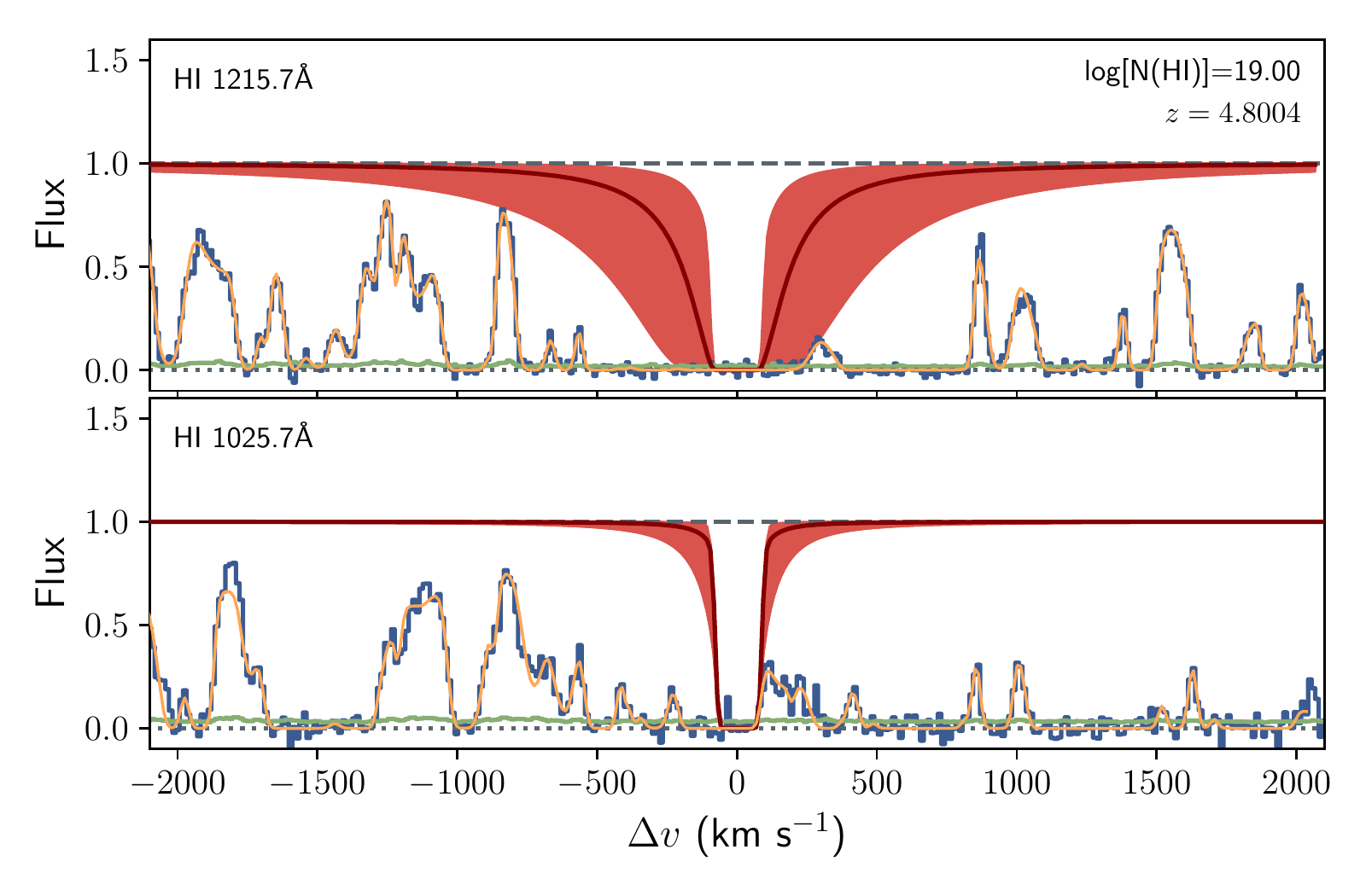}
    \includegraphics[width=0.90\columnwidth]{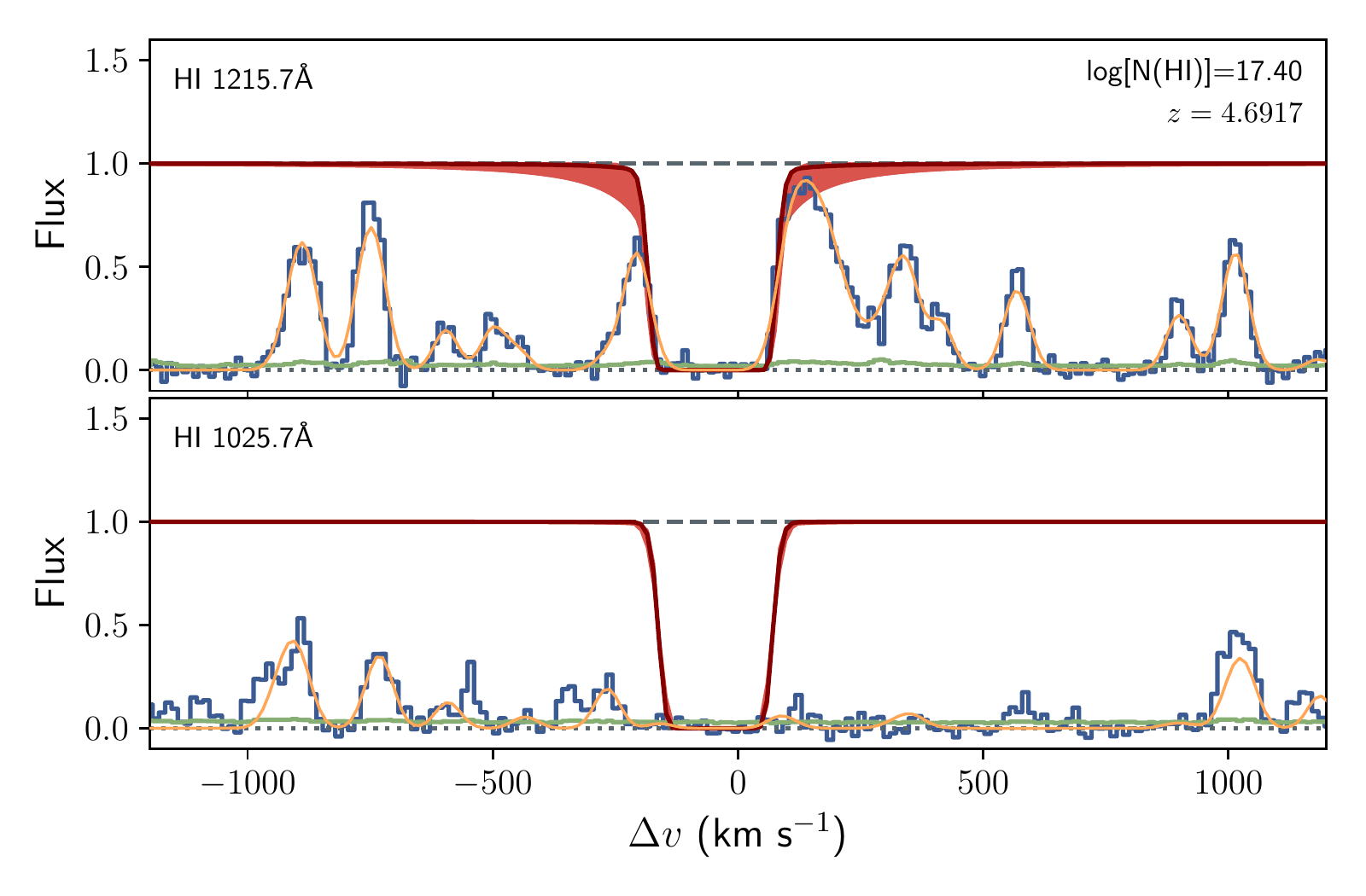}
    \includegraphics[width=0.90\columnwidth]{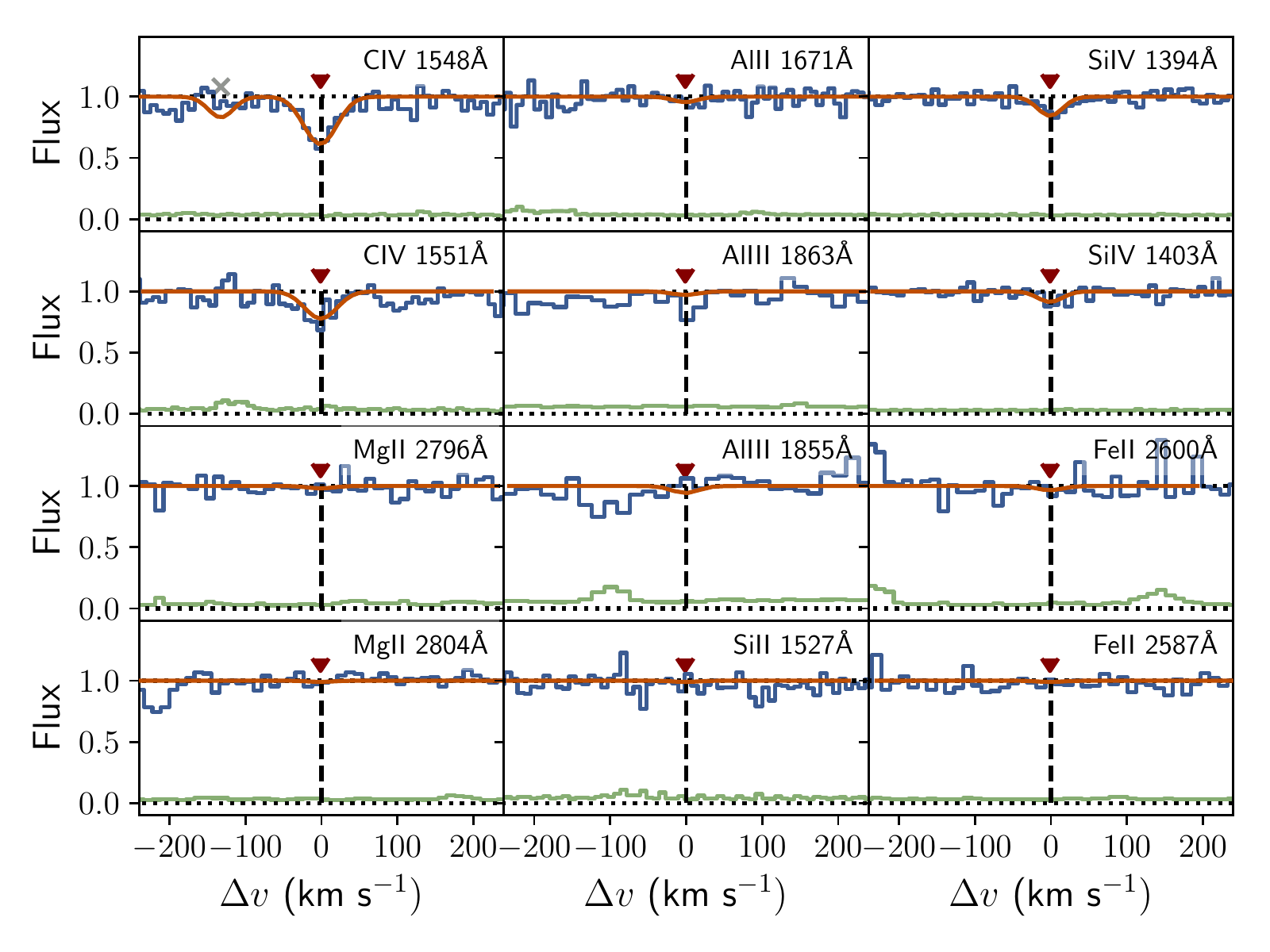}
    \includegraphics[width=0.90\columnwidth]{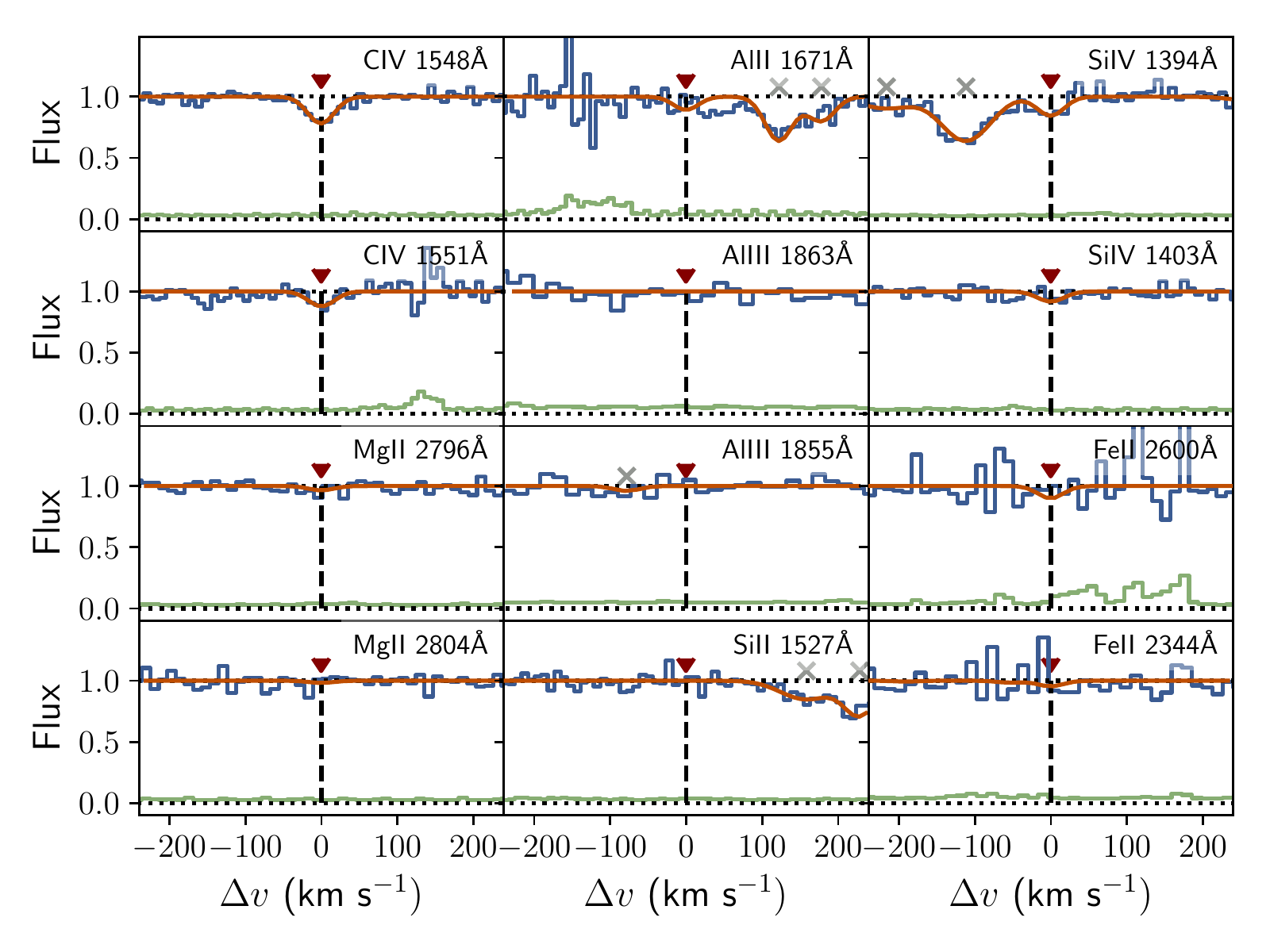}
    \caption{As in Fig.~\ref{fig:DLA-ALS_a}, but for systems at $z=4.80$ (left) and $z=4.69$ (right).}
    \label{fig:LLS-ALS_c}
\end{figure*}

\begin{figure*}
    \centering
    \includegraphics[width=0.90\columnwidth]{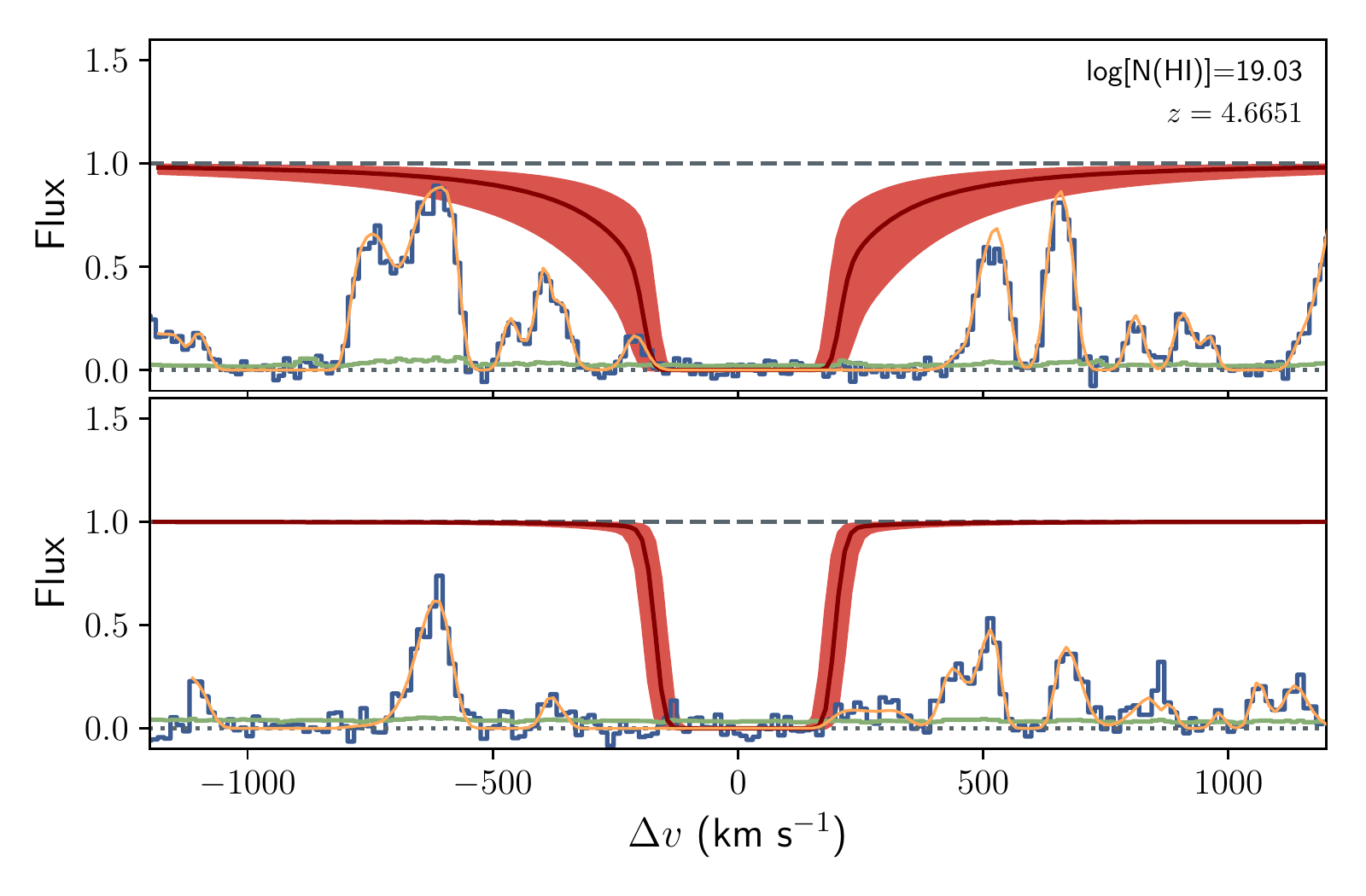}
    \includegraphics[width=0.90\columnwidth]{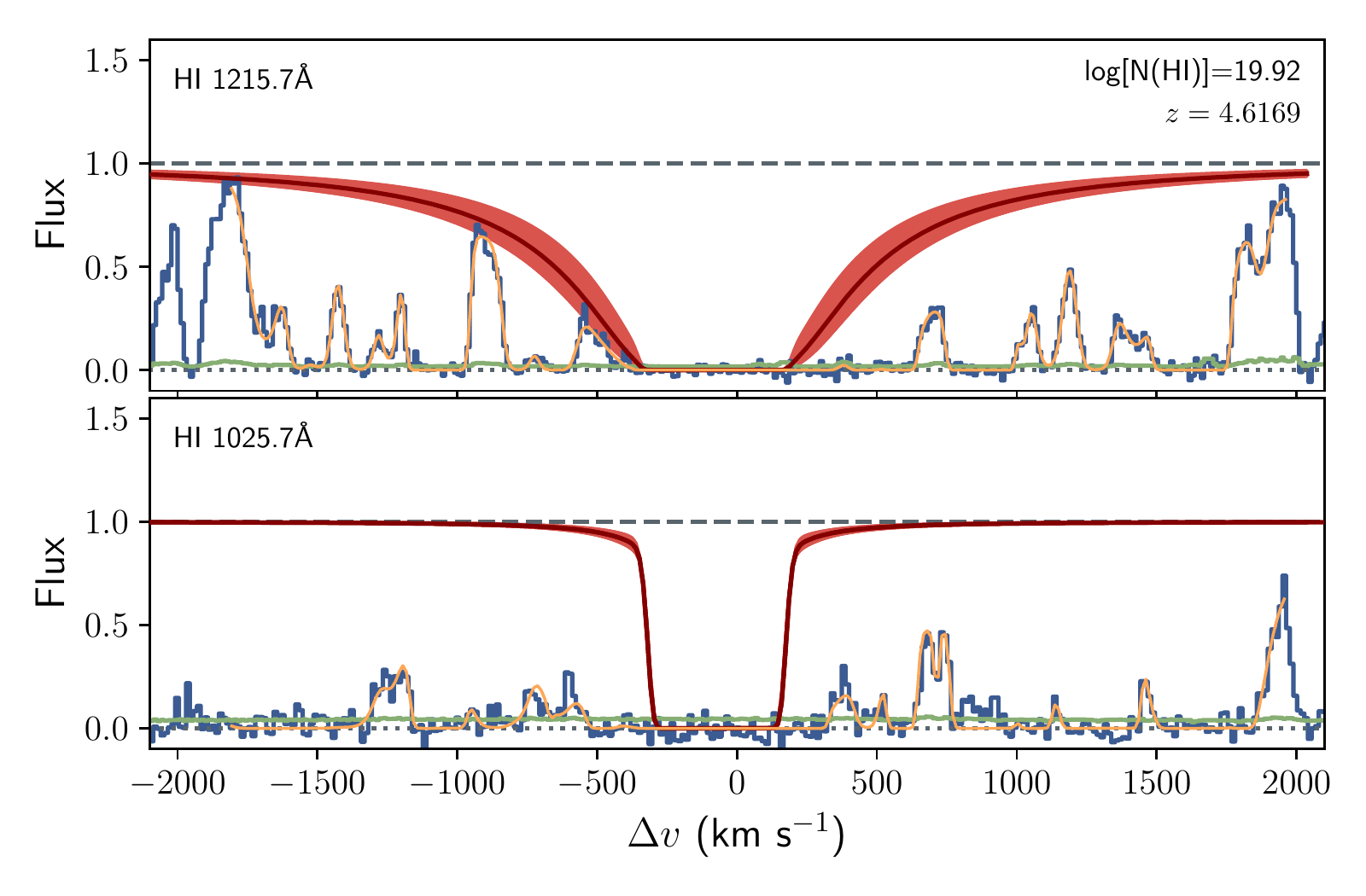}
    \includegraphics[width=0.90\columnwidth]{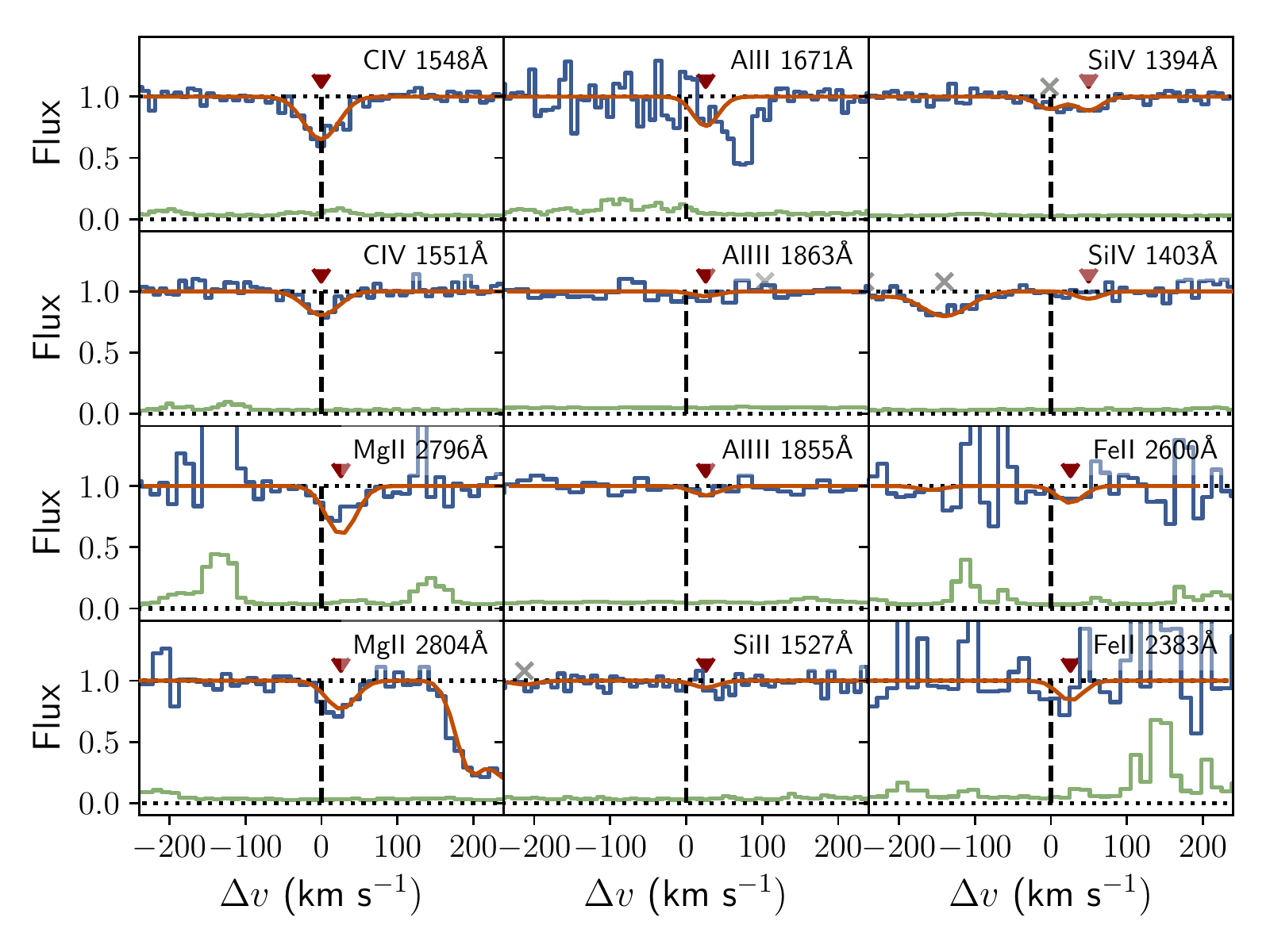}
    \includegraphics[width=0.90\columnwidth]{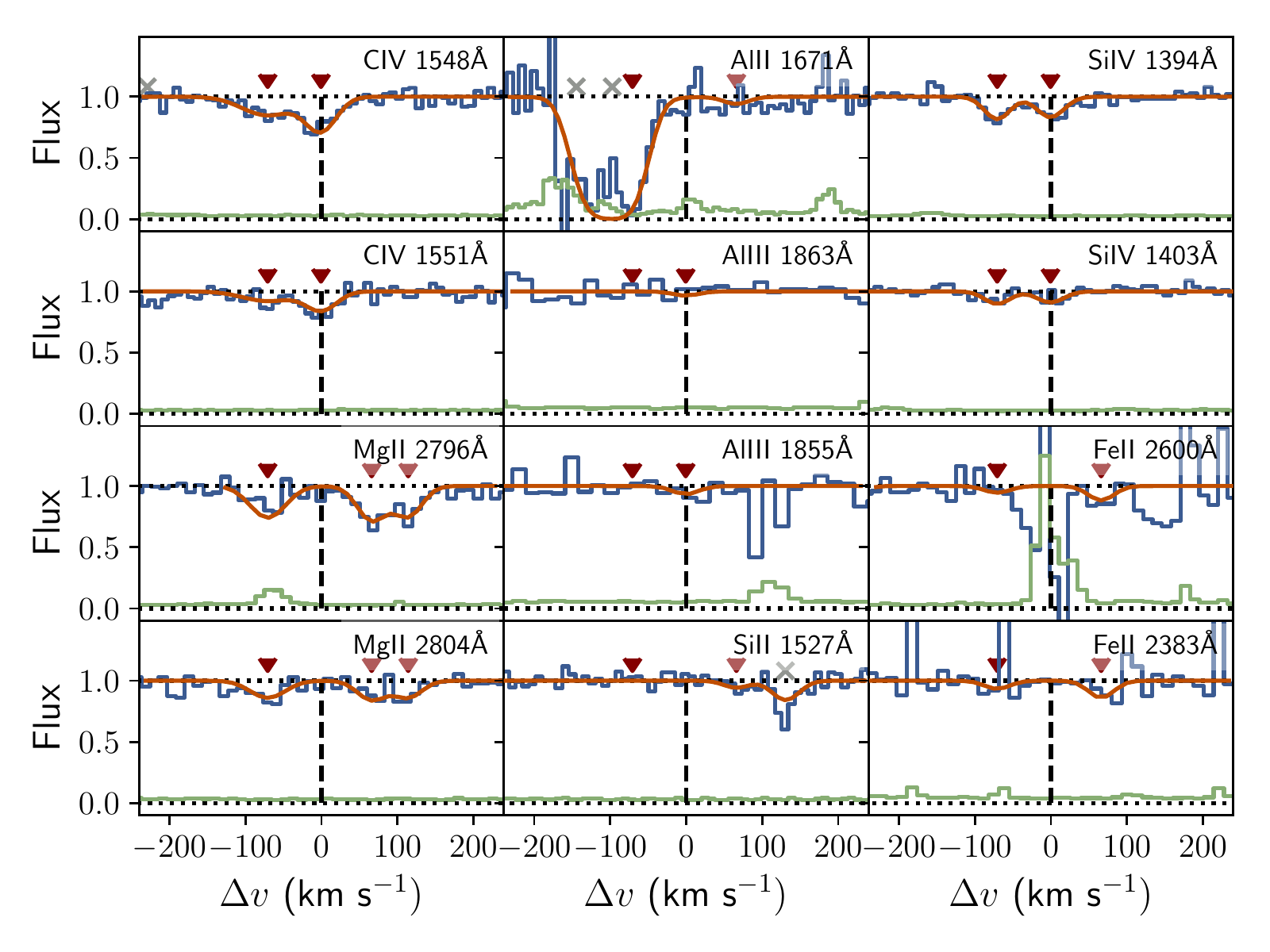}
    \caption{As in Fig.~\ref{fig:DLA-ALS_a}, but for systems at $z=4.67$ and $z=4.62$.}
    \label{fig:LLS-ALS_d}
\end{figure*}

\begin{figure*}
    \centering
    \includegraphics[width=0.90\columnwidth]{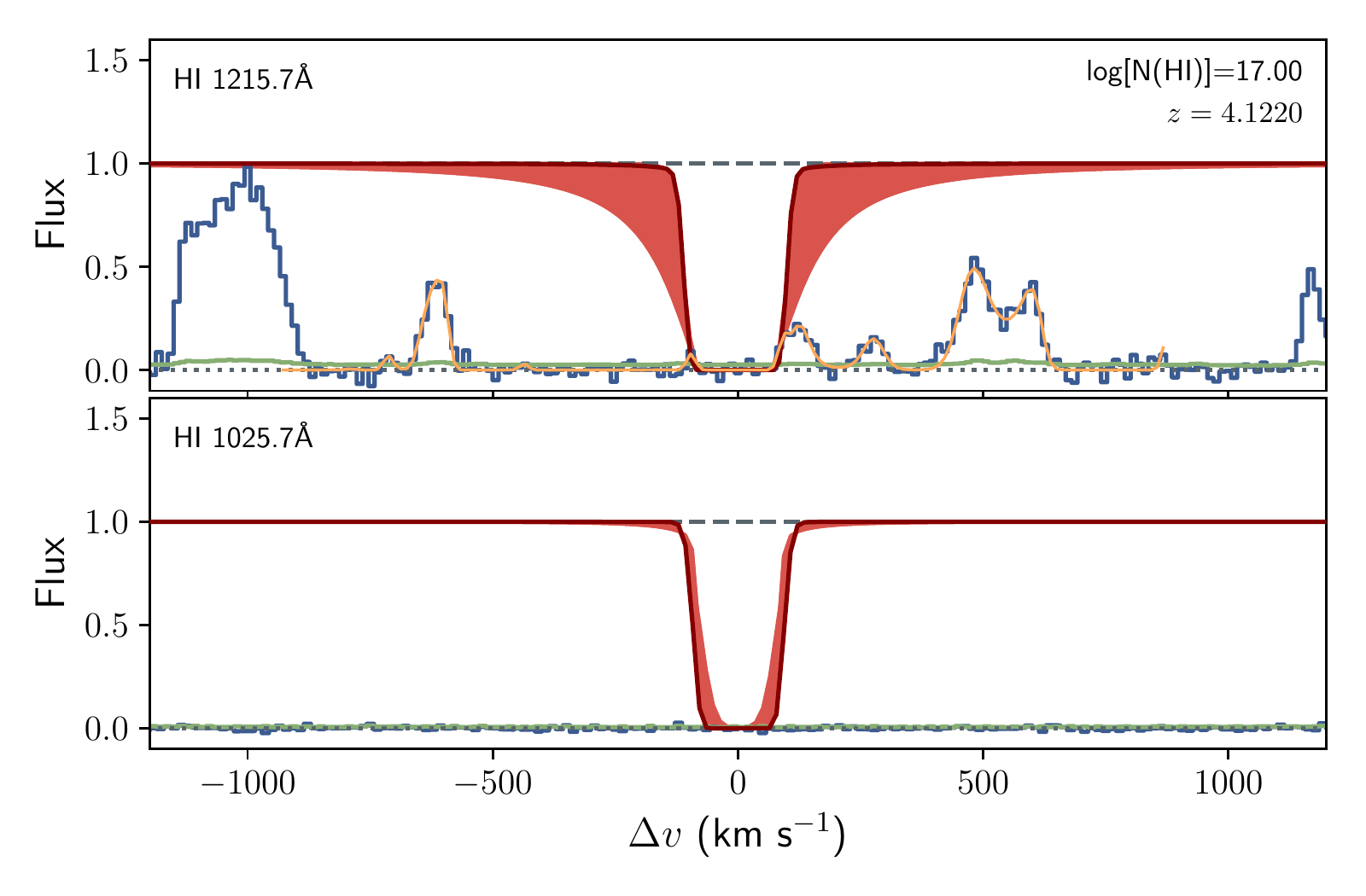}
    \includegraphics[width=0.90\columnwidth]{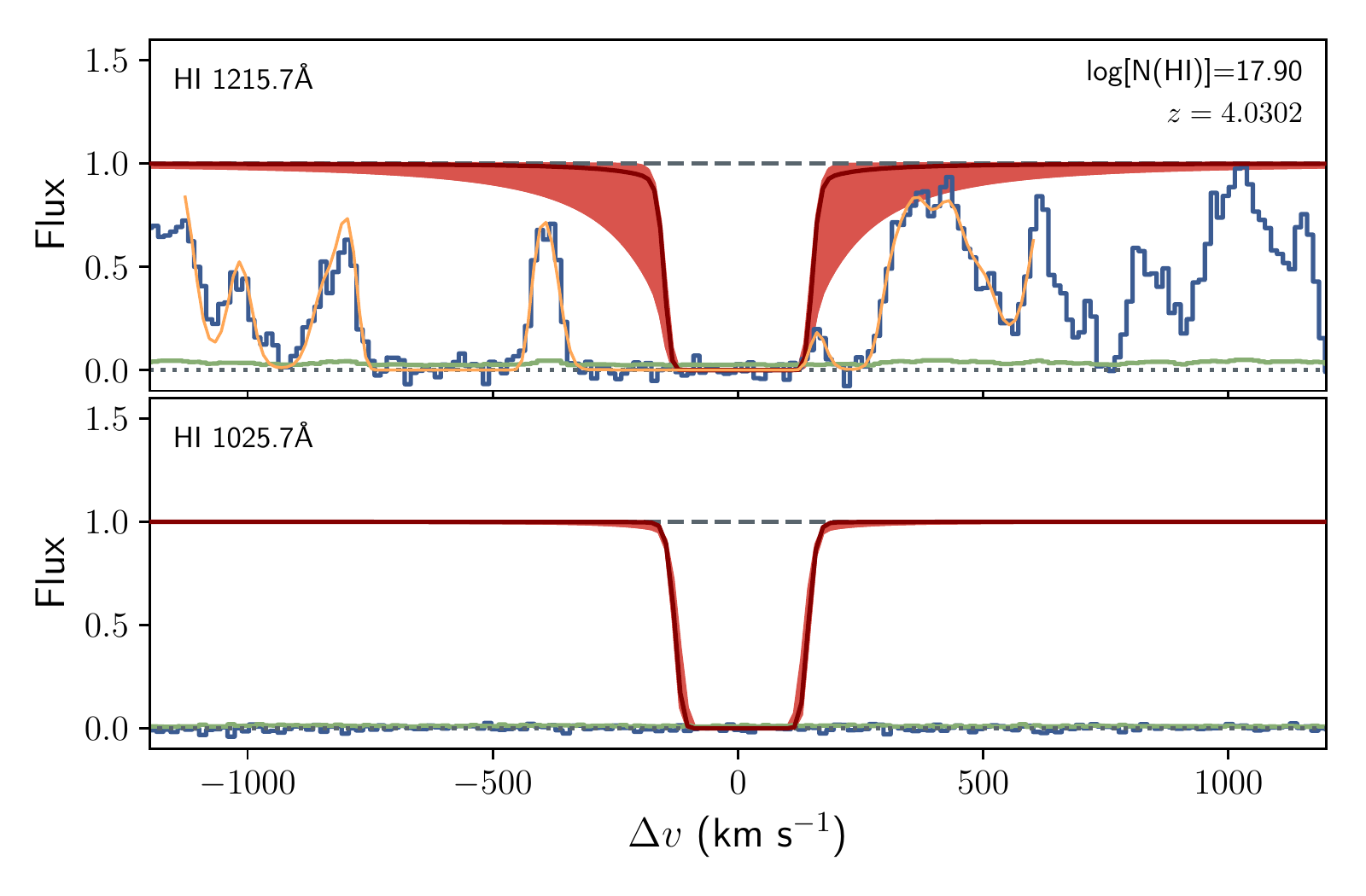}
    \includegraphics[width=0.90\columnwidth]{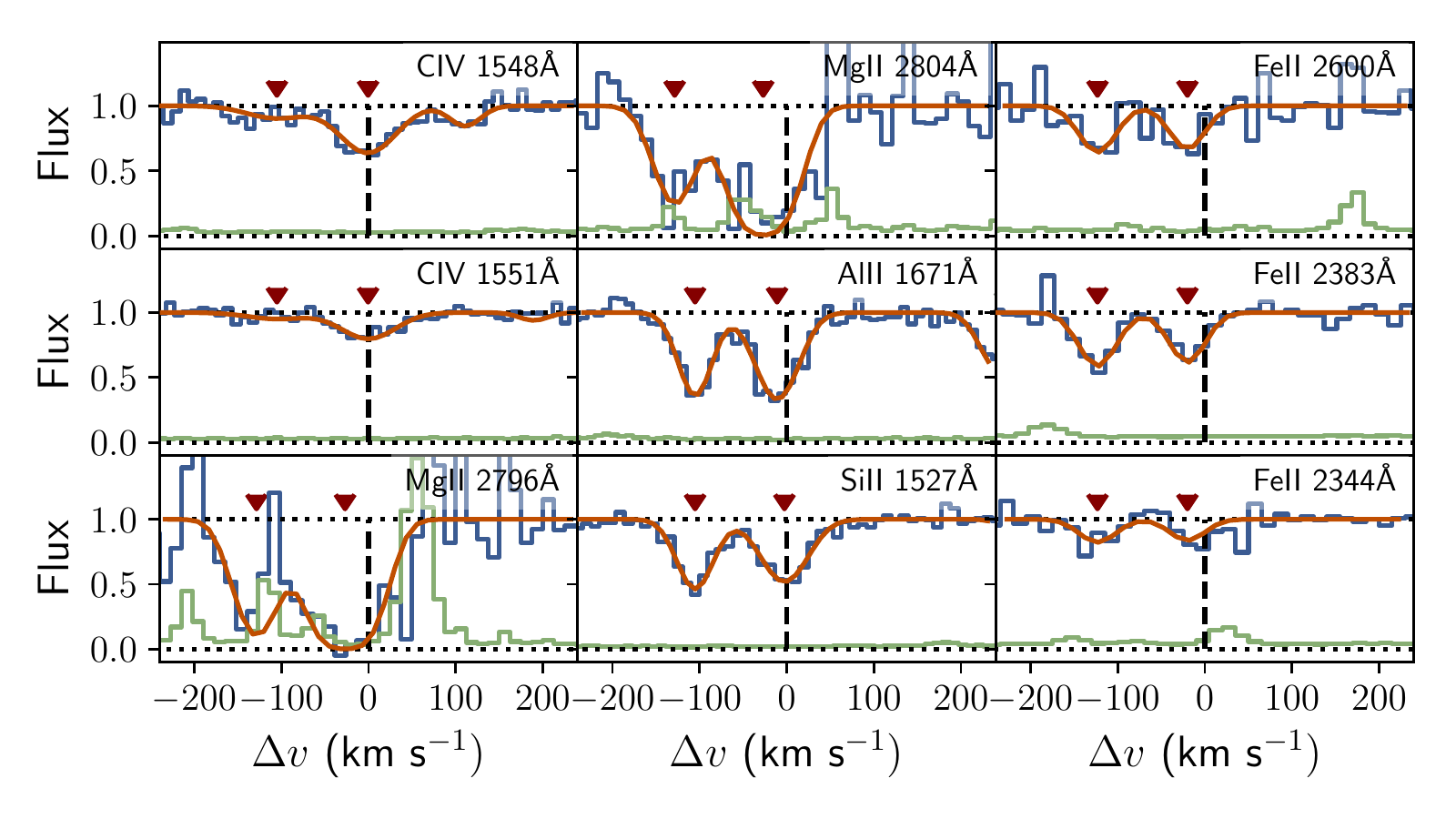}
    \includegraphics[width=0.90\columnwidth]{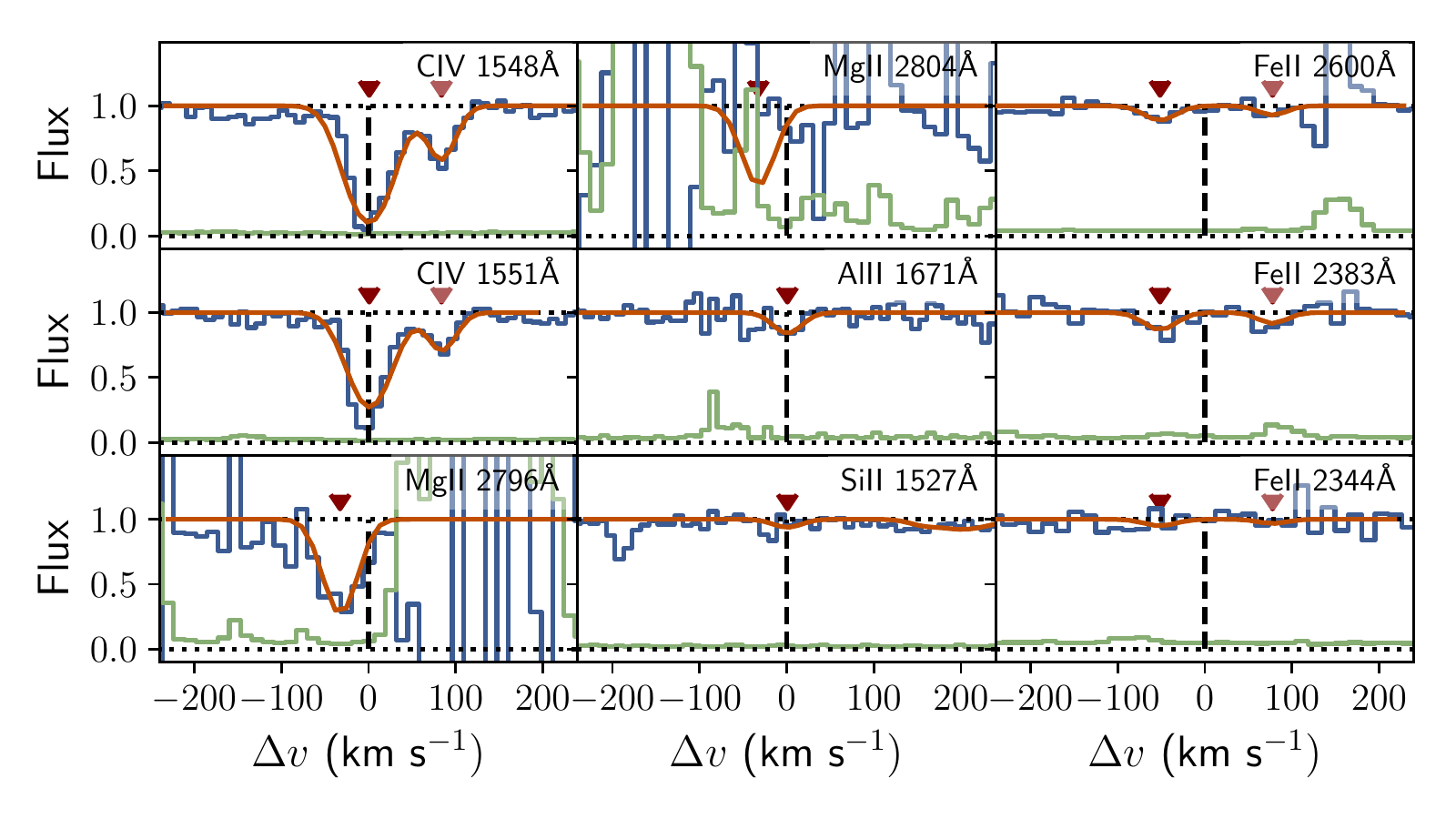}
    \caption{As in Fig.~\ref{fig:DLA-ALS_a}, but for systems at $z=4.12$ and $z=4.03$.}
    \label{fig:LLS-ALS_e}
\end{figure*}


\bsp	
\label{lastpage}
\end{document}